\documentclass{imsart}

\usepackage{amsthm,amsmath,amsfonts,amssymb,marginnote}
\usepackage{caption,subcaption}
\usepackage{graphicx,epstopdf}
\usepackage{multirow}
\usepackage[numbers,sort]{natbib}
\usepackage{fullpage}

\startlocaldefs
\theoremstyle{plain}

\newtheorem{cor}{Corollary}
\newtheorem{theorem}{Theorem}[section]
\newtheorem{lemma}[theorem]{Lemma}
\newtheorem{prop}{Proposition}
\newtheorem{res}{Result}

\theoremstyle{remark}
\newtheorem{definition}[theorem]{Definition}

\newtheorem{rmk}{Remark}
\newcommand{\RR}{\mathbb{R}}
\newcommand{\ZZ}{\mathbb{Z}}
\newcommand{\EE}{\mathbb{E}}
\newcommand{\PP}{\mathbb{P}}
\newcommand{\tr}{\mathrm{tr}}

\endlocaldefs

\begin{document}

\begin{frontmatter}
\title{Random line graphs and edge-attributed network inference}

\runtitle{Random Line Graphs}

\begin{aug}
\author[A]{\fnms{Zachary} \snm{Lubberts}\ead[label=e1]{zlubberts@virginia.edu}},
\author[B]{\fnms{Avanti} \snm{Athreya}\ead[label=e2,mark]{dathrey1@jhu.edu}},
\author[C]{\fnms{Youngser} \snm{Park}\ead[label=e3,mark]{youngser@jhu.edu}},
\and 
\author[B]{\fnms{Carey E.}
\snm{Priebe}\ead[label=e4,mark]{cep@jhu.edu}}
\address[A]{Department of Statistics, University of Virginia, \printead{e1}}

\address[B]{Department of Applied Mathematics and Statistics, Johns Hopkins University, \printead{e2,e4}}

\address[C]{Center for Imaging Science, Johns Hopkins University, \printead{e3}}

\end{aug}

\begin{abstract}
We extend the latent position random graph model to the {\em line graph of a random graph}, which is formed by creating a vertex for each edge in the original random graph, and connecting each pair of edges incident to a common vertex in the original graph. We prove concentration inequalities for the spectrum of a line graph, as well as limiting distribution results for the largest eigenvalue and the empirical spectral distribution in certain
settings. For the stochastic blockmodel, we establish that although naive spectral decompositions can fail to extract necessary signal for edge clustering, there exist signal-preserving singular subspaces of the line graph that can be recovered through a carefully-chosen projection. Moreover, we can consistently estimate edge latent positions in a random line graph, even though such graphs are of a random size, typically have high rank, and possess no spectral gap. Our results demonstrate that the line graph of a stochastic block model exhibits underlying block structure, and in simulations, we synthesize and test our methods against several commonly-used techniques, including tensor decompositions, for cluster recovery and edge covariate inference. By naturally incorporating information encoded in both vertices and edges, the random line graph improves network inference.

\end{abstract}



\end{frontmatter}

\section{Introduction}
\label{S:intro} 

Modern networks encode far more information than a simple list of vertex connections. Indeed, real-world network data include relevant context for the vertices in the network and information directly associated to the edges themselves. 
Such signal arises naturally, for example, in brain graphs \cite{durante2014bayesian,sporns_complex,MBStructure,takemura13:_drosop}. 
In such settings, though, it is less clear how to model unobserved characteristics of these random edges, estimate such characteristics or combine edge covariate information with graph structural information for broader inference.

In the current work, we address these questions by considering the random line graph, a graph structure that effectively transposes the roles of edge and vertex, and allows us to define the notion of an unobserved {\em latent position} for each edge in a random graph. We demonstrate how such edge latent positions can be consistently estimated from a single observation of the line graph, and show how to modify these estimates to incorporate edge covariates. 

Our approach  translates Hoff's well-known latent position model \cite{hoff_raftery_handcock,diaconis08:_graph_limit_exchan_random_graph,asta_cls} to the setting where both the edges and the vertices have latent positions.
Latent position random graphs enjoy a wide array of applications because they model an intrinsic feature of most networks: that connections between vertices 
depend on characteristics of the vertices themselves.
These characteristics are often well-represented as low-dimensional vectors, and the probabilities of connections between nodes are computed by the {\em link function} or {\em kernel} $\kappa(x,y)$ of the associated vectors. However, since these underlying vertex characteristics are typically unobserved, it is necessary to infer them from a realization of the adjacency matrix of the graph itself. 
Such inference depends on the complexity of the link function, but one widely flexible and geometrically interpretable model is the {\em random dot product graph} (RDPG), in which the link function is the inner product.  RDPGs of suitably high dimension can approximate any independent-edge random graph \cite{wolfe13:_nonpar}. Moreover, 
with appropriate constraints on the set of possible latent positions, spectral decompositions supply highly accurate and computationally simple latent position estimation for single and multi-layer networks \cite{sussman12:_univer,STFP-2011,airoldi13:_stoch,lyzinski13:_perfec, athreya2013limit,tang14:_semipar,tang14:_nonpar, paul2020spectral}.

Because latent positions are specifically formulated to encompass unobserved properties of vertices, the resulting networks lend themselves to models of graphs with vertex covariates. Spectral methods for networks with vertex covariates are considered in \cite{Rohe_bink_vogel}; ongoing research suggests that the methodology specified in \cite{arroyo2019inference}, which is a joint embedding for multiple networks with common underlying spectral structure, may prove even more effective for identifying signal in vertex covariates. 

 There are settings in which the inferential quantities of interest are instead properties of the (random) {\em edges} themselves, or where important graph covariates are more naturally related to the edges than to the vertices; see, for example, \cite{lelarge2013reconstruction}. Since our approach is to construct a graph in which the notion of latent positions for network edges is well-defined, we can exploit methodology for vertex attributes to address meaningful challenges in edge-related inference.
This is not the only option for addressing edge-associated signal: one could consider a tensor incorporating adjacency and edge covariate information into different layers, and study the decompositions of such an object as in \cite{zhang2018tensor}. This merits further study, and in our simulations, we compare the proposed methods to this tensor-based approach.

The central component of our edge-related inference is that of the \emph{line graph} \cite{gross2005graph}, which is constructed from the original graph by creating a vertex for each edge in the original graph, and connecting each pair of edges which were incident to the same vertex in the original graph. Because the vertices in the line graph correspond to the edges in the original graph, the line graph necessarily has a random vertex set, and represents an inversion of the traditional perspective---its adjacency matrix describes which vertices are incident to an edge, instead of which edges connect vertices. Since the data for any fixed edge in the original graph is encoded by the associated row of the adjacency matrix of the line graph, this naturally leads us to consider edges as vectors in a high-dimensional Euclidean space. Defining a latent position for each edge then amounts to finding a low-dimensional representation which preserves important characteristics of the edges, and more precisely, finding projections onto special subspaces which retain this information. 

Line graphs are an interesting object of study in their own right. Deterministic line graphs, and some of their important spectral properties, are considered in \cite{KSSC99, evans2009line,skiena_discrete}. Random line graphs are examined in \cite{liu2013random,manka2010clustering}, though these have no theoretical results for the spectra of such graphs. To the best of our knowledge, the present work is the first theoretical exploration of the spectra of random line graphs.

To any random graph, there is an associated line graph; the converse of this statement is also true for sufficiently large connected line graphs \cite{whitney1932line,lehot1974optimal}. As such, there is very minor loss of generality in considering the case where we observe a random graph and construct its line graph, rather than considering the case of observing a random line graph directly. On the other hand, if one has a reason to consider such an object, our results will still elucidate its spectrum and help explain the discrepancies between these and other kinds of random graphs.

Random line graphs and latent position random graphs on a fixed vertex set differ in important ways. First, random line graphs are of {\em random size}; this presents immediate challenges for the interpretation of mean matrices for the adjacency matrix of a random line graph and subsequent analysis of their spectral properties, such as singular subspaces. Second, adjacency matrices of random line graphs are typically of high rank and not easily approximated by lower-rank decompositions.  Therefore, many low-rank estimation techniques that work quite serviceably for latent position random graphs---such as latent position estimation or vertex clustering via truncated singular value decomposition---will fail if applied naively for random line graphs. In fact, the top few singular vectors of the adjacency matrix of the random line graph are typically not useful for accurate edge clustering (see Section~\ref{S:sim}).

One way to overcome the challenges of a line graph's random size is to regard its adjacency matrix as a {\em random projection} of the adjacency matrix of the line graph of the complete graph on the same set of vertices onto the rows and columns corresponding to those edges which appear in the original graph. This makes precise the notion of the mean of a matrix with random size. To address the complexities of rank, we consider projections of the random line graph adjacency matrix onto certain lower-dimensional singular subspaces. In the case of line graphs arising from a stochastic blockmodel, we find a projection onto a lower-dimensional subspace that preserves important signal and produces consistent estimates for the edge latent positions.

These methods represent a necessary step in the incorporation of edge-related data into statistical network estimation. 
Given the ubiquity of such data and the critical signal it contains, no comprehensive graph inference program can neglect such a data modality. Our theoretical framework of random line graphs and our computational techniques for analyzing their spectral decompositions pave the way towards a principled, robust, and genuinely data-driven paradigm for network inference. 

\subsection{Structure and overview}
\label{s:structure}

The paper is organized as follows. We open by discussing the spectra of adjacency matrices of random line graphs. Such spectral analysis requires contending with the inherently random size of these adjacency matrices.  The random size and accompanying lack of fixed order for the mean matrix render inapplicable off-the-shelf results on matrix concentration. We address this by formalizing a generative model for the random line graph with the key feature that a fixed order for the mean of the adjacency matrix arises as an immediate consequence. For this generative model, we prove foundational concentration inequalities for the spectrum of random line graphs (Theorem~\ref{theorem:concentrate}).

In Section~\ref{sec:er}, we prove a central limit theorem for the top eigenvalue of the line graph of an Erd\"os-R\'enyi random graph, replicating the classical F\"uredi-Koml\'os Central Limit Theorem in the setting of line graphs. Furthermore, we determine the asymptotic empirical spectral distribution of a broader class of random matrices with dependent entries, as a corollary of which we obtain the limiting empirical spectral distribution for an ER line graph. Our proof of Theorem~\ref{theorem:limitdist} resolves through elementary methods that the limiting spectral distribution of the line graph matches the free convolution given in \cite{ding2010spectral} for the limiting spectral distribution of the Laplacian in this setting.

In Section~\ref{S:sbm}, we make precise the notion that the line graph of a stochastic blockmodel (which we term an {\em SBM line graph}) is itself stochastic blockmodel-like: Specifically, the mean of its adjacency matrix has an invariant subspace spanned by a matrix whose columns indicate the cluster memberships for the edges, just like the stochastic blockmodel (Proposition~\ref{prop:sbmlike}). We also discuss the important differences between the stochastic blockmodel and its line graph, explaining the meanings of two kinds of large eigenvalues of the line graph adjacency matrix. Since this particular invariant subspace of the mean exactly identifies the cluster memberships of the edges, when we find a consistent estimator for this subspace, this will be a signal-preserving subspace in the sense of Section~\ref{s:not}. Furthermore, identifying this particular invariant subspace of the mean of the line graph adjacency matrix allows us to take the first step in addressing the issue of high rank for random line graph adjacency matrices.

Next, we consider the corresponding projection for the adjacency matrix of an SBM line graph, assuming the true cluster memberships are known (we later address the case where these are only imperfectly known). It turns out that the nonzero entries of this random projection matrix are dependent random variables which follow what we coin the ``damped binomial" distribution; in the supplement, we prove several properties of this distribution which may be of independent interest. We then demonstrate that the projected random line graph adjacency matrix concentrates around the corresponding projection of its mean (Proposition~\ref{prop:hmat}), and that the first few left singular vectors of the projected adjacency matrix are consistent estimators for those of the mean (Theorem~\ref{theorem:singvecs}). This implies that the left singular vectors of this projected random line graph adjacency matrix span a \emph{signal-preserving subspace for edge clustering}, a term which we define precisely in Section~\ref{s:not}. This is not a property enjoyed by most of the subspaces we might typically employ for the goal of clustering the vertices of a random graph: for instance, this property does not hold for the subspace spanned by the top few eigenvectors of the random line graph adjacency matrix, as we demonstrate in Section~\ref{S:sim}. It is precisely because we have this signal-preserving property that we think of these left singular vectors of the projected adjacency matrix as estimates for the edge latent positions, one of the fundamental goals of this work.

In Section~\ref{S:sim}, we show simulations of our methods for clustering the edges of a stochastic blockmodel. In particular, we show how even an imperfect initial estimate of the node memberships, combined with our signal-recovering projection and the edge covariate information, leads to better performance over using only the adjacency information; using only the edge covariate information; or using a combination of these {\em without} the signal-recovering projection. We integrate the information from the adjacency matrix of the line graph with that of the edge covariates by following  \cite{arroyo2019inference}, which provides an estimate for the eigenvectors of random matrices whose means share a common invariant subspace. Proofs of all major results are contained in the supplement\cite{lubberts_et_al_supplement}. We conclude with a discussion of future work.

\subsection{Notation}
\label{s:not}

We denote the space of matrices of size $n_1\times n_2$ with entries in $S$ by $M_{n_1,n_2}(S).$ Where appropriate, we denote the eigenvalues and singular values of a matrix $A$ in nonincreasing order by $\lambda_1(A)\geq \lambda_2(A)\geq\cdots,$ $\sigma_1(A)\geq \sigma_2(A)\geq\cdots$, respectively. We denote the set of eigenvalues by $\sigma(A)$. For a positive integer $n$, we denote by $[n]$ the set $\{1,\ldots,n\}$.

Given a graph $G=([n],E)$, where $E\subseteq\binom{[n]}{2}=\{\{i,j\}: i,j\in[n], i\neq j\}$ denotes the edges, we denote its adjacency matrix as $A(G)$.

\begin{definition}
Let $G=([n],E)$ be a graph. The \emph{line graph of }$G$ is the graph $L(G)=(E,\mathcal{E}),$ where $\{\{i,j\},\{r,s\}\}\in\mathcal{E}\subseteq\binom{E}{2}$ whenever $|\{i,j\}\cap \{r,s\}|=1$. That is, the vertices of the line graph are the edges in the original graph, with edges in the line graph between any two edges which are incident to the same vertex in the original graph.
\end{definition}

As such, line graphs are a case of random intersection graphs \cite{KSSC99}, though they have additional structure in our setting. We have a particular interest in the case where the original graph comes from the following model:

\begin{definition}
A random graph $G$ is said to come from a \emph{stochastic blockmodel}, or \emph{SBM}, when edges appear independently in $G$, and there is a partition of $[n]$ as $C_1,\ldots,C_k$ such that $\PP[\{i,j\}\in E]=\mathcal{P}_{i,j}=B_{r,s}$ when $i\in C_r$ and $j\in C_s$, for some symmetric matrix $B\in M_{k}([0,1])$. Letting $Z\in M_{n,k}(\{0,1\})$ be a matrix with one nonzero entry in each row, this means that $\mathcal{P}$ factorizes as $ZBZ^T$.
\end{definition}

This definition of clustering concerns the vertices of $G$, but we might instead consider a clustering of the edges. As we will soon show, not all of the singular subspaces of the adjacency matrix for a random line graph preserve the signal necessary for identification of these edge clusters. To distinguish the subspaces which do preserve this signal, we use the following definition:

\begin{definition}
Let $U$ be a matrix whose rows are constant for edges belonging to the same cluster, and distinct for any two edges which belong to different clusters. In other words, clustering the rows of $U$ leads to a perfect identification of the edge cluster memberships. We say that $\hat{U}$ is \emph{signal-preserving (for edge clustering)} when there exist $c_1,c_2,\alpha,\beta>0$ such that with probability at least $1-\alpha n^{-c_1}$, there exists an orthogonal matrix $\mathcal{O}$ such that $$\|\hat{U}-U\mathcal{O}\|_F\leq \beta n^{-c_2}.$$
\end{definition}

When $\hat{U}$ is signal-preserving and $n$ is large enough, we see that clustering based on the rows of $\hat{U}$ will (with high probability) give a perfect assignment of the edge cluster memberships.

\section{Eigenvalues of random line graph adjacency matrices}
\label{S:linegraphs}

Our approach for incorporating edge covariate information, as discussed in the introduction, is to extend the notion of latent positions for vertices of a random graph to the case where the edges are allowed to have latent positions. We can then combine edge covariate information with these latent positions to get more informative positions associated to each edge, empowering further inference. This depends on a clear understanding of the singular subspaces of the adjacency matrices of random line graphs. To prove any kind of spectral convergence results requires a well-defined notion of the mean of the necessarily randomly-sized adjacency matrix of the random line graph. 
Once we define the mean of the random line graph, we show that certain of the eigenvalues of the random line graph adjacency matrix concentrate. This is the foundation of our theory for the spectra of random line graphs, which we will build on in subsequent sections. Our results on the eigenvalues of the random line graph adjacency matrix may be proven without assuming any special structure on the probabilities of edge formation in the graph: We only assume that edges appear in the graph independently of one another. As such, these results have broad applicability even when the original graph of interest does not exhibit stochastic blockmodel structure.

In later sections, we elucidate the structure of the singular vectors of the random line graph adjacency matrix when the random graph comes from a stochastic blockmodel.

\subsection{The mean of the adjacency matrix of a random line graph}
\label{s:mean}

Given that the set of edges which appear in $G$ is random, it is clear that the vertex set of the line graph is random, and therefore the adjacency matrix of $L(G)$ will have a random size. Even conditioning on a certain size will not resolve this issue completely, since the rows and columns for one vertex set and another will typically not be in correspondence, even when they are of the same order.

To address this, we consider the adjacency matrix of the line graph as a random submatrix of the adjacency matrix of the line graph of the complete graph on the same set of vertices as the original, $A(L(K_n))$. This means that every pair $\{i,j\}$ has a given row and column associated to it, which will always appear even if $\{i,j\}$ happens not to appear as an edge in $G$. In this case, we simply zero out the associated row and column of $A(L(K_n))$. This operation may be carried out using a diagonal projection matrix $P=\mathrm{diag}(A(G)_{i,j}),$ which on its diagonal lists all of the entries of $A(G)$ with $1\leq i<j\leq n$ in the same order as the indices of $A(L(K_n))$. Since $A(G)$ has binary entries, when $\{i,j\}\in E$, $A(G)_{i,j}=1$, and otherwise this entry is 0. Thus, $$[PA(L(K_n))P]_{\{i,j\},\{r,s\}}=\begin{cases} A(L(K_n))_{\{i,j\},\{r,s\}}& \{i,j\},\{r,s\}\in E\\0&\text{otherwise}.\end{cases}$$ Moreover, $A(L(K_n))_{\{i,j\},\{r,s\}}=1$ if and only if $\{i,j\}$ and $\{r,s\}$ are incident to a common vertex in $[n]$, which means they will be incident to a common vertex in $G$. As such, the submatrix of $PA(L(K_n))P$ with row and column indices given by the set of entries with $A(G)_{i,j}=1$ is nothing more than $A(L(G))$, the adjacency matrix of the line graph of the random graph $G$.

Having represented $A(L(G))$ by $PA(L(K_n))P$ for the corresponding diagonal projection matrix $P$, we can now compute its mean. Independence of edge formation in the graph $G$ ensures that the random variables $A(G)_{i,j}$ and $A(G)_{r,s}$ are independent when $\{i,j\}\neq\{r,s\}$. Since $A(L(K_n))$ has 0 entries along its diagonal, this independence implies that  $\EE[PA(L(K_n))P]=\EE[P]A(L(K_n))\EE[P],$ where $\EE[P]=\mathrm{diag}(\EE[A(G)_{i,j}])=\mathrm{diag}(\mathcal{P}_{i,j}).$ This matrix, then, is the mean of the adjacency matrix for the random line graph.

\subsection{Eigenvalue concentration of the random line graph adjacency matrix}
\label{s:eigs}

There are many possible models one could consider for $A(L(G))$; our model is a judicious choice that makes spectral analysis both tractable and inferentially useful. 
We begin by connecting the spectra 
of $A(G)$ and $A(L(G))$ when $G$ is a fixed graph. It turns out that the nonzero eigenvalues of $A(L(G))+2I$ are the same as the nonzero eigenvalues of $A(G)+D$, where $D$ is the diagonal matrix with the degrees of the vertices of $G$ on its diagonal. This means that determining the nontrivial eigenvalues of $A(L(G))$ is immediate when we have a fixed graph $G$. In this lemma, we denote the multiplicity of an eigenvalue by \emph{mult}.

\begin{prop}
\label{prop:lgeigs}
Let $G=([n],E)$ be a graph, $L(G)=(E,\mathcal{E})$ its line graph, and $\widehat{m}=|E|$. Then $$\sigma(A(L(G)))\begin{cases}=(\sigma(A(G)+D)-2)\cup\{-2\,(\text{mult. }\widehat{m}-n)\}&\text{if }\widehat{m}\geq n,\text{ and}\\\subset \sigma(A(G)+D)-2&\text{if }\widehat{m}<n,\end{cases}$$ where $D$ is the diagonal matrix with $D_{i,i}=\mathrm{deg}_{G}(i), 1\leq i\leq n$. In the latter case, $$(\sigma(A(G)+D)-2)\setminus\sigma(A(L(G)))=\{-2\,(\text{mult. }n-\widehat{m})\}.$$
\end{prop}

This result follows from the known decomposition $A(L(G))=B^TB-2I,$ where $B$ is the incidence matrix of the graph $G$ \cite[pg.~136]{skiena_discrete}, but we include its proof in the in the supplement for completeness.

Of course, given our definition for the mean of $A(L(G))$, one case that is of particular interest is when $G=K_n$. The next corollary describes this matrix's eigenvalues, and is proved alongside the preceding proposition.

\begin{cor}
\label{c:lgcompleteeigs}
$$\sigma(A(L(K_{n})))=\left\{2n-4,n-4\,(\text{mult. }n-1),-2\,(\text{mult. }\binom{n}{2}-n)\right\}.$$
\end{cor}

This proposition has important implications. First, the nontrivial eigenvalues of the line graph are exactly a translation of the eigenvalues of $A(G)+D$. In the case when $\widehat{m}\geq n$, this means that if $\widehat{m}$ grows in size, none of the new eigenvalues in the spectrum of $A(L(G))+2I$ will be nonzero, but all of the important convergence occurs in the relationship between $\sigma(A(G)+D)$ and $\sigma(A(K_{n})+(n-1)I).$ Second, we see that comparing $A(G)$ and $A(L(G))$ is difficult, but this difficulty can be ameliorated by looking at $A(G)+D$ and $A(L(G))+2I$. In particular, $A(L(G))+2I$ is a rank-$n$ positive semidefinite matrix as long as $\widehat{m}\geq n$, and this rank does not change as $\widehat{m}$ does. Were we to instead consider $A(L(G))$ without this addition, we would have a rank-$\widehat{m}$ matrix, where the rank changes as the size of $E$ does.

 This means that for $\widehat{m}\geq n$, the rank of $A(L(G))+2I$ is not just the same as that of $P(A(L(K_n))+2I)P$ (which is trivially true), but is actually the same as that of $\EE[P](A(L(K_n))+2I)\EE[P]$, the mean matrix to which we would like to compare $A(L(G))+2I$. This allows us to state a concentration result for the nonzero eigenvalues of $A(L(G))+2I$, comparing these to the nonzero eigenvalues of the mean matrix we have defined.

\begin{theorem}
\label{theorem:concentrate}
Let $G=([n],E)$ be a graph with adjacency matrix $A$ whose entries are independently distributed $A_{i,j}\sim Bernoulli(\mathcal{P}_{i,j})$ for all $1\leq i< j\leq n$, where $\mathcal{P}\in M_{n}([p_{\mathrm{min}},p_{\mathrm{max}}])$, for $0<p_{\mathrm{min}}\leq p_{\mathrm{max}}\leq 1.$ Then if we define $\lambda_{\ell}(A(L(G)))=-2$ whenever $\ell>|E|=:\widehat{m}$, we have \begin{align*}
\EE\left[\lambda_{n}(A(L(G)))\right]&\geq 0.63 p_{\mathrm{min}}(n-2)-2(\log(n)+1),\text{ and}\\
\EE\left[\lambda_{1}(A(L(G)))\right]&\leq 3.44 p_{\mathrm{max}}(n-1)+2(\log(n)-1).
\end{align*}
Moreover,
\begin{align*}
\PP\left[\lambda_{n}(A(L(G)))\leq tp_{\mathrm{min}}(n-2)-2\right]&\leq ne^{-(1-t)^{2} p_{\mathrm{min}}(n-2)/4},\quad t\in[0,1),\text{ and}\\
\PP\left[\lambda_{1}(A(L(G)))\geq tp_{\mathrm{max}}\,2(n-1)-2\right]&\leq n\left(\frac{e}{t}\right)^{t p_{\mathrm{min}}(n-1)},\quad t\geq e.
\end{align*}
If $\widehat{m}\geq n$, $\lambda_{n+1}(A(L(G))),\ldots,\lambda_{\widehat{m}}(A(L(G)))=-2$.
\end{theorem}

Since the nonzero eigenvalues of $A(L(K_n))+2I$ are a simple eigenvalue $2(n-1)$, and the eigenvalue $n-2$ with multiplicity $n-1$, we see that these bounds tell us that the nonzero eigenvalues of $A(L(G))+2I$ rarely exceed $p_{\mathrm{max}}\lambda_1(A(L(K_n))+2I),$ nor go below $p_{\mathrm{min}}\lambda_n(A(L(K_n))+2I)$. These look like the nonzero eigenvalues of $A(L(K_n))+2I$, discounted by a factor which depends on the probabilities that edges appear in $G$. On the other hand, were we to bound the corresponding eigenvalues of $\EE[P](A(L(K_n))+2I)\EE[P]$, the natural lower bound would be $p_{\mathrm{min}}^2\lambda_n(A(L(K_n))+2I)$, and we would similarly have a factor of $p_{\mathrm{max}}^2$ appearing in the upper bound. The bounds from Theorem~\ref{theorem:concentrate} may seem, at first glance, to be missing a factor of $p_{\mathrm{min}}$ or $p_{\mathrm{max}}$.

The resolution of this apparent disagreement lies in the nonlinearity of the maps $\lambda_i: M_n(\RR)\rightarrow\RR$ which take a symmetric matrix to its $i$th largest eigenvalue, ordered nonincreasingly. This means that the expected values $\EE[\lambda_i(P(A(L(K_n))+2I)P)]$ are not the same as $\lambda_i(\EE[P(A(L(K_n))+2I)P])$. To account for this difference and obtain our concentration results, we need to consider both $\EE[P]A(L(K_n))\EE[P]$ and $\sqrt{\EE[P]}A(L(K_n))\sqrt{\EE[P]}$, depending on the quantity of interest.

\begin{rmk}
\label{rmk:concentrate}
Suppose in the statement of Theorem~\ref{theorem:concentrate} that instead $\mathcal{P}_{i,j}\in[p_{\mathrm{min}},p_{\mathrm{max}}]$ for all $1\leq i<j\leq n$ such that $\{i,j\}\in S$, where $S\subseteq\binom{[n]}{2}$, and otherwise $\mathcal{P}_{i,j}=0.$ Let $K=([n],S)$ denote the graph for which all edges in $S$ appear, and let $\mu_1=\lambda_1(A(L(K)))+2, \mu_n=\lambda_n(A(L(K)))+2.$ Then the same argument shows that
\begin{align*}
\PP\left[\lambda_{n}(A(L(G)))\leq tp_{\text{min}}\mu_n-2\right]&\leq ne^{-(1-t)^{2} p_{\text{min}}\mu_n/4},\quad t\in[0,1),\text{ and}\\
\PP\left[\lambda_{1}(A(L(G)))\geq tp_{\text{max}}\,\mu_1-2\right]&\leq n\left(\frac{e}{t}\right)^{t p_{\text{min}}\mu_1/2},\quad t\geq e.
\end{align*}
\end{rmk}

\subsection{Limit results for Erd\"os-R\'enyi line graphs}
\label{sec:er}
In this section, we consider limiting results for the top eigenvalue and empirical spectral distribution for the line graph of an Erd\"{o}s-R\'enyi random graph (\cite{Erdos_Renyi_1960_original}), wherein each (undirected) edge in a graph forms independently with some fixed probability $p \in (0, 1)$. We first describe the limiting behavior of its top eigenvalue:

\begin{theorem}
\label{theorem:topeig}
Let $G\sim \mathrm{ER}(p)$, and let $L(G)$ be its line graph. Then with probability 1, 
$$ \frac{1}{n}\lambda_1(A(L(G))) \rightarrow 2p\text{ as }n\rightarrow\infty. $$
\end{theorem}

\begin{theorem}
\label{theorem:topeigclt}
Let $G\sim \mathrm{ER}(p)$, and let $L(G)$ be its line graph. Then as $n\rightarrow\infty$, we have the following convergence in distribution:
$$ \frac{\lambda_1(A(L(G)))-2(n-1)p-4(1-p)-2}{\sqrt{8p(1-p)}}\rightarrow \mathcal{N}(0,1). $$
\end{theorem}

To describe the limiting behavior of the bulk eigenvalues, we consider the empirical spectral distribution (ESD) of the matrix $A(L(G))$. Since this matrix will always have $\hat{m}-n$ eigenvalues equal to $-2$, we remove these from consideration. We prove that the limiting distribution is neither a semicircle nor a normal distribution, but rather a free convolution, as described in \cite{ding2010spectral}. This limiting distribution applies to empirical spectral distributions for matrices in the class of \emph{Wigner matrices with Dependent Diagonal (WignerDD)}, which we define below. 

\begin{definition}
\label{def:wdd}
Let $W\in M_n(\RR)$ be a symmetric random matrix with all entries above the main diagonal being independent, zero-mean random variables with common variances $\sigma^2$. Suppose that for each $k\geq 1,$ we have the uniform bound on $k$th moments $$\max_{1\leq i<j\leq n} \EE[|(W_n)_{i,j}|^k]\leq \Omega_k,$$ and suppose that $(W_n)_{ii}=\sum_{j\neq i}(W_n)_{ij}$ for each $i$. We say that $W_n$ is a \emph{Wigner matrix with Dependent Diagonal (WignerDD)}.
\end{definition}

We do not require identical distribution of the entries above the diagonal, but simply that they all have the same variance and satisfy the uniform moment bound. While the assumptions of a WignerDD matrix are more general, for the special case of graph adjacencies, the requirement that diagonal entries be the sum of the off-diagonal entries connects  WignerDD matrices to graph Laplacians. Namely, if $A$ is the adjacency matrix of a loop-free network and $D$ the diagonal matrix of degrees, then, then $D-A$ is its Laplacian and $D+A$ is a WignerDD matrix.  We show the empirical spectral distribution of any WignerDD matrix satisfies the following limit, which matches the results given in \cite{ding2010spectral} for the case in which the diagonal entries are the {\em negations} of the row sums for the off-diagonal. 

\begin{theorem}
\label{theorem:esd}
Let $W_n\in M_n(\RR)$ be a WignerDD matrix as in Definition~\ref{def:wdd}. Define the scaled empirical spectral distribution for $x\in\RR$ as 
$$ \hat{F}_n(x) = \frac{\#\{i: \lambda_i(W_n)\leq x\sigma\sqrt{n}\}}{n}.$$ Then $\hat{F}_n(x)\rightarrow L(x)$ almost surely as $n\rightarrow\infty$, where $L(x)$ is the CDF of the unique distribution $\mathcal{L}$ with all odd moments equal to 0, and even moments given by
$$ L_{2k} = \sum_{m=0}^{k-1} \frac{1}{m+1} \sum_{\substack{j_1,\ldots,j_{m+1}\geq 0\\ \sum j_i = m}}\;\sum_{\substack{n_1,\ldots,n_{m+1}\geq 0 \\ \sum n_i = k-m}} \;\prod_{i=1}^{m+1} \binom{2n_i+j_i}{j_i} \frac{(2n_i)!}{2^{n_i} n_i!}. $$
This sequence starts as $L_0=1, L_2=2, L_4=9, L_6=56, L_8=431, L_{10}=3852,$ and for $k \geq 1$, satisfies $0\leq L_{2k}\leq (4e)^k k^{2k}$; hence the moments satisfy Carleman's condition. This distribution may also be characterized as the free additive convolution of the semicircle and normal distributions.
\end{theorem}

The case of the eigenvalues of the line graph adjacency matrix in the setting of an Erd\"{o}s-Renyi graph follows from this general theorem:

\begin{cor}
\label{cor:esd}
Let $G\sim \mathrm{ER}(p)$, and let $L(G)$ be its line graph. Define the scaled, centered empirical spectral distribution of the eigenvalues of the matrix $A(L(G))$ not equal to $-2$ as follows: for $x\in\RR$, let
$$ W_n(x):=\frac{\#\{i:\lambda_i(A(L(G)))\neq-2, \lambda_i(A(L(G)))-(n-2)p-2\leq x\}}{n}.$$
Then as $n\rightarrow\infty$, the empirical spectral distribution converges weakly to $\mathcal{L}$ almost surely. That is, $W_n(x\sqrt{np(1-p)})\rightarrow L(x)$ with probability 1.
\end{cor}

It turns out that this matches the limiting ESD for the Laplacian of an Erd\"{o}s-Renyi random graph \cite{ding2010spectral}. These results give tighter bounds for the top eigenvalue and spectral distribution of the line graph of an ER graph. Both of them are reminiscent of corresponding results for the spectrum of ER graphs, a consequence of the homogeneous connection probability in such a network, though we see that the distributional limit for the bulk eigenvalues is quite different in our setting, more closely resembling that of the Laplacian than the adjacency matrix. In the next section, we consider a significant extension of these results to the case of a stochastic blockmodel, whose line graph requires materially more sophisticated analysis.

\section{Stochastic blockmodel line graphs}
\label{S:sbm}

To obtain concentration of the nontrivial eigenvalues, we require only very weak assumptions on the probabilities of edge formation. We now explore the implications of additional structure in the original graph, namely when the original graph is a stochastic blockmodel. 

\begin{figure}
    \centering
    \begin{subfigure}{0.45\hsize}
    \includegraphics[width=\linewidth]{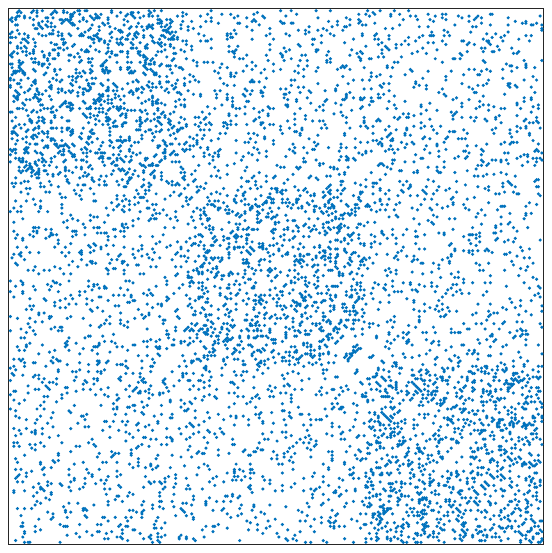}
    \caption{Adjacency matrix of an SBM graph}
    \end{subfigure}
    \begin{subfigure}{0.45\hsize}
    \includegraphics[width=\linewidth]{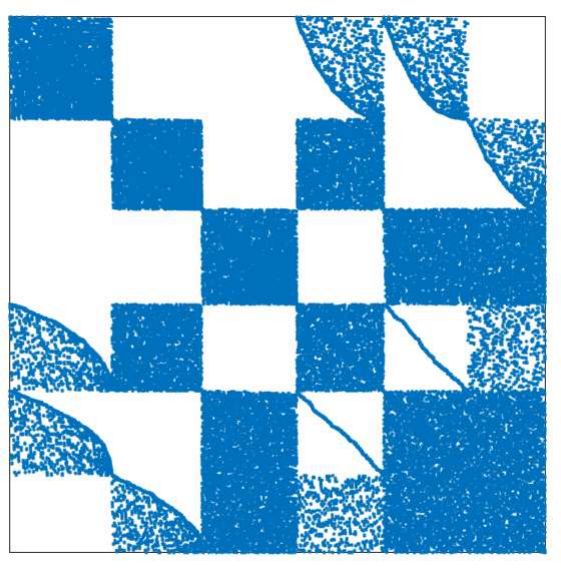}
    \caption{Adjacency matrix of an SBM line graph}
    \end{subfigure}
    \caption{A stochastic blockmodel graph and its line graph.}
    \label{fig:sbm}
\end{figure}

Stochastic block model line graphs exhibit similar structure to SBM graphs themselves, as can be seen in Figure~\ref{fig:sbm}. The mean matrix for an SBM line graph and its invariant subspace carry all the relevant signal for determining edge cluster memberships. An important difference between SBM graphs and SBM line graphs is that for line graphs, this signal subspace is {\em not} simply the eigenspace corresponding to the dominant eigenvalues of $A(L(G))$. Rather, the signal subspace is a proper subspace of this eigenspace, and must be carefully identified lest it be conflated with the other eigenspace of $A(L(G))$ with large eigenvalues, which encodes the line graph structure, but does not reveal cluster memberships. However, there is no spectral gap separating these two subspaces, which makes a projection imperative for obtaining useful information from the line graph adjacency matrix. The importance of this projection is illustrated very clearly by the simulations of Section~\ref{S:sim}.

We first define the notion of the edge partition induced by the partition of the vertices, which gives us the ``blocks" in the SBM line graph. These are the memberships we estimate when we obtain a clustering of the edges.
We then analyze the mean of SBM line graphs and this signal subspace. In the next section, we look at the corresponding (random) subspace of $A(L(G))$, and prove that it is a signal-preserving subspace. 

\begin{definition}
Let $[n]$ be partitioned into $C_{1},\ldots,C_{k}$, where $|C_{r}|=:n_r\geq 2$ for all $r$. We call the partition of $\binom{[n]}{2}$ given by $C_{(r,s)},$ $1\leq r\leq s\leq k$, where $C_{(r,s)}=\{\{i,j\}\in \binom{[n]}{2}: i\in C_{r}, j\in C_{s}\}$, the \emph{partition induced by }$C_{1},\ldots,C_{k}$. For each $r$, $|C_{(r,r)}|=\binom{n_r}{2}=:m_{r,r},$ and for all $r<s,$ $|C_{(r,s)}|=n_r n_s=:m_{r,s}.$ The condition $n_r\geq 2$ ensures that none of the blocks in the induced partition are empty.
\end{definition}

The following proposition describes invariant subspaces of $A(L(K_n))$ corresponding to any induced partition of the edges, giving a decomposition of this matrix into its actions on a signal subspace for that particular clustering, and on the remainder of the space. When we consider a particular SBM structure for $G$, this corresponds to structure in the entries of $\EE[P]$, where $P$ is the random projection matrix taking $A(L(K_n))$ to its random submatrix $A(L(G))$, as described in Section~\ref{s:mean}. With this additional probabilistic structure, only some of these invariant subspaces of $A(L(K_n))$ will remain invariant subspaces of $\EE[PA(L(K_n))P]$. In particular, there is a unique invariant subspace among these which has minimum dimension, and this is the signal subspace we estimate.

\begin{prop}
\label{prop:qproj}
Let $[n]$ be partitioned into $C_{1},\ldots, C_{k}$, such that $n_{r}\geq2$ for all $r$. Let \\
$Q\in M_{\binom{n}{2},\binom{k+1}{2}}(\RR)$ be defined by 
$$
Q_{\{i,j\},(r,s)}=\begin{cases}m_{r,s}^{-1/2}&\text{ if }i\in C_{r}, j\in C_{s},\\ 0&\text{ otherwise,}\end{cases}
$$
for all $\{i,j\} \in \binom{[n]}{2}$ and $1\leq r\leq s\leq k$. Then $QQ^{T}$ is an orthogonal projection onto an invariant subspace of $A(L(K_n))$, and letting $M=Q^{T}A(L(K_{n}))Q\in M_{\binom{k+1}{2}}(\RR),$ $$\sigma(M)=\{2n-4, n-4 (\text{mult. }k-1), -2 (\text{mult. }\binom{k}{2})\}.$$
Letting $[Q,V]$ be a unitary matrix, we have the following decomposition of $A(L(K_n))$:
$$A(L(K_{n}))=QMQ^{T}+VLV^{T},$$ where $\sigma(L)=\{n-4(\text{mult. } n-k),-2(\text{mult. } \binom{n}{2}-n-\binom{k}{2})\}.$
\end{prop}

\begin{rmk}
\label{rmk:projstructure}

The proof of this proposition reveals that the space of eigenvectors of $M+2I$ with nonzero eigenvalues is the range of a certain matrix $F\in M_{\binom{k+1}{2},k}(\RR).$

The matrix $F$ clearly reveals the structure of the clustered line graph. The $k$ blocks $C_{(i,i)}$ correspond to the first $k$ rows of $F$, and are just the standard unit vectors in $\RR^k$. The remaining rows correspond to $C_{(i,j)}$ for $i\neq j$, and in the case that all of the $n_j$ are equal and large, these are very close to $\frac{1}{\sqrt{2}}e_i^T+\frac{1}{\sqrt{2}}e_j^T.$ These points all belong to the unit sphere in $\RR^k$ intersected with the positive orthant. For each $i$, $C_{(i,i)}$ appears at the corners of this set, and for each $i< j$, $C_{(i,j)}$ appears on the midpoint of the arc connecting the latent positions of $C_{(i,i)}$ and $C_{(j,j)}$. Considered together, the matrix $QF$ gives a point in $\RR^k$ associated to every edge, which depends only on its assigned block. When we turn to the invariant subspaces of $A(L(G))$, this is the defining characteristic of the signal subspaces we wish to estimate. This property holds in that setting as a consequence of the structure of $QF$ we observe here. 
\hfill$\square$
\end{rmk}


Now we consider the case of a specific SBM structure for $G$. If $Z\in M_{n,k}(\{0,1\})$ is the matrix whose rows indicate the cluster memberships of the vertices, we know that the edge-probability matrix for the stochastic blockmodel is given by $\mathcal{P}=ZBZ^{T},$ where $B\in M_{k}([0,1])$ is symmetric. This corresponds to $\EE[P]$ with block-constant entries along its diagonal. This being the case, only certain of the projectors $Q$ from Proposition~\ref{prop:qproj} will still project onto an invariant subspace of $\EE[PA(L(K_n))P]$. We describe these in the following proposition.

\begin{prop} 
\label{prop:sbmlike}
Suppose $G$ is an SBM graph with $\EE[A(G)]=ZBZ^T$. Let $Z$ be fixed, let $C_1,\ldots,C_k$ be the corresponding clustering of the vertices, and let $C_{(r,s)}, 1\leq r\leq s\leq k$ be the induced partition of the edges. 

A projection matrix $\tilde{Q}$ from Proposition~\ref{prop:qproj} corresponding to clusters $\tilde{C}_1,\ldots,\tilde{C}_{k'}$ satisfies $\EE[P]\tilde{Q}=\tilde{Q}D$ where $D$ is a diagonal matrix for every symmetric matrix $B\in M_{k}([0,1])$ if and only if for each $\tilde{C}_i$ there is a cluster $C_j$ such that $\tilde{C}_i\subseteq C_j$; that is, the clustering defining $\tilde{Q}$ is a refinement of the clustering of the vertices given by $Z$. In this case, $\tilde{Q}$ projects onto an invariant subspace of $\EE[PA(L(K_n))P]$.

Among these, the unique projection $Q$ having minimum rank is the one with $\tilde{C}_i=C_i$ for each $1\leq i\leq k$. $Q$ enjoys the following properties, where $V, M, L$ are defined as in Proposition~\ref{prop:qproj}:
$$\EE[P]Q=Q\mathrm{diag}([B_{i,j}]_{i\leq j}),\quad  \EE[P]V=V\mathrm{diag}([B_{i,j} I_{m_{i,j}-1}]_{i\leq j}),$$ and the mean matrix $\EE[PA(L(K_n))P]$ has the following decomposition:
\begin{equation}
\label{eq:meandecomp}
Q\mathrm{diag}([B_{i,j}])M\mathrm{diag}([B_{i,j}])Q^T+V\mathrm{diag}([B_{i,j}I_{m_{i,j}-1}])L\mathrm{diag}([B_{i,j}I_{m_{i,j}-1}])V^{T}.
\end{equation}
\end{prop}


\begin{rmk}
\label{rmk:sbmlike}

This proposition highlights the similarities between SBM line graphs and SBM graphs. One of the defining features of the SBM is that the edge probability matrix $\mathcal{P}=ZBZ^T$ is block-constant. This block-constant structure appears in the mean of the SBM line graph adjacency matrix in the first term of the decomposition (\ref{eq:meandecomp}), that is, in $T_1=Q\mathrm{diag}([B_{i,j}])M\mathrm{diag}([B_{i,j}])Q^T$. As in the case of an SBM, where we wish to estimate the block memberships encoded in $Z$ and the connection probabilities $\mathcal{P}$, in this setting we wish to estimate the edge clusters encoded in $Q$ and $T_1$. \hfill$\square$
\end{rmk}

A natural question is what the latter term $T_2$ encodes. This matrix has nonzero blocks in precisely the same places as $T_1$, but within each block, the sum of the entries in every row and column is 0. $T_2$ is what gives $A(L(G))$ the structure of a line graph in terms of its pattern of zero and nonzero entries. In that sense, both terms are informative, but only $T_1$ contains the edge cluster signal we are interested in for the task at hand. This too is analogous to the case of an SBM: An observed $A(G)$ has a pattern of zero and nonzero entries which are informative for the connections between vertices. But for the purposes of recovering the blockmodel structure, we want to estimate the block-constant $\mathcal{P}$. In our setting, the specific pattern of zero and nonzero entries tells us that our graph is a line graph, but for the purposes of estimating the edge clusters, we want to estimate the block-constant matrix $T_1$.

\begin{figure}
\begin{subfigure}{0.24\textwidth}
\includegraphics[trim={2cm 2.55cm 1cm 1cm},clip,height=1.5in,width=\textwidth]{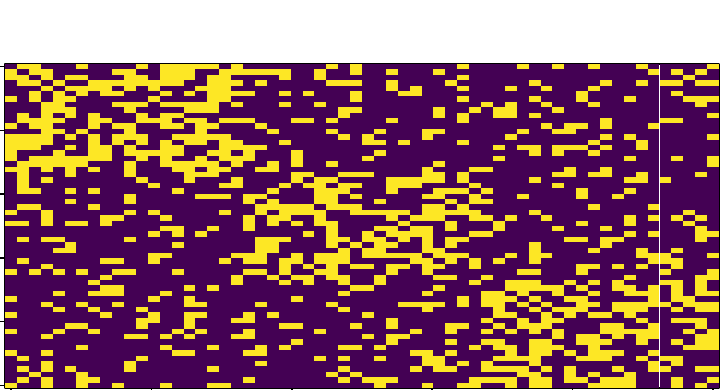}
\end{subfigure}
\begin{subfigure}{0.24\textwidth}
\includegraphics[trim={2cm 2.55cm 1cm 1cm},clip,height=1.5in,width=\textwidth]{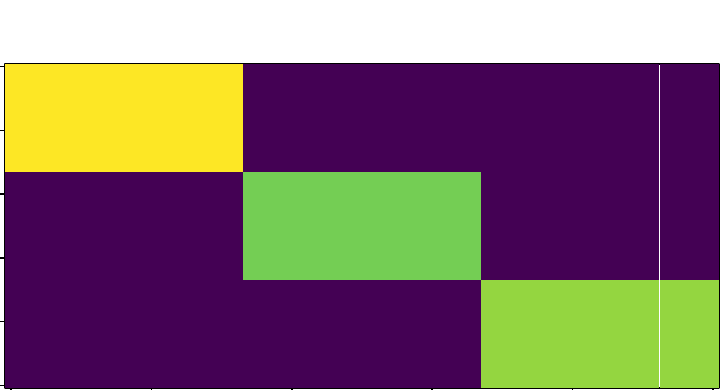}
\end{subfigure}
\begin{subfigure}{0.24\textwidth}
\includegraphics[trim={2cm 2.55cm 1cm 1cm},clip,height=1.5in,width=\textwidth]{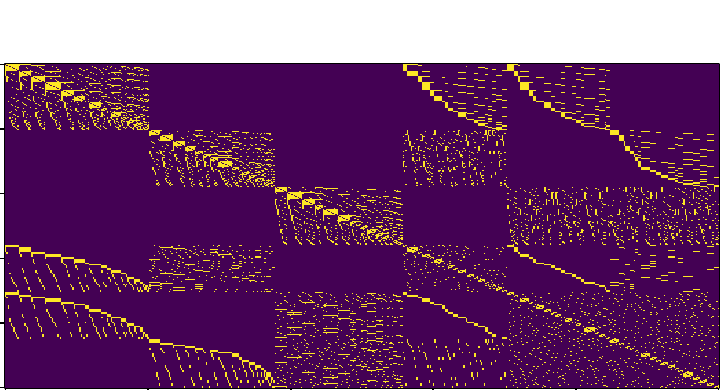}
\end{subfigure}
\begin{subfigure}{0.24\textwidth}
\includegraphics[trim={2cm 2.55cm 1cm 1cm},clip,height=1.5in,width=\textwidth]{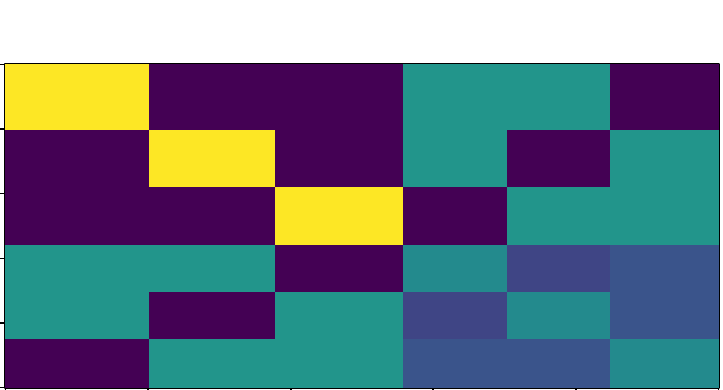}
\end{subfigure}
\caption{A three-block SBM adjacency matrix with its best block-constant estimate. The line graph of this matrix with its best block-constant estimate, corresponding to $T_1$.}
\label{fig:cartoon}
\end{figure}

The difference in the two cases is that for an SBM, the noise matrix has much smaller eigenvalues than the signal matrix; for an SBM line graph, the noise and signal matrices have equally large eigenvalues. Since there is no spectral gap between the leading eigenspaces of these two terms, a straightforward calculation of the singular vectors of $A(L(G))$ without making use of any projection is unlikely to recover a useful subspace, since it will include a mixture of information from both $T_1$ and $T_2$. It may clearly be seen in the simulations of Section~\ref{S:sim} that this behavior occurs not just in theory, but in practice.

\section{Estimating edge latent positions}
\label{S:estimate}

The principal result of this section is Theorem \ref{theorem:singvecs}, in which we show that a judicious projection leads us to the signal-preserving subspace for edge clustering. The bedrock of this theorem, in turn, is the explicit construction of this projection matrix $\hat{Q}$ and its action on $A(L(G))$. We show the action of $\hat{Q}$ on $A(L(G))$ is analogous to the action of $Q$ on $A(L(K_n))$, in that the random projection $\hat{Q}$ extracts from $A(L(G))$ the same signal as the fixed projection $Q$ extracts from its mean. This presents a striking illustration of the importance of isolating the proper subspace and the sensitivity of this choice to the subsequent inference task itself. More specifically, when it comes to edge clustering, the signal-preserving subspace is not the eigenspace associated to the largest eigenvalues of $A(L(G))$. Our explicit construction not only reveals how to locate such a subspace, but further describes associated distributional limits for the projection that defines it.


We require some notation at the outset.
Let $\widehat{m}_{i,j}=|C_{(i,j)}\cap E|$ for all $1\leq i\leq j\leq k$. We define 
\begin{equation}
\label{eq:qmat}
\overline{\hat{Q}}=PQ\mathrm{diag}\left(\sqrt{\frac{m_{i,j}}{\widehat{m}_{i,j}}}\right)=\mathrm{diag}\left(\delta_e\sqrt{\frac{m_e}{\widehat{m}_e}}\right)Q=\mathrm{diag}(Y_e)Q\in M_{\binom{n}{2},\binom{k+1}{2}},
\end{equation}
where $m_e$ is the size of the block of the induced partition $C_{(i,j)}$ to which the edge $e$ belongs, and $\widehat{m}_e$ is the corresponding observed quantity. $\overline{\hat{Q}}$ is then the same size as $Q$, but has a zero row whenever the corresponding edge is missing from $G$; the second factor renormalizes so that the columns of $\overline{\hat{Q}}$ have norm 1 when they are nonzero (and when $\delta_e=0$ for all of the edges $e$ in a certain block of the induced partition, we will multiply by 1 instead of $\sqrt{m_e/\widehat{m}_e}$, since the latter does not make sense in this case. The specific choice is not important, since the corresponding column is just zero). 

Consider a single diagonal entry from the second expression. If $e\in C_{(i,j)}$, $Y_e=\sqrt{m_{i,j}} W$, where $W\sim\mathrm{DampedBinomial}(1,m_{i,j}-1,B_{i,j})$, which is defined as follows:
\begin{definition}
\label{def:dampbinom}
Let $n,m$ be positive integers, and let $p\in(0,1)$. We say that\\ $W\sim\mathrm{DampedBinomial}(n,m,p)$ when $$W=\frac{B_n}{\sqrt{B_n+B_m}},$$ where $B_n$ and $B_m$ are independent Binomial random variables with $n$ and $m$ trials respectively, and common success probability $p$. We take $W=0$ in the case $B_n=B_m=0$.
\end{definition}
In Theorem~\ref{theorem:dampedbinomial} it is shown that $\mu_e=\EE[Y_e]$ satisfies
\begin{equation}\label{eq:mubounds}
\sqrt{B_{i,j}}-\frac{1-B_{i,j}}{2\sqrt{B_{i,j}}m_{i,j}}\leq \mu_e\leq \sqrt{B_{i,j}}.
\end{equation}
When all $m_{i,j}$ are large, this gives $$\EE[\overline{\hat{Q}}]=\mathrm{diag}(\mu_e)Q\approx \sqrt{\EE[P]}Q=Q\mathrm{diag}(\sqrt{B_{i,j}}).$$ As such, we see that on average, $\hat{Q}$ (after being padded with zeroes so that it has a consistent size) just looks like a re-scaling of the projection $Q$. This scaling ensures that both matrices are isometries despite their different sizes. In particular, the span of $\EE[\overline{\hat{Q}}]$ is the same as that of $Q$.

Proposition 4 shows that $A(L(G))\hat{Q}$ approximates $A(L(K_n))Q\mathrm{diag}(\mu_{i,j})$ after padding this matrix with zeros so that it has a consistent size. Since the latter matrix will be used to define the edge latent positions, this proposition is a necessary step for showing that the left singular vectors of $A(L(G))\hat{Q}$ span a signal-preserving subspace.

\begin{prop}
\label{prop:hmat}
Let $H=A(L(K_n))(\overline{\hat{Q}}-Q\mathrm{diag}(\mu_{i,j})).$ The columns of $H$ are independent. When all clusters have sizes $n_i\propto n$, then there is a constant $C$ such that with probability at least $1-O(1/n)$, for $n$ sufficiently large, $$\|H\|_F\leq C n^{3/4}.$$
\end{prop}

\begin{rmk}
\label{rmk:hmat}
The proof of Proposition~\ref{prop:hmat} shows that the entries of $H$ may be expressed as
 $$H_{\{i,j\},(r,s)}=\frac{1}{\sqrt{m_{r,s}}}\sum_{\substack{\{\ell,p\}\in C_{(r,s)}\\ |\{\ell,p\}\cap\{i,j\}|=1}}(Y_{\{\ell,p\}}-\mu_{r,s}),$$ which is a sum of dependent, mean-zero random variables. Such a sum resists straightforward analysis, but letting $\eta(\{i,j\},(r,s))=:\eta$ denote the number of terms in this sum, we may view $H_{\{i,j\},(r,s)}$ as a centered DampedBinomial($\eta,m_{r,s}-\eta,B_{r,s}$) random variable. We carefully analyze this distribution in the supplement.
 
This expression also reveals that while the columns of $H$ are independent, there is dependence between any two rows of $H$. This dependence can be very strong, since certain collections of entries of $H$ will be identical. This further complicates the analysis of $\|H\|_F$, requiring very tight bounds on the DampedBinomial distribution in order to prove Proposition~\ref{prop:hmat}. 
 
  \hfill$\square$
\end{rmk}

The following corollary shows that the block-constant part of $A(L(G))$ looks like a re-scaled version of the block-constant part of $A(L(K_n))$ as described in Remark~\ref{rmk:sbmlike}, with the scaling determined by the probabilities of edges appearing in each block of the induced partition:

\begin{cor}
\label{c:mhat}
With probability at least $1-O(1/n)$,
$$\|\hat{Q}^T A(L(G)) \hat{Q}-\mathrm{diag}(\mu_{i,j}) M\mathrm{diag}(\mu_{i,j})\|_F\leq C n^{3/4}.$$
\end{cor}

We may now define the latent positions of the edges. Since $\mu_e\approx \sqrt{\mathcal{P}_{e}}$, the matrix $$\mathrm{diag}(\mu_e)A(L(K_n))\mathrm{diag}(\mu_e)$$ is similar in spirit to $\EE[PA(L(K_n))P]$, but appears to be missing a factor of roughly $\sqrt{\EE[P]}$ on either side. As we mentioned in Section~\ref{s:eigs}, this is nevertheless the correct object to consider for concentration because of the nonlinearities involved. When we consider projecting this matrix, we find that
\begin{align*}\mathcal{R}&=\mathrm{diag}(\mu_e)A(L(K_n))\mathrm{diag}(\mu_{e})Q=Q\mathrm{diag}(\mu_{i,j})M\mathrm{diag}(\mu_{i,j})\\&=Q\mathrm{diag}(\mu_{i,j})(FDF^T-2I)\mathrm{diag}(\mu_{i,j}),\end{align*} where $F\in M_{\binom{k+1}{2},k}(\RR)$ spans the nonzero eigenspace of $M$, as discussed in Remark~\ref{rmk:projstructure}. As such, the $k$ dominant left singular vectors of $\mathcal{R}$ span the same subspace as $Q\mathrm{diag}(\mu_{i,j})FD^{1/2}$, which has rows that depend only on the cluster to which the edge belongs. These are the latent positions of the edges that we wish to estimate.

\begin{rmk}
Inspection of these latent positions shows that $C_{(i,i)}$ has latent position $(2\mu_{i,i}/\sqrt{n_i}) e_i \approx 2\sqrt{\frac{B_{i,i}}{n_i}} e_i \in\RR^k$, and for $i<j$, $C_{(i,j)}$ has latent position $\mu_{i,j}(\frac{1}{\sqrt{n_i}}e_i+\frac{1}{\sqrt{n_j}} e_j)\approx \sqrt{\frac{B_{i,j}}{n_i}}e_i+\sqrt{\frac{B_{i,j}}{n_j}}e_j\in\RR^k.$ 
\end{rmk}

Having defined the edge latent positions, we now show that the left singular vectors of $A(L(G))\hat{Q}$ are consistent estimators for them. We cannot compare the left singular vectors of $A(L(G))\hat{Q}$ to $\mathcal{R}$ directly, since these matrices do not have the same size. But if we instead look at the matrix $\hat{Q}\mathrm{diag}(\mu_{i,j})M\mathrm{diag}(\mu_{i,j})$, we see that this has rows which depend only on the cluster to which the corresponding edge belongs, and (up to a normalization), these rows are the same as those of $\mathrm{diag}(\mu_e)A(L(K_n))\mathrm{diag}(\mu_e)Q$ for any rows corresponding to edges appearing in $G$. But $\hat{Q}\mathrm{diag}(\mu_{i,j})M\mathrm{diag}(\mu_{i,j})$ has the same size as $A(L(G))\hat{Q}$, so we may compare the left singular vectors of these two matrices.

Since clustering the edge latent positions leads to perfect recovery of the edge memberships, the consistent estimator for these given by the left singular vectors of $A(L(G))\hat{Q}$ defines a signal-preserving subspace.

\begin{theorem}
\label{theorem:singvecs}
Let $U,\hat{U}\in M_{\widehat{m},k}(\RR)$ be the left singular vectors of $\hat{Q}\mathrm{diag}(\mu_{i,j})M\mathrm{diag}(\mu_{i,j})$ and $A(L(G))\hat{Q},$ respectively, corresponding to their dominant $k$ singular values. Then there exists $C>0$ such that with probability $1-O(1/n)$, for $n$ sufficiently large there exists an orthogonal matrix $\mathcal{O}$ such that $$\|\hat{U}-U\mathcal{O}\|_F\leq \frac{C}{n^{1/4}}.$$
\end{theorem}

\begin{rmk}
The previous theorem shows us what each row of $A(L(G))\hat{Q}$ centers around, but because of the random size of this matrix, requires us to compare to a random object $\hat{Q}\mathrm{diag}(\mu_{i,j})M\mathrm{diag}(\mu_{i,j}).$ We may avoid this by embedding all of the possible edges in the graph so that we have an estimator with a consistent size.

Formally, let $U$ and $\hat{U}\in M_{\binom{n}{2},k}(\RR)$ be the left singular vectors of, respectively, 
$$A(L(K_n))Q\mathrm{diag}(\mu_{i,j})=QM\mathrm{diag}(\mu_{i,j}) \textrm{ and } A(L(K_n))\bar{\hat{Q}},$$ corresponding to their dominant $k$ singular values. Then there exists $C>0$ such that with probability $1-O(1/n)$, for $n$ sufficiently large there exists an orthogonal matrix $\mathcal{O}$ such that $$\|\hat{U}-U\mathcal{O}\|_F\leq \frac{C}{n^{1/4}}.$$

The rows of $\hat{U}$ give an embedding of all of the edges in $E$. Note that the mean matrix to which we compare has one fewer factor of $\mathrm{diag}(\mu_{i,j})$: this accounts for the different scalings in $Q$ and $\hat{Q}$, since both are isometries but do not have the same number of rows. \hfill$\square$
\end{rmk}

\section{Simulations}
\label{S:sim}

In this section, we demonstrate the utility of the line graph for inference in a stochastic block model with edge attributes. As Figure \ref{fig:sbm} shows, the line graph of an SBM has SBM-like structure. However, when incorporating edge attributes, it is important to project onto the correct eigenspace for edge clustering, rather than projecting onto the eigenspace of leading singular vectors without using a projection. We further show that even when we do not know the true subspace onto which we should project but merely estimate it, the results are superior to methods which do not make use of an estimated projection. When combining edge covariate information with the adjacency information as observed through the line graph, we find that our approach that combines both sources of information (while using an estimated projection) outperforms methods that use only one or the other source of signal, as well as those that combine these different sources of information naively.

\begin{figure}[h]
\centering
 \includegraphics[width=0.48\hsize]{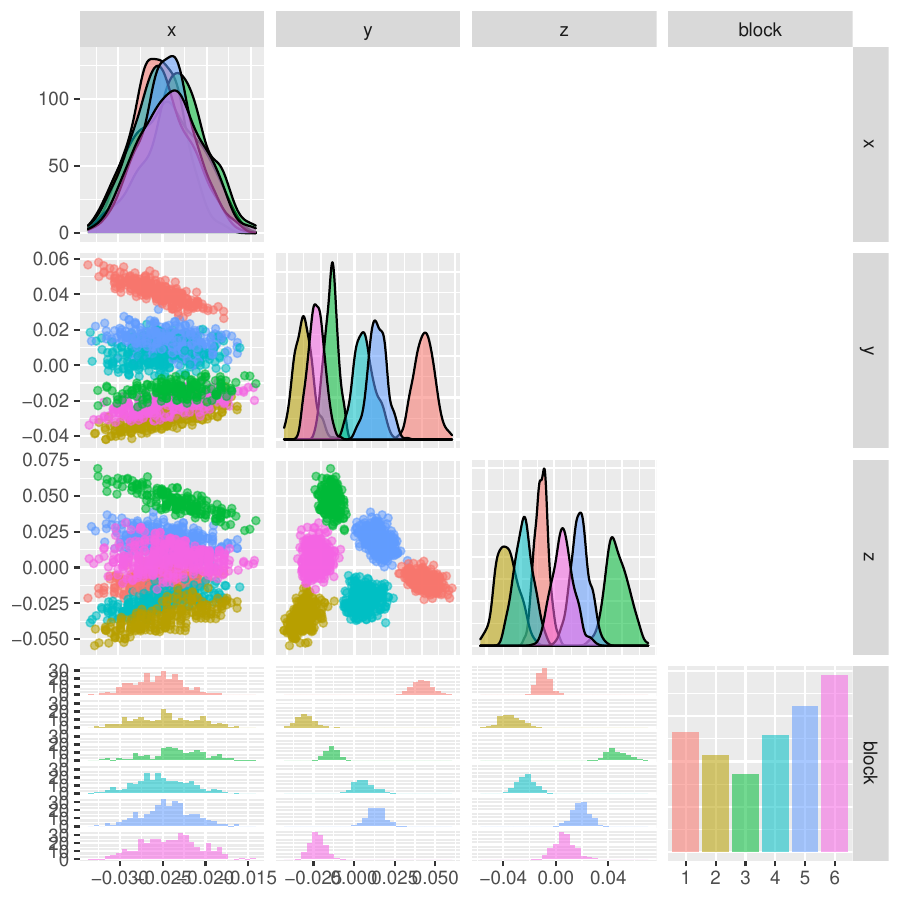}
 \caption{Depiction of the embedding obtained by using the left singular vectors of $A(L(G))\hat{Q}$ corresponding to the largest three singular values. GMM clustering using the first two singular vectors results in an ARI of 0.5 when compared to the true clusters, but GMM clustering using the second two singular vectors results in an ARI of 1, indicating perfect recovery of the clusters. Our method reveals the underlying edge communities.}
 \label{fig:qplot}
\end{figure}

\begin{figure}
\centering
 \includegraphics[width=0.48\hsize]{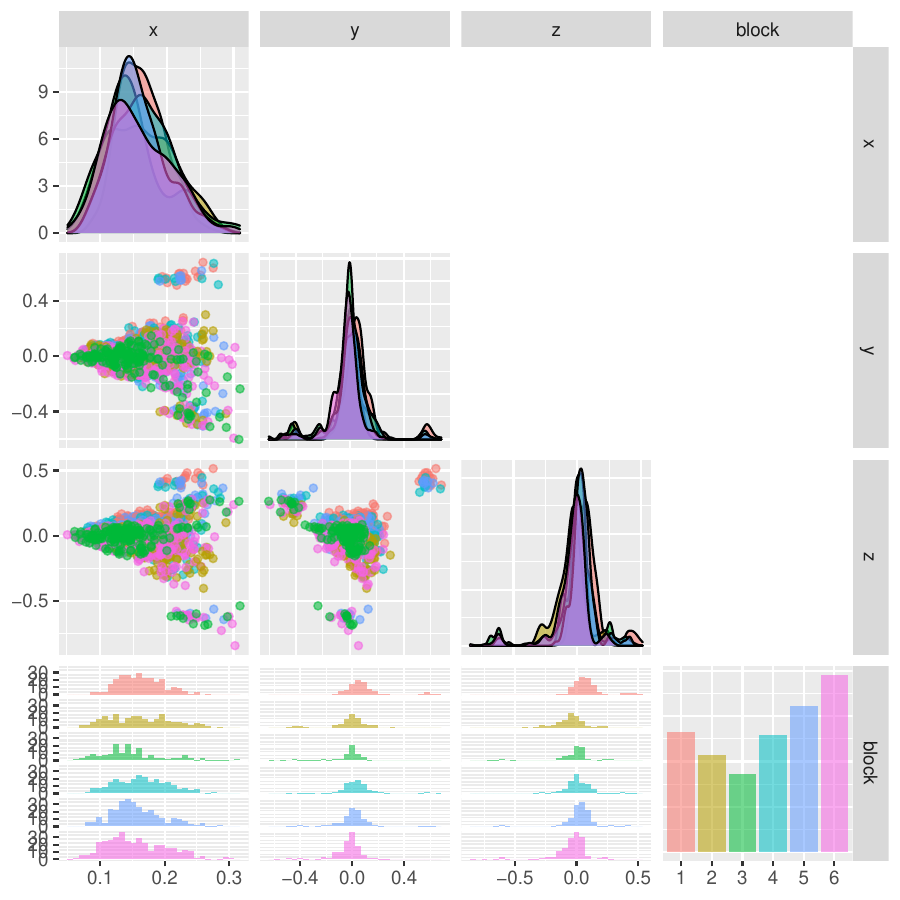}
 \caption{Depiction of the embedding obtained by using the left singular vectors of $A(L(G))$ corresponding to the largest three singular values. GMM clustering using any two of these singular vectors results in an ARI of 0 when compared to the true clusters, indicating a clustering with no relation to the true memberships. This naive method fails to extract important signal, namely the edge cluster memberships.}
 \label{fig:aseplot}
\end{figure}

We consider a 3-block SBM with $n_1=n_2=n_3=50$. After generating $G$ and computing $L(G)$, we find $\hat{Q}$, and compute the singular value decomposition of $A(L(G))\hat{Q}=\hat{U}\hat{S}\hat{V}^T.$ When we retain only the largest three singular vectors, we get the plot in Figure~\ref{fig:qplot}. We compare this to the left singular vectors of $A(L(G))$ computed directly without using $\hat{Q},$ which is shown in Figure~\ref{fig:aseplot}. Using the adjusted Rand index, we can compare clusterings obtained from these embeddings using Gaussian mixture modeling (GMM). When we do, we find that using the embedding given by the rows of the left singular vectors of $A(L(G))\hat{Q}$ corresponding to the second and third-largest singular values gives an ARI of 1, indicating perfect clustering. Using the left singular vectors of $A(L(G))\hat{Q}$ corresponding to the first and second largest singular values gives an ARI of 0.5, since the first singular vector is relatively constant across all groups. On the other hand, using any combination of two of the three left singular vectors of $A(L(G))$ corresponding to the largest three singular values results in an ARI of 0, indicating a clustering with essentially no relationship to the truth. From this, we conclude that the projection $\hat{Q}$ introduced here is absolutely essential to the determination of the clustering. Failing to include the projection $\hat{Q}$ results in an embedding which retains too much information about the line graph structure, at the cost of information about the cluster structure, and leads to a far worse embedding for the task of clustering the edges. We note that in these experiments, we use $\hat{Q}$ corresponding to the clustering of the edges induced by the ground truth clustering of the vertices, which in some ways makes the comparison of these two unfair, as we can see in the stark differences between their results. In the next experiments, we will consider an approximate version of $\hat{Q}$ obtained only using information in the observed graph, and show that the results are similar in this case, indicating the importance of the projection even when this does not perfectly reflect the true cluster memberships of the edges.

\subsection{Incorporating edge covariates}
\label{s:edgecov}

\begin{figure}
\begin{subfigure}{0.49\textwidth}
    \includegraphics[width=\textwidth]{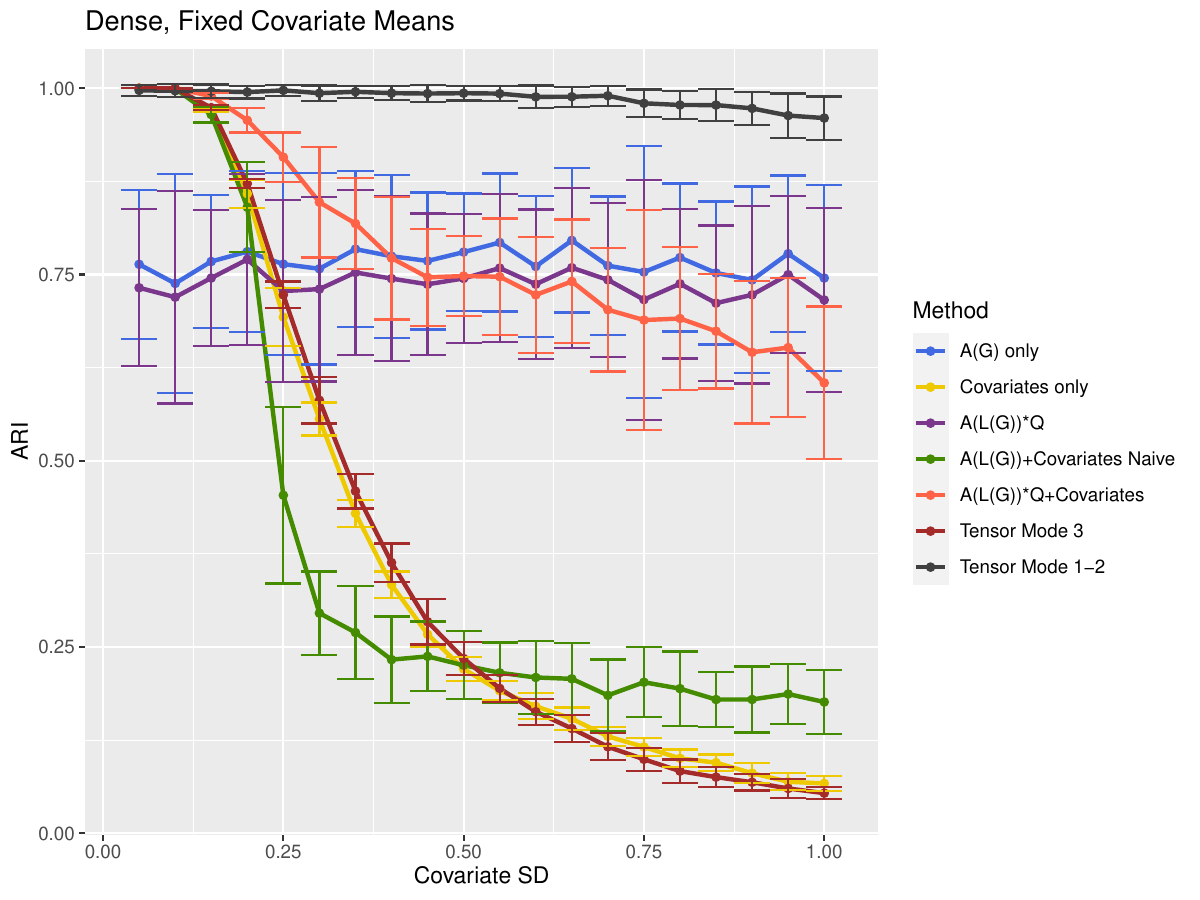}
    \caption{Dense network, fixed covariate means.}
\end{subfigure}
\begin{subfigure}{0.49\textwidth}
    \includegraphics[width=\textwidth]{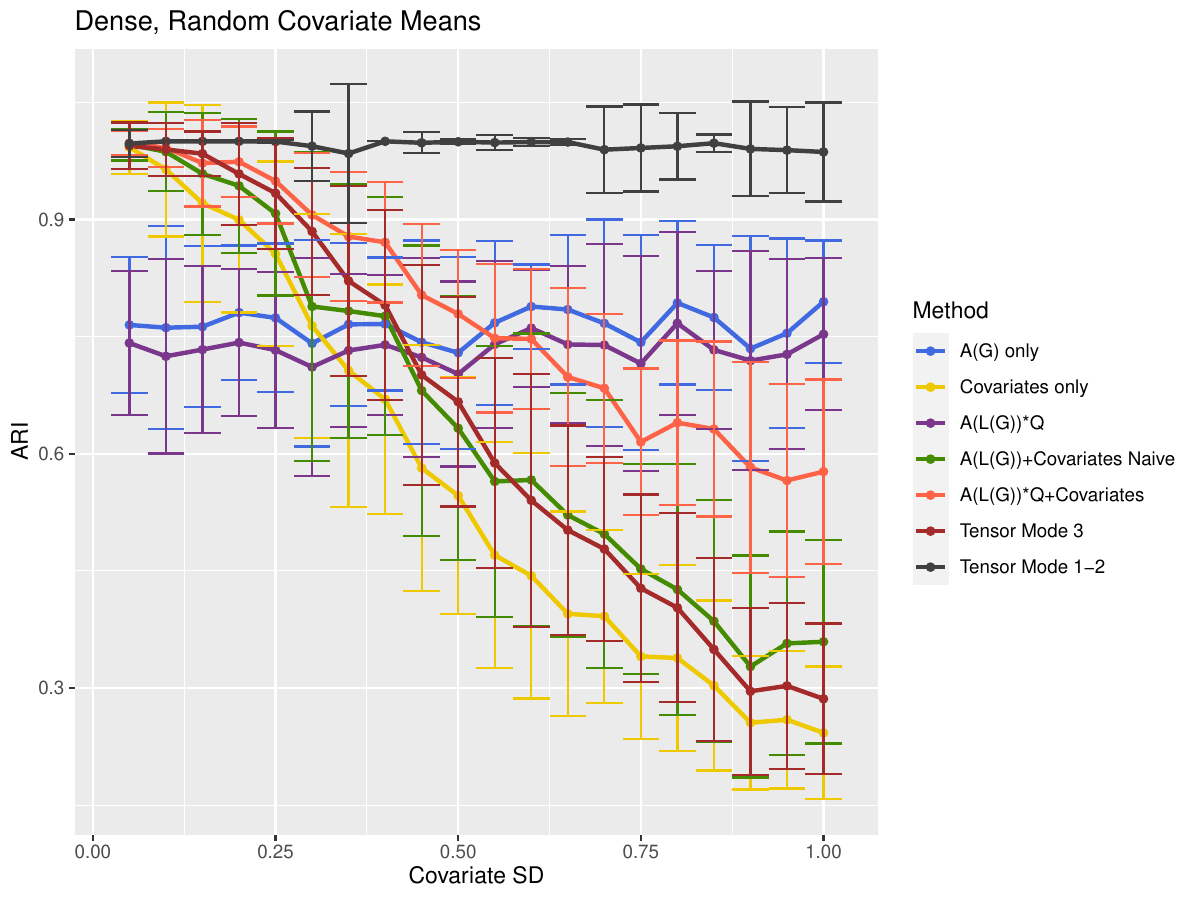}
    \caption{Dense network, random covariate means.}
\end{subfigure}

\begin{subfigure}{0.49\textwidth}
    \includegraphics[width=\textwidth]{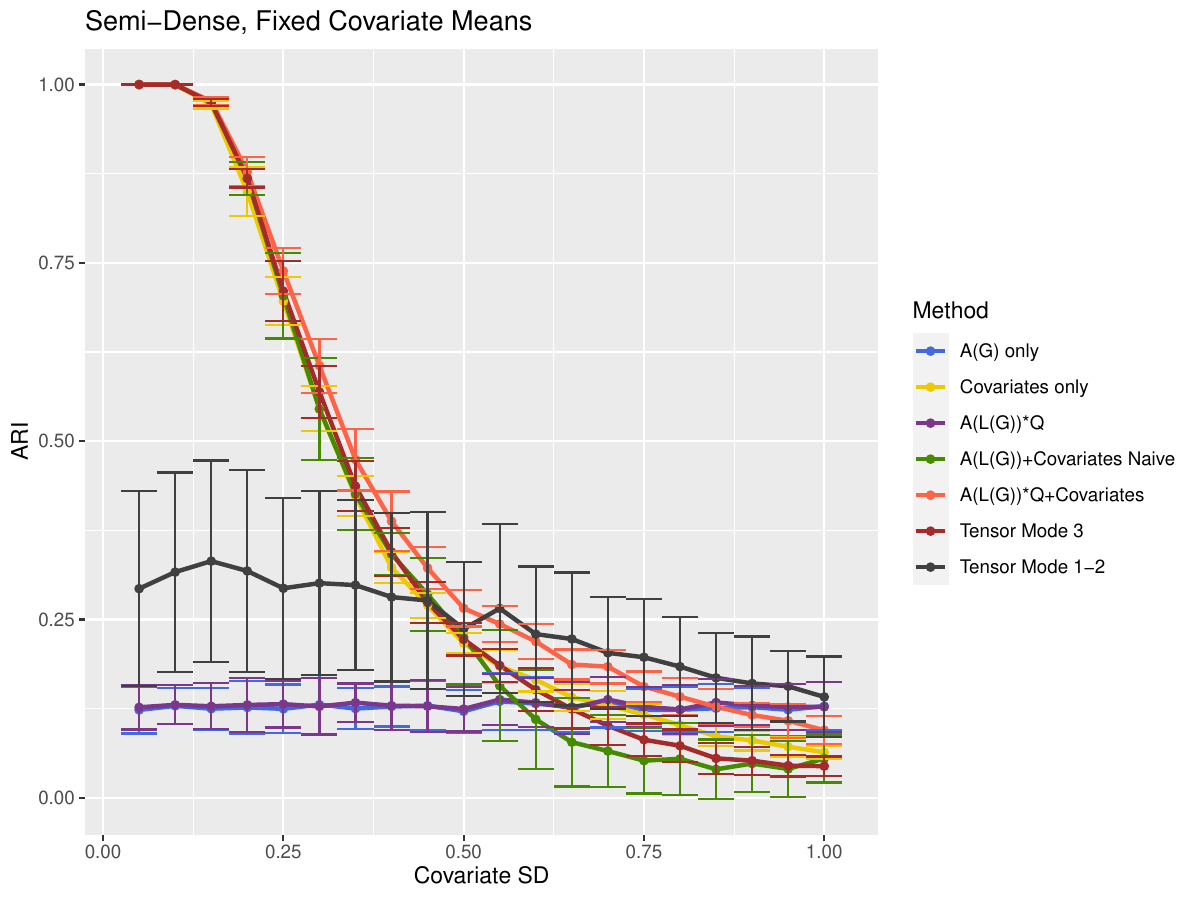}
    \caption{Semi-Dense network, fixed covariate means.}
\end{subfigure}
\begin{subfigure}{0.49\textwidth}
    \includegraphics[width=\textwidth]{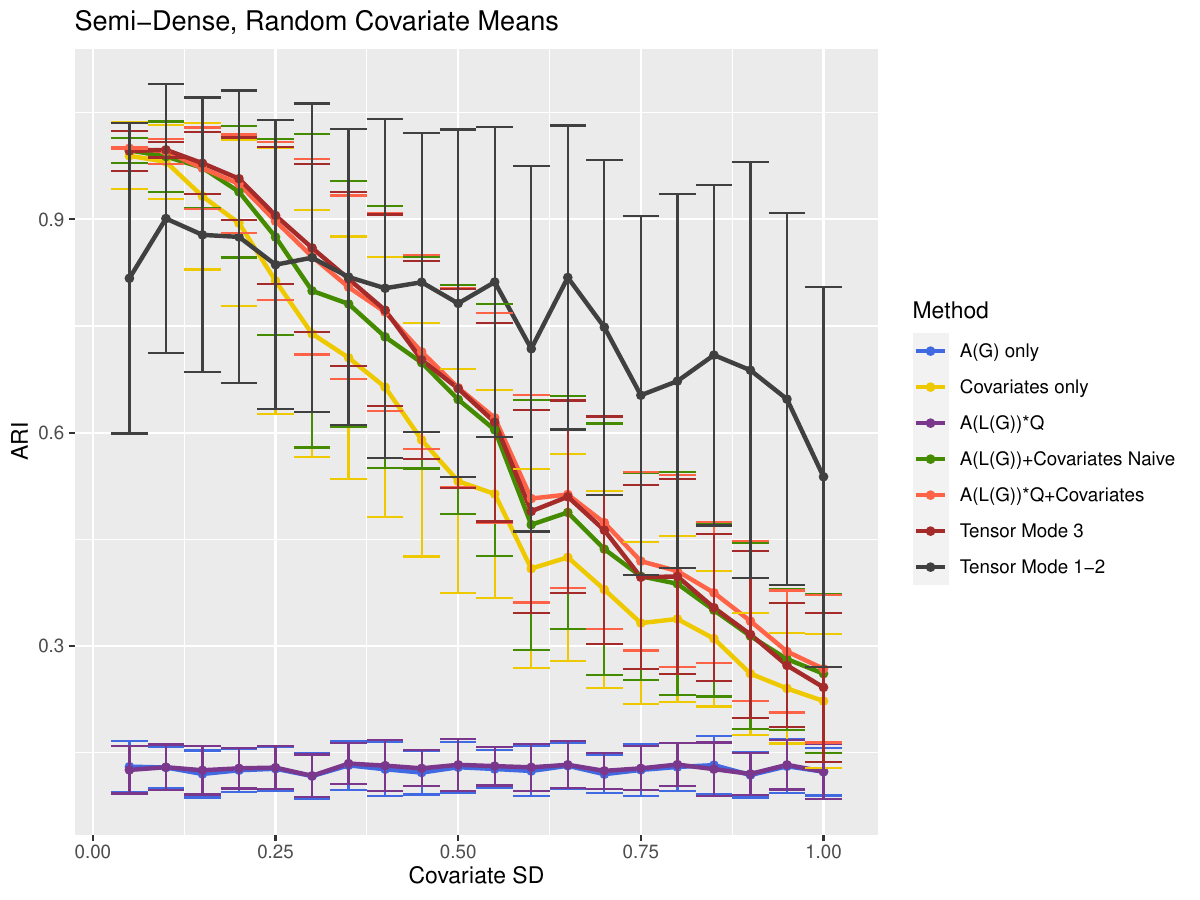}
    \caption{Semi-Dense network, random covariate means.}
\end{subfigure}

\begin{subfigure}{0.49\textwidth}
    \includegraphics[width=\textwidth]{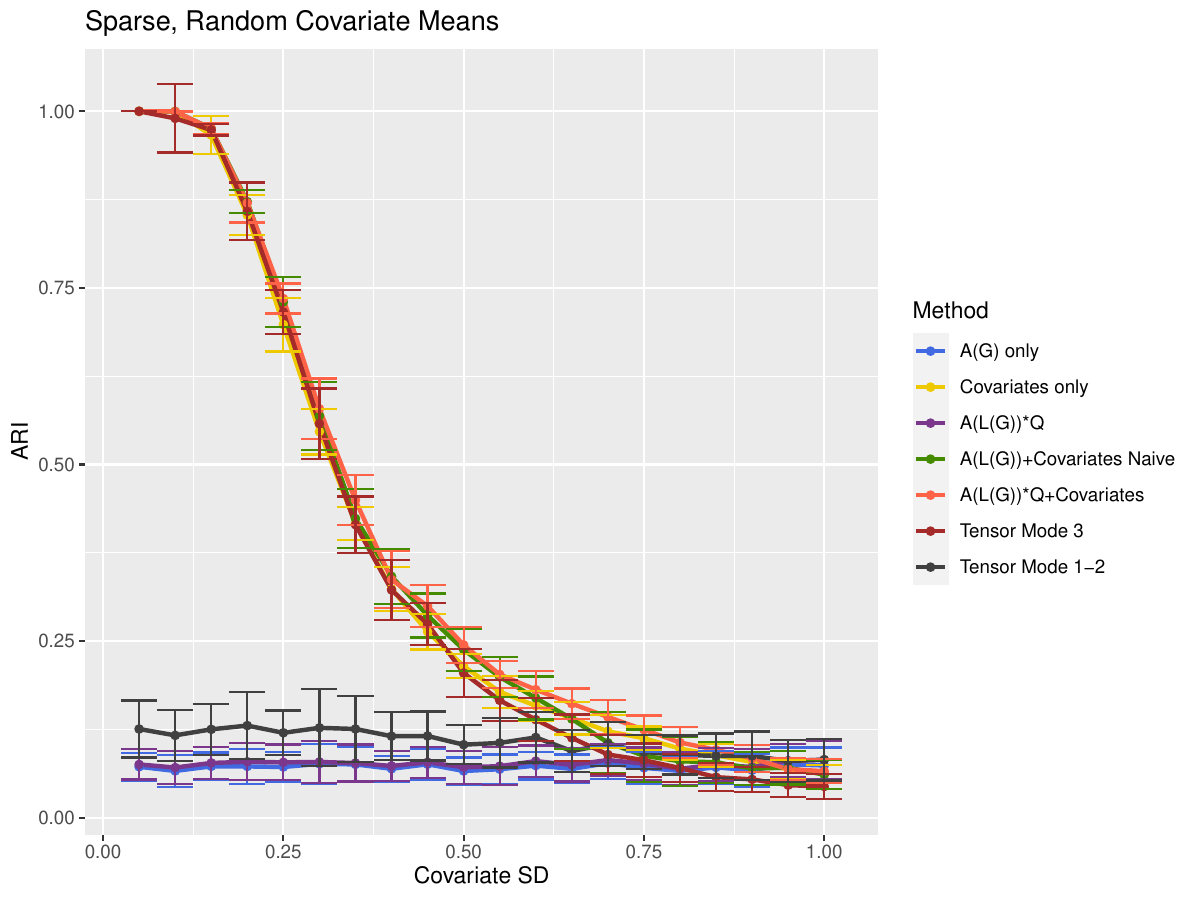}
    \caption{Sparse network, fixed covariate means.}
\end{subfigure}
\begin{subfigure}{0.49\textwidth}
    \includegraphics[width=\textwidth]{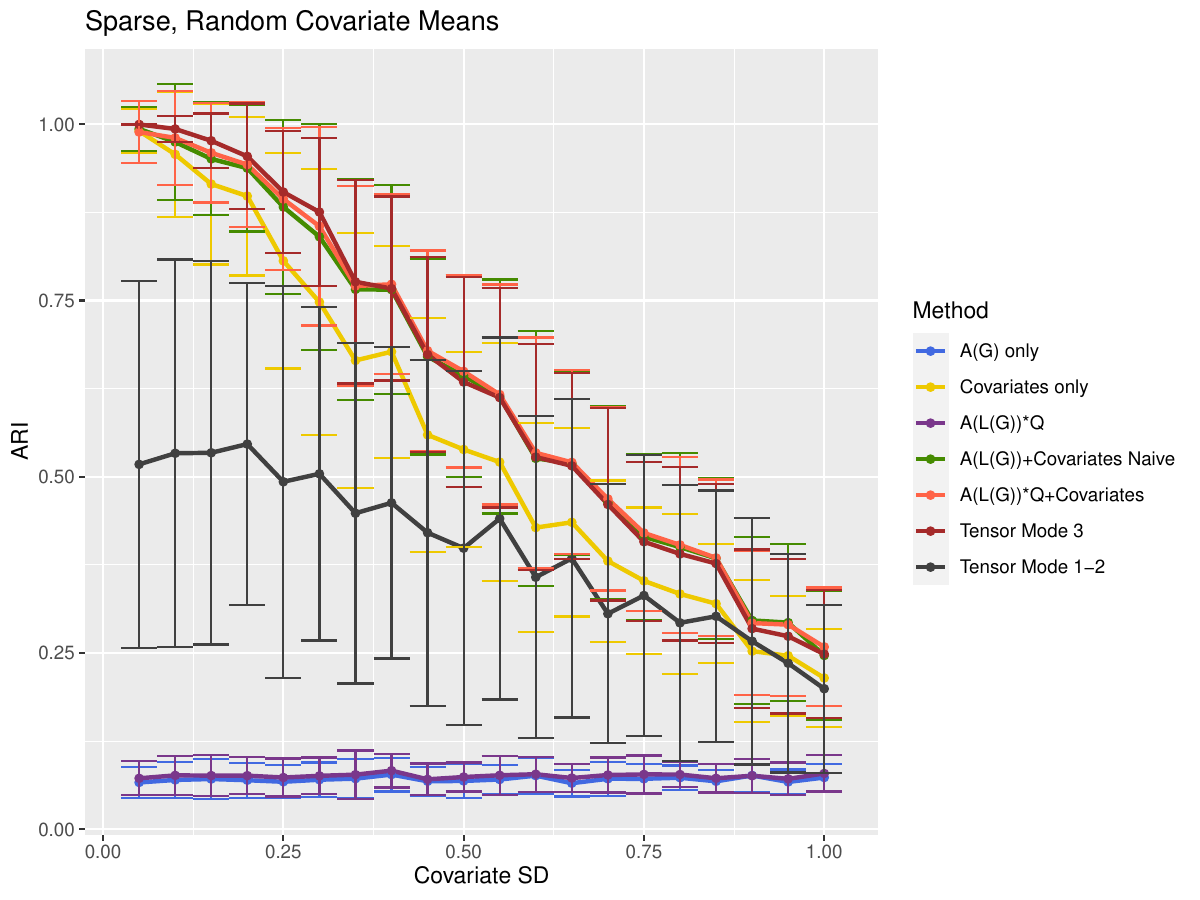}
    \caption{Sparse network, random covariate means.}
\end{subfigure}
\caption{Comparison of edge clusterings based on (i) Adjacency matrix of the original graph only, using induced clustering (blue); (ii) Edge covariates only (yellow) (iii) Adjacency matrix of the line graph, using a projection matrix, with no covariates (purple); (iv) Adjacency matrix of the line graph with covariates, but no projection matrix (green); (v) Adjacency matrix of the line graph with covariates and projection matrix (orange); (vi) Tensor decomposition mode 3 embeddings; (vii) Tensor decomposition modes 1 and 2 embeddings. All clusterings are compared to ground truth, measured using adjusted Rand index. We see that combining the information from both sources using scMASE with the projection yields results comparable or better than the maximum of those approaches which only make use of one source of data. All experiments consider a 3-block SBM with $n=300$ and 50 trials. See the text for a full description of the procedure used to generate these plots.}
\label{fig:simulations_wtensor}
\end{figure}

In the previous simulations, we used the true projection matrix $\hat{Q}$, which requires knowledge of the true clusters. We now consider the case where $\hat{Q}$ is estimated from the observed graph. In these experiments, we consider a 3-block SBM $G$ with $n=300$ vertices arranged into equally-sized blocks, with block connection probability matrix given by 
$$
B=\rho\begin{bmatrix}0.5&0.25&0.1\\0.25&0.4&0.15\\0.1&0.15&0.3\end{bmatrix},
$$
where $\rho\in\{1/10, 1/5, 1/2\},$ which we call the \emph{sparse, semi-dense,}  and \emph{dense} settings, respectively. We compute the adjacency spectral embedding of the adjacency matrix of the original graph, $\mathrm{ASE}(A(G)),$ using 3 dimensions, and obtain an approximate clustering using GMM on the rows of this matrix. From this approximate clustering, we find the induced clustering of the edges and approximate $\hat{Q}$ based on this (imperfect) clustering. This induced clustering gives us the blue curve in Figure~\ref{fig:simulations_wtensor}, which only relies on information in the observed graph. We also generate edge covariates in $\RR^3$ using two approaches:
\begin{itemize}
    \item For the fixed covariate means, we set the vectors to be $(1,0,0), (0,1,0), (0,0,1)$ for the edge memberships between vertices in the same cluster, and $(1,1,0)/\sqrt{2}, (1,0,1)/\sqrt{2}, (0,1,1)/\sqrt{2}$ for the edge memberships between vertices in different clusters.
    \item For the random covariate means, we choose 6 vectors from a $\mathcal{N}(0,I_3)$ distribution.
\end{itemize}
These covariate means are then perturbed with $\mathrm{Normal}(0,\sigma^2 I_3)$ noise for each edge in the corresponding cluster to obtain a matrix of edge covariates, $\mathcal{C}\in M_{\hat{m},3}(\RR).$ Clustering based on the rows of the singular value decomposition of $\mathcal{C}$ gives the yellow curve in Figure~\ref{fig:simulations_wtensor}. For the green and orange curves, we combine the information from the first two sources using a scaled Multiple Adjacency Spectral Embedding (scMASE) \cite{arroyo2019inference}. When deciding the scaling on the edge covariates, we make use of the variance estimate obtained from GMM from clustering the edge covariates only. The difference between these curves is the method used for obtaining the information coming from the observed graph. The green curve uses the 3 scaled singular vectors of $A(L(G))$ corresponding to the largest singular values. The orange curve uses the corresponding quantity for $A(L(G))\hat{Q}$ (where $\hat{Q}$ is computed only using observed information as described earlier). The purple curve just uses the scaled singular vectors of $A(L(G))\hat{Q}$ without using the edge covariates. We also consider a tensor-based approach which arranges the adjacency matrix and covariates as a $300\times 300\times 4$ tensor, and computes a low-rank decomposition via symmetric higher-order orthogonal iteration. A clustering of the edges is then obtained by applying GMM to the 3rd-mode tensor embeddings (red curve). We also compare this to finding a vertex clustering based on the first two modes of the tensor decomposition, then obtaining the induced clustering of the edges based on these vertex labels, which yields the black curve. All of these experiments were repeated with 50 times, and error bars are plotted for all of the approaches in all of these settings.

Comparing these approaches, we see that there is a marked improvement from using even the approximate projection over using the singular vectors of $A(L(G))$ directly.  Over the range of different variances for the edge covariates, the combined method making use of the approximate projection outperforms either of the methods that only make use of one source of information. Moreover, comparing the green and orange curves, this approximate projection clearly leads to a better extraction of the clustering information from $A(L(G))$. A tensor-based approach to edge clustering may be preferred in certain settings, but we see that the performance of this approach depends quite strongly on the density of the network.

We note that a clustering $\{\hat{C}_{(i,j)}\}_{1\leq i\leq j\leq m}$ arising from GMM applied to the rows of scMASE of $A(L(G))\hat{Q}$ and $\mathcal{C}$ may not be an induced clustering corresponding to an underlying clustering of the vertices in the original graph $G$. In the case that such a clustering of the vertices is desired, we propose maximizing $\mathrm{ARI}(\{\hat{C}_{(i,j)}\},\{\tilde{C}_{(i,j)}\})$ over all clusterings of the vertices, where $\{\tilde{C}_{(i,j)}\}$ is the induced clustering of the edges corresponding to the given clustering of the vertices. A computationally-attractive heuristic approach to solving this problem would be to cluster vertices according to a voting scheme among all edges incident to the given vertex and their clusters $\hat{C}_{(i,j)}.$ We leave a detailed analysis of this optimization for future work.

\subsection{Computational considerations}

The adjacency matrix of the line graph is typically much larger than that of the original graph, since the number of edges $m$ is often on the order of $n^2$ when the number of vertices is $n$. However, since the proposed embedding method only relies on a singular value decomposition of $A(L(G))\hat{Q}$, we may circumvent the need to construct the full line graph adjacency matrix, significantly reducing computational costs.

When we have edge covariates $v_e\in \RR^c$, we would expect to be given an edge list $\mathcal{L}\in M_{m,2+c}$, where the $\{i,j\}$th row is given by $$\mathcal{L}_{\{i,j\},\cdot}=[i,j,v_{\{i,j\}}^T].$$ From this, we can easily construct the adjacency matrix $A(G)$, and apply whichever method we choose to obtain an initial clustering of the vertices, $\hat{Z}\in M_{n,k}(\RR)$. We now observe that for the matrix with unnormalized columns $\hat{Q}_{\text{un}}\in M_{m,\binom{k+1}{2}}(\RR)$, and the incidence matrix $B\in M_{n,m}(\{0,1\})$, we get for $i\in \hat{C}_r$:
$$ (B\hat{Q}_{\text{un}})_{i,(r,s)}= \#\{j\in \hat{C}_s: \{i,j\}\in E\}= (A(G)\hat{Z})_{i,s}. $$
This immediately gives $\hat{m}_{r,s}= [(\hat{Z}^TA(G)\hat{Z})_{r,s}+(\hat{Z}^TA(G)\hat{Z})_{s,r}]/2$, so we can determine the normalizations of the columns in $\hat{Q}$. This computation can be completed in time proportional to matrix multiplication for the original adjacency matrix. Now for an edge $\{i,j\}\in E$, $i\in \hat{C}_r$, $j\in \hat{C}_s$, we get 
$$ (A(L(G))\hat{Q})_{\{i,j\},(t,u)}=\begin{cases}
[(A(G)\hat{Z})_{i,s}+(A(G)\hat{Z})_{j,r}-2]/\hat{m}_{r,s}^{1/2}&\text{if }(t,u)=(r,s)\\
(A(G)\hat{Z})_{i,u}/\hat{m}_{r,s}^{1/2}&\text{if }t=r\neq s\neq u\\
(A(G)\hat{Z})_{j,t}/\hat{m}_{r,s}^{1/2}&\text{if }t\neq r\neq s=u\\
[(A(G)\hat{Z})_{i,u}+(A(G)\hat{Z})_{j,u}]/\hat{m}_{r,r}^{1/2}&\text{if }r=s=t\neq u\\
0&\text{otherwise}.
\end{cases}$$
This computation is linear in $m$, which is true for any algoritheorem that makes use of all the edge covariates. We can now compute the singular value decomposition of $A(L(G))\hat{Q}$ to obtain the edge latent positions, all without ever having constructed the full matrix $A(L(G))$.

\section{Discussion}

In real networks, relevant information is often associated to network edges. The line graph of a random graph effectively exchanges the roles of vertex and edge, so spectral decompositions of the line graph can yield consistent inference for random graphs with edge covariates. They are therefore a valuable tool for broader graph inference. 

Our work here establishes foundational components of spectral inference in a line graph: first, rendering well-defined the mean matrix for such a graph, which has a necessarily random vertex set; second, probabilistic concentration of the spectrum of the adjacency matrix of the line graph; third, detailed analysis of the largest eigenvalue and empirical spectral distribution for the ER line graph, mirroring results for the adjacency matrix and Laplacian in this setting; and finally, the determination of certain signal-preserving singular subspaces of this adjacency matrix and their extraction through an appropriate projection. The first of these is an important departure from analogous results for random graphs on a fixed vertex set; the second and third are natural parallels to similar results for random matrices; and the fourth is a key contribution of this work: without the ``correct" singular subspace, inference with edge covariates can be significantly corroded, but an appropriate projection can isolate the signal-preserving subspace, with which edge covariate information can be integrated for more accurate inference. 

More specifically, we establish that under mild conditions, the line graph of a stochastic block model has block-model structure. That is, the mean matrix of the line graph of an SBM can be decomposed into its actions on certain orthogonal invariant subspaces, one of which is spanned by block-constant vectors. Since the entries of these vectors are constant for edges in the same cluster, this is a signal-preserving subspace for edge clustering. Estimating the corresponding projection operator onto this subspace allows us to estimate important signal dimensions. Our simulations show that (a) the signal dimensions cannot be obtained without a projection, and (b) even when the estimated projection matrix is only approximately correct, it still enables us to extract useful signal from the line graph. Incorporating edge covariate information with the approximate projection is a vast improvement over discarding the covariates altogether, or using a naive singular value decomposition and projecting onto an inapt subspace. 

There are several theoretical and practical questions that remain, even in the case of SBM graphs. In Section~\ref{s:edgecov}, we consider estimating the projection $\hat{Q}$ using the induced clustering arising from an estimated clustering of the vertices in the underlying graph. From Figure~\ref{fig:simulations_wtensor}, we see that an estimated projection $\hat{Q}$ relying on a (likely imperfect) initial clustering of the vertices is still useful for capturing the signal in $A(L(G))$. Rigorous concentration bounds for $A(L(G))\hat{Q}$, when $\hat{Q}$ itself relies on an estimated clustering, is important for both performance guarantees and robust implementation on real data. 
A related question is that of choosing optimal scalings for scMASE, which we use to combine the information from adjacency and edge covariates. A plausible heuristic is to consider a range of scalings for the edge covariates, generate multiple embeddings corresponding to different values of this parameter, and choose the value that maximizes the clustering coefficient for this embedding. 

Our analysis of line graphs overcomes major hurdles, including random graph size, high rank, and a nonexistent spectral gap. The very weak conditions in our key concentration inequality (discussed in Remark~\ref{rmk:concentrate}) show the utility of viewing the observed line graph as a random subgraph of the line graph of the complete graph on the same set of vertices. For extensions to random graphs beyond SBM graphs---for example, a random dot product graph---Theorem \ref{theorem:concentrate} still guarantees concentration of the spectrum. If, in addition, the mean of the random line graph exhibits identifiable matrix structure, we may be able to find a projection that permits estimation of a signal-preserving subspace for the specified inference task. This holds real promise for inference in a wider class of graph models. 

\begin{supplement}
\stitle{Proofs}
\sdescription{The supplement contains all proofs, both for the main results and for the damped binomial distribution.}
\end{supplement}

\section*{Acknowledgements}

This work was supported in part by
the Defense Advanced Research Projects Agency under the LogX program (N6523620C8008- 01) and the D3M program administered through contract FA8750-17-2- 0112,
the Naval Engineering Education Consortium (NEEC), Office of Naval Research (ONR) Award Number N00174-19-1-0011,
and by funding from Microsoft Research.

The authors would like to thank Benjamin Pedigo for his insightful comments on the manuscript. 

\bibliographystyle{imsart-number} 
\bibliography{biblio_summary_feb2224}    


\newpage

\title{Supplement to ``Random line graphs and edge-attributed network inference''}


\begin{aug}
\author[A]{\fnms{Zachary} \snm{Lubberts}}
\author[B]{\fnms{Avanti} \snm{Athreya}\
}
\author[C]{\fnms{Youngser} \snm{Park}
}
\and 
\author[B]{\fnms{Carey E.}
\snm{Priebe}
}
\address[A]{Department of Statistics, University of Virginia}

\address[B]{Department of Applied Mathematics and Statistics, Johns Hopkins University, 
}

\address[C]{Center for Imaging Science, Johns Hopkins University,
}

\end{aug}

\vspace{0.2in}

\noindent {\large {\bf Supplementary material: proofs of main and supporting results.}}\\
Here, we provide proofs of main and supporting results for our paper ``Random line graphs and edge-attributed network inference." In the proofs we give here, we rely on the following decomposition and representation for the adjacency matrix of a line graph. Let $G$ be an undirected graph on $n$ vertices with edge set $E$ with size $\hat{m}$.
Let $B\in M_{n,\widehat{m}}(\ZZ)$ be defined by 
\begin{equation}\label{eq:B-def}
B_{i,e}=\chi_{e}(i)
\end{equation}
where $\chi_{e}$ is the indicator function of $e\in E\subseteq \binom{V}{2}$. Then
\begin{equation}\label{B-trans-b-def}
(B^{T}B)_{e_{1},e_{2}}=\sum_{i=1}^{n}\chi_{e_{1}}(i)\chi_{e_{2}}(i)=\begin{cases}2&e_{1}=e_{2}\\ 1&|e_{1}\cap e_{2}|=1\\ 0&e_{1}\cap e_{2}=\varnothing,\end{cases}
\end{equation} so $A(L(G))=B^{T}B-2I$. On the other hand, \begin{equation}\label{eq:B-b-trans-def}
(BB^{T})_{i,j}=\sum_{e\in E}\chi_{e}(i)\chi_{e}(j)=\begin{cases}\mathrm{deg}_{G}(i)&i=j\\ \chi_{E}(e)&i\neq j,\end{cases}\end{equation} 
so $A(G)=BB^{T}-D$, with $D$ the diagonal matrix of degrees. In summary, we derive that
\begin{align}\label{eq:line-graph-bb-representation}
A(L(G))&=B^{T}B-2I\\
A(G)&=BB^{T}-D
\end{align}
The following is the proof of Proposition~\ref{prop:lgeigs} and its corollary.
\begin{proof}
Let $B\in M_{n,\widehat{m}}(\ZZ)$ be defined by Equation~\ref{eq:B-def}.
If $\widehat{m}<n$, $A(G)+D=BB^{T}$ has the same eigenvalues as $B^{T}B=A(L(G))+2I$ along with $n-\widehat{m}$ zeros, so subtracting $2$ from the eigenvalues of both sides gives the result. On the other hand, when $\widehat{m}\geq n$, $A(L(G))+2I=B^{T}B$ has the same eigenvalues as $BB^{T}=A(G)+D$ along with $\widehat{m}-n$ zeros, so subtracting $2$ from the eigenvalues of both sides gives the result.\\

Finally, when $G=K_{n}$, $\widehat{m}=\binom{n}{2}$, $D=(n-1)I$, and $A(G)=J-I$, so $A(G)+D=J+(n-2)I,$ which has eigenvalues $\{2n-2,n-2\,(\text{mult. }n-1)\}$. Plugging this into the proposition gives the corollary.
\end{proof}

We will need the following results in what follows. The first may be found in \cite{Tropp2015}:

\begin{res}[Matrix Chernoff Bound]
\label{res:matchernoff}
Consider a finite sequence $\{X_{\ell}\}$ of independent, random, Hermitian matrices with common dimension $d$. Assume that $$0\leq \lambda_{\text{min}}(X_{\ell})\text{ and }\lambda_{\text{max}}(X_{\ell})\leq L\text{ for each index }\ell.$$ Introduce the random matrix $Y=\sum_{\ell}X_{\ell},$ and define the extreme eigenvalues $\mu_{\text{min}},\mu_{\text{max}}$ of $\EE Y$: \begin{align*}\mu_{\text{min}}&=\lambda_{\text{min}}(\EE Y)=\lambda_{\text{min}}\left(\sum_{\ell}\EE X_{\ell}\right),\text{ and}\\ \mu_{\text{max}}&=\lambda_{\text{max}}(\EE Y)=\lambda_{\text{max}}\left(\sum_{\ell}\EE X_{\ell}\right).\end{align*} Then \begin{align*} \EE\lambda_{\text{min}}(Y)&\geq 0.63\mu_{\text{min}}-L\log d,\text{ and}\\ \EE\lambda_{\text{max}}(Y)&\leq 1.72\mu_{\text{max}}+L\log d,\end{align*} and \begin{align*} \PP\left[\lambda_{\text{min}}(Y)\leq t\mu_{\text{min}}\right]&\leq de^{-(1-t)^{2}\mu_{\text{min}}/2L}\text{ for }t\in[0,1),\text{ and}\\ \PP\left[\lambda_{\text{max}}(Y)\geq t\mu_{\text{max}}\right]&\leq d\left(\frac{e}{t}\right)^{t\mu_{\text{max}}/L}\text{ for }t\geq e.\end{align*}
\end{res}

The following result may be found in \cite{HJ85_matrix}:

\begin{res}[Ostrowski]
\label{res:ostrowski}
Let $A\in M_{n}$ be Hermitian, $S\in M_{n}$ be nonsingular. Then for each $\ell=1,\ldots,n$, $$\lambda_{\ell}(SAS^{*})=\theta_{\ell}\lambda_{\ell}(A),$$ where $\theta_{\ell}\in[\sigma_{n}(S)^{2},\sigma_{1}(S)^{2}]$. 
\end{res}

\begin{proof}[Proof of Theorem~\ref{theorem:concentrate}]
The crucial step is finding the right matrix to express as a sum of random Hermitian matrices so that we may apply Result~\ref{res:matchernoff}. We observe that $A(L(G))$ is the submatrix of $A(L(K_{n}))$ with rows and columns indexed by $e\in E$. Then if we let $P=\mathrm{diag}([\delta_{e}]_{e\in \binom{[n]}{2}}),$ where $\delta_{\{i,j\}}\sim\text{Bernoulli}(\mathcal{P}_{i,j})$ are all independent, any particular observation of the matrix $P$ yields $A(L(G))$ as the nonzero submatrix of $PA(L(K_{n}))P$, since this matrix just zeroes out all rows and columns of $A(L(K_{n}))$ for which $\delta_{e}=0$, which is equivalent to $e\not \in E$. So as in the proof of Proposition~\ref{prop:lgeigs}, let $B=[\chi_{e}(i)]_{i\in[n],e\in\binom{[n]}{2}},$ so that $A(L(K_{n}))+2I=B^{T}B.$ Now if we consider $BP,$ letting $B_{e}$ be the $e$th column of $B$, and $j_{e}$ the $e$th standard unit vector, we see that $BP=\sum_{e}\delta_{e}B_{e}j_{e}^{T}.$ Then the matrix $P(A(L(K_{n}))+2I)P=PB^{T}BP$ has $\binom{n}{2}-n$ zero eigenvalues, plus the eigenvalues of $$Y=BP(BP)^{T}=\sum_{e_{1},e_{2}}\delta_{e_{1}}\delta_{e_{2}}(B_{e_{1}}j_{e_{1}}^{T})(j_{e_{2}}B_{e_{2}}^{T})=\sum_{e}\delta_{e}B_{e}B_{e}^{T},$$ since $\delta_{e}^{2}=\delta_{e}$, being a $\{0,1\}$-valued random variable. This is a sum of independent, random, Hermitian, positive semidefinite matrices with common dimension $n$, so we can apply the matrix Chernoff bounds of Result~\ref{res:matchernoff} so long as we can get estimates for $\mu_{\text{min}},\mu_{\text{max}},$ and $L$.\\

The bound on $L$ follows since each column of $B$ has exactly 2 nonzero entries, meaning that $\text{max}_{e}\|B_{e}\|_{2}^{2}=2,$ so $\|\delta_{e}B_{e}B_{e}^{T}\|_{2}\leq 2$ for all $e$, which means $L=2$ suffices. Considering $\EE Y= \EE BPB^{T}=B(\EE P)B^{T},$ we see that $\EE P=\mathrm{diag}(\mathcal{P}_{i,j}),$ which we will call $R^{2}$, for the diagonal matrix $R$. Now $\EE Y=BR(BR)^{T}$ has the same nonzero eigenvalues as $(BR)^{T}BR,$ which is a $\ast$-congruence of the matrix $B^{T}B$, so by Result~\ref{res:ostrowski}, for $1\leq \ell\leq n$, $\lambda_{\ell}(\EE Y)=\lambda_{\ell}(RB^{T}BR)=\theta_{\ell}\lambda_{\ell}(B^{T}B)$ for some $\theta_{\ell}\in [\sigma_{\text{min}}^{2}(R),\sigma_{\text{max}}^{2}(R)]=[p_{\text{min}},p_{\text{max}}].$ Combining this with the information from Corollary~\ref{c:lgcompleteeigs}, since $B^{T}B=A(L(K_{n}))+2I$, we see that \begin{align*} \mu_{\text{min}}&=\lambda_{n}(\EE Y)\in [p_{\text{min}},p_{\text{max}}](n-2),\text{ and}\\ \mu_{\text{max}}&=\lambda_{1}(\EE Y)\in [p_{\text{min}},p_{\text{max}}]2(n-1).\end{align*}

Plugging this information into Result~\ref{res:matchernoff}, we get
\begin{align*} \EE\lambda_{n}(Y)&\geq 0.63 p_{\text{min}}(n-2)-2\log n,\text{ and}\\ \EE\lambda_{1}(Y)&\leq 1.72(p_{\text{max}}2(n-1))+2\log n,\end{align*} and \begin{align*} \PP\left[\lambda_{n}(Y)\leq t\mu_{\text{min}}\right]&\leq n e^{-(1-t)^{2}\mu_{\text{min}}/4}\text{ for }t\in[0,1),\text{ and}\\ \PP\left[\lambda_{1}(Y)\geq t\mu_{\text{max}}\right]&\leq n \left(\frac{e}{t}\right)^{t\mu_{\text{max}}/2}\text{ for }t\geq e.\end{align*} In the probability statements, we decrease the size of the first event by using the lower bound on $\mu_{\text{min}},$ $p_{\text{min}}(n-2)$, and increase the bound on the right hand side of the first line by using the same bound on this quantity. For the second line, we decrease the size of the event by using the upper bound on $\mu_{\text{max}},$ $p_{\text{max}}2(n-1),$ and increase the bound on the right hand side by using the lower bound on $\mu_{\text{max}},$ $p_{\text{min}}2(n-1).$\\

Now the eigenvalues of $A(L(G))+2I$ are the top $\widehat{m}$ eigenvalues of $P(A(L(K_{n}))+2I)P$, which has $\binom{n}{2}-n$ zero eigenvalues, along with the eigenvalues of $Y$. Then the eigenvalues of $A(L(G))$ are just $(\sigma(Y)-2)\cup\{-2\},$ where we observe that if $\widehat{m}<n$, $Y$ must have zero eigenvalues, so $\lambda_{n}(A(L(G)))=-2$ is the correct assignment for the values in this case.
\end{proof}

\begin{proof}[Proof of Theorem~\ref{theorem:topeig}]
Recalling Equation~\ref{eq:line-graph-bb-representation}, observe that the adjacency matrix of the line graph, $A(L(G))$, satisfies
$A(L(G))+2I=B^TB$. Thus $\sigma(A(L(G))$, the spectrum of the line graph, corresponds to the spectrum of $B^TB-2I$. Of course, the spectrum of $B^TB$ is the same as that of $BB^T$, and again from Equation~\ref{eq:line-graph-bb-representation}, we find that $BB^T=A(G)+D$ where $A(G)$ is the adjacency matrix of the graph $G$ itself and $D$ is its diagonal matrix of degrees. Importantly, $BB^T$ is an $n \times n$ matrix of fixed size, and its mean is straightforward to compute:
\begin{equation}\label{eq:mean_of_B-btrans}
\mathbb{E}[BB^T]=p(J-I)+(n-1)pI
\end{equation}
where $J$ is the matrix of all ones. The largest eigenvalue of this matrix is $2(n-1)p$, and all other eigenvalues are $(n-2)p$. Applying the matrix Bernstein inequality \cite{Tropp2015}, we see that $$\PP[\|BB^T-\EE[BB^T]\|\leq c\sqrt{n\log(n)}]\geq 1-C n^{-2}.$$ So by Weyl's inequality, on this high-probability event, $|\lambda_1(BB^T)-\lambda_1(\EE[BB^T])|\leq c\sqrt{n\log(n)}$. In other words, with high probability, $\lambda_1(A(L(G)))-[2(n-1)p-2]\sim O(\sqrt{n\log(n)})$. Applying the Borel-Cantelli lemma, we see that almost surely, $$\frac{\lambda_1(A(L(G)))}{n}-2p\rightarrow0\text{ as }n\rightarrow\infty.$$
\end{proof}

To prove Theorem~\ref{theorem:topeigclt}, we require the following lemma. While the statements are quite similar to Lemma 3 in \cite{furedi1981eigenvalues}, our setting is made significantly more complicated by the introduction of diagonal terms which are dependent on all of the off-diagonal entries in the corresponding row and column, while \cite{furedi1981eigenvalues} assumes independence between the diagonal and off-diagonal entries. For ease of exposition, we defer the proof of Lemma~\ref{lem:lemma3analogue} until later in the supplement. For now, we state the needed bounds and then use them in the proof of Theorem~\ref{theorem:topeigclt}.
\begin{lemma}
\label{lem:lemma3analogue}
\begin{align}
\PP &[ |\|S-Lj\|^2-4n^2p(1-p)|> 8p(1-p)n^{3/2}x] < 1/x^2 \label{eq:clt1}\\
\PP&\left[ \left|\sum_{i,j} \tilde{A}_{i,j}(S_i-L)(S_j-L)-4n^2p(1-p)(2-3p)\right|>Cn^2x\right] <1/x^2 \label{eq:clt2}\\
\EE&\left[ \sum S_i /n\right] = (n-1)p, \quad \mathrm{Var}(\sum S_i/n) = 8p(1-p) -8p(1-p)/n \label{eq:clt3}\\
\PP&\left[ \left| \frac{\sum S_i^2}{\sum S_i} - \frac{\sum S_i}{n}-4(1-p)\right| > Cp(1-p)x^2/n\right] \leq 1/x^2 \label{eq:clt4}
\end{align}
Moreover, $\sum S_i/n$ is a scaled, translated Binomial random variable, so centering and scaling by the mean and variance given in (\ref{eq:clt3}) as $n\rightarrow\infty$, this quantity will converge in distribution to a Normal random variable.
\end{lemma}

We are now ready to prove Theorem~\ref{theorem:topeigclt}.

\begin{proof}[Proof of Theorem~\ref{theorem:topeigclt}]
The argument begins as the proof of Theorem~\ref{theorem:topeig}, though we now require tighter control on the top eigenvalue of $BB^T=A(G)+D$. To obtain such control, we follow \cite{furedi1981eigenvalues}. Define $\tilde{A}= A+D-(n-1)pI, \tilde{P}=\EE[\tilde{A}]=p(J-I),$ where $J$ is the $n\times n$ matrix of all ones. Let $S= \tilde{A}j$, where $j$ is the vector of all ones. Note that $\EE[S]= (n-1)pj=:Lj$. 

We may write $\tilde{A}-\tilde{P}= \sum_{i<j} (A_{ij}-p)(e_i+e_j)(e_i+e_j)^T$ as a sum of independent, mean-0 Hermitian matrices, with each term bounded by $2$ in spectral norm. The matrix variance is given by 
\begin{align*}
\EE[\sum_{i<j} 2(A_{ij}-p)^2 (e_i+e_j)(e_i+e_j)^T] &= 2p(1-p)\sum_{i<j} (e_i+e_j)(e_i+e_j)^T\\
&= 2p(1-p)(J-I)+2(n-1)p(1-p)I,
\end{align*}
which has spectral norm $4(n-1)p(1-p).$ Then by the matrix Bernstein inequality, 
$$\PP[\|\tilde{A}-\tilde{P}\|>t] \leq 2n\exp\left(\frac{-t^2/2}{4(n-1)p(1-p)+2t/3}\right),$$ so with high probability, $\|\tilde{A}-\tilde{P}\|\leq C\sqrt{n\log(n)p(1-p)}.$ Since $\tilde{P}$ has one eigenvalue $(n-1)p$ and the remaining eigenvalues are just $-p$, Weyl's inequality tells us that on this high-probability event, 
\begin{equation}
\label{eq:clt6}
\max_{i\geq 2} |\lambda_i(\tilde{A})+p| \leq C\sqrt{n\log(n)p(1-p)}.
\end{equation}

The rest of the argument now follows as in \cite{furedi1981eigenvalues}, utilizing Lemma~\ref{lem:lemma3analogue} in place of Lemma 3 in \cite{furedi1981eigenvalues}, and Equation~(\ref{eq:clt6}) in place of Theorem 2 from \cite{furedi1981eigenvalues}.
\end{proof}

We prove Theorem~\ref{theorem:esd} on the weak limit of the empirical spectral distribution of a Wigner matrix. The following lemma, a variant of a result in \cite{van2000asymptotic} on weak limits and moment sequences, is helpful.

\begin{lemma}
\label{lem:momentsequence}
Let $\mu_n$ be measures on $\RR$ for all $n\geq 1$, and suppose the following:
\begin{enumerate}
\item for all $k\geq 0$, 
\begin{equation}\label{eq:unif_integ_moments}
\int x^k\, \mathrm{d}\mu_n\rightarrow m_k\in \RR\text{ as }n\rightarrow\infty,
\end{equation}
\item $m_k$ satisfy Carleman's condition.
\end{enumerate}
Then the sequence $\mu_n$ is tight; there exists a unique measure $\mu$ on $\RR$ satisfying $m_k = \int x^k\,\mathrm{d}\mu$ for all $k\geq0$; and $\mu_n\overset{\mathrm{w}}{\rightarrow}\mu$ as $n\rightarrow\infty$. Condition (iii) may be replaced by any sufficient condition for determinacy of the Hamburger moment problem.
\end{lemma}
\begin{proof}[Proof of Lemma~\ref{lem:momentsequence}]

Note that Eq. \ref{eq:unif_integ_moments} guarantees uniform integrability of $|x|^k$ for all $k$  By Cauchy-Schwarz, it suffices to show this for even $k$, so observe that for any $A>0$
\begin{equation}\label{eq:Lemma_2_unif_integ}\int_{[-A,A]^c} x^{2k}A^2\,\mathrm{d}\mu_{n} \leq \int_{[-A,A]^c} x^{2k+2}\,\mathrm{d}\mu_{n}\leq \int_\RR x^{2k+2}\,\mathrm{d}\mu_{n}.
\end{equation}
Since the right hand side is a convergent sequence of real numbers, it is bounded, and thus
\begin{equation}\label{eq:lemma2_unif_integ_final_bound}\int_{[-A,A]^c} x^{2k}\,\mathrm{d}\mu_{n} \leq \frac{B_k}{A^2},
\end{equation}
Putting $k=0$ implies that
$$\mu_n([-A, A]^c) \leq \frac{B_0}{A^2}$$
which can be made arbitrarily small by choosing $A$ sufficiently large. Hence $\mu_n$ is a tight sequence. By Prohorov's Theorem, there exists a subsequence $\mu_{n_j}$ that converges weakly to a probability measure $\mu^{S}$, where $S$ denotes the potential dependence of this limit on the choice of subsequence. By weak convergence of $\mu_{n_j}$ and uniform integrability of $|x|^k$,
$$m_k=\lim\limits_{j \rightarrow \infty}\int_{\mathbb{R}} |x|^k \mathrm{d}\mu_{n_j}=\int |x|^k \mathrm{d}\mu^{S}.$$
Thus $\mu^S$ has $k$th moment given by $m_k$ for all $k\geq 0$, and therefore $m_k$ is a moment sequence. Since $m_k$ satisfies Carleman's condition, there is only one measure with these moments. As a result, $\mu^S$ is independent of the subsequence itself, so we can write $\mu^S\equiv \mu$. This guarantees weak convergence of the full sequence $\mu_n$ to $\mu$.


\end{proof}


Our proof of Theorem \ref{theorem:esd} relies on moment convergence for the empirical spectral distribution, bounds sufficient to guarantee convergence with probability one, and detailed counting arguments. We establish, using elementary methods, that the empirical spectral distribution coincides with that given in Theorem 2 of \cite{ding2010spectral}, where the authors consider, by contrast, diagonal entries that are the {\em negation} of the row sums of the diagonal.
\begin{proof}[Proof of Theorem \ref{theorem:esd}] For any given WignerDD matrix $W_n$, scaling the off-diagonal entries by $\sigma$ produces a Wigner DD matrix with unit variance, so without loss of generality we consider the case $\sigma=1$. For such a WignerDD matrix $W_n$,  we order the necessarily real eigenvalues by
$$\lambda_1(W_n) \leq \cdots \leq \lambda_n(W_n).$$
Let $X_n=\frac{W_n}{\sqrt{n}}$.
Let $\hat{\mu}_n:=\hat{\mu}_{X_n}$ denote the empirical spectral distribution of $X_n$. This is a random measure on $\mathbb{R}$ defined by
$$\hat{\mu}_n((-\infty, x])=\frac{1}{n} \sum_{j=1}^n I_{(-\infty, x]}(\lambda_j(X_n)).$$
The $k$th moment of this random measure is the random variable
\begin{equation}\label{eq:hatuk}
\hat{L}^{k}_n=\int_{\mathbb{R}} x^k\, d\hat{\mu}_n(x)=\frac{1}{n} \sum_{j=1}^n \lambda_j^k(X_n)
=\frac{1}{n} \tr\left(\left(\frac{W}{\sqrt{n}}\right)^k\right)
\end{equation}
We show that the sequence of random measures $\hat{\mu}_n$ converges weakly almost surely, as $n \rightarrow \infty$, to a unique and deterministic measure $\mu$ on $\mathbb{R}$ with cumulative distribution function $L(x)$. For $k$ a positive integer, the odd moments $L_{2k+1}$ of this limiting distribution are zero, and the even moments are given by
$$ L_{2k} = \sum_{m=0}^{k-1} \frac{1}{m+1} \sum_{\substack{j_1,\ldots,j_{m+1}\geq 0\\ \sum j_i = m}}\;\sum_{\substack{n_1,\ldots,n_{m+1}\geq 0 \\ \sum n_i = k-m}} \;\prod_{i=1}^{m+1} \binom{2n_i+j_i}{j_i} \frac{(2n_i)!}{2^{n_i} n_i!}. $$
To establish this, we proceed as follows. First, we prove a moment convergence for the expected values of the $k$th random moments under $\hat{\mu}_n$, namely that  
\begin{equation}\label{eq:moment_convergence}
\lim\limits_{n \rightarrow \infty}\mathbb{E}[\hat{L}_n^k]=\lim\limits_{n \rightarrow \infty}\mathbb{E}\left[\frac1n\tr\left(\left(\frac{W}{\sqrt{n}}\right)^k\right)\right]=L_k
\end{equation}
Second, we show that for any $k$, 
\begin{equation}\label{eq:random_moment_close_to_mean}
\PP\left[\lim_{n\rightarrow\infty}\bigg|\hat{L}_n^k-\mathbb{E}[\hat{L}_n^k]\bigg|=0\right]=1.
\end{equation}

The limiting results in Equations~(\ref{eq:moment_convergence}) and (\ref{eq:random_moment_close_to_mean}) imply that 
for each $k\geq 1$, the set $A_k=\{\omega: \lim\limits_{n\rightarrow \infty}\hat{L}_n^k(\omega)= L_k\}$ has full measure. Next, we show that the deterministic sequence of real numbers $L_k$ satisfies Carleman's condition. With this in hand,
put $\mathcal{A}=\cap A_k$; then $\mathcal{A}$ has full measure, and for any $\omega \in \mathcal{A}$,  $\hat{L}_n^k(\omega)\rightarrow L_k$ for all $k \geq 0$. Lemma~\ref{lem:momentsequence} establishes that there exists a unique, deterministic probability measure $\mu$ such that for all $\omega \in \mathcal{A}$, $\hat{\mu}_n(\omega)$ converges weakly to $\mu$. 

Thus the proof is complete once we verify Equations~(\ref{eq:moment_convergence}) and (\ref{eq:random_moment_close_to_mean}) and demonstrate that $L_k$ satisfy Carleman's condition.
We focus now on Equation~(\ref{eq:moment_convergence}), which is the key claim. Since the trace of a matrix is the sum of its diagonal elements, and since the $r$th diagonal element of $W_n^k$ is given by
$$(W_n^k)_{rr}= \left(\sum_{i_2, \cdots, i_k} \left[W_{r i_2} W_{i_2 i_3} \cdots W_{i_kr}\right]\right)$$
where, for notational ease on the right-hand side, we suppress the dependence on $n$ of the entries of $W_n$, and adopt the convention that $W_{i_l i_{l+1}}=(W_n)_{i_l i_{l+1}}$. Note that the sum on the right-hand side is over all $k-1$ tuples $(i_2,\cdots i_k)$ whose components range from 1 to $n$. Hence
\begin{align}
\mathbb{E}\left[\frac{1}{n} \tr\left(\left(\frac{W_n}{\sqrt{n}}\right)^k\right)\right]&=\frac{1}{n^{k/2+1}} \sum_{i_1=1}^n \left(\sum_{i_2, \cdots, i_k} \mathbb{E}\left[W_{i_1i_2} W_{i_2 i_3} \cdots W_{i_ki_1}\right]\right)\\
&=\frac{1}{n^{k/2}+1} \sum_{(i_1, i_2, \cdots i_k)}\mathbb{E}\left[W_{i_1i_2} W_{i_2 i_3} \cdots W_{i_ki_1}\right]\label{eq:product_of_matrix_entries_expected_k_moment}
\end{align}
where, again, the sum is over all possible $k$ tuples $(i_1, \cdots, i_k)$ whose individual components range from $1$ to $n$. 
Consider the entries $W_{i_l i_{l+1}}$ for any $1 \leq l \leq k$. Suppose there exists some $l$, $1 \leq l \leq k-1$, with $i_l=i_{l+1}$, or $i_k=i_1$. In any such instance, there is at least one diagonal entry of $W_n$ among the factors that comprise the product in Eq. \ref{eq:product_of_matrix_entries_expected_k_moment}. Because a WignerDD  matrix $W$ has the property that
$W_{ii}=\sum_{j \neq i} W_{ij}$, we can replace any diagonal entry $W_{i_l i_l}$ in Eq. \ref{eq:product_of_matrix_entries_expected_k_moment} by $W_{i_l i_l}=\sum_{j_l \neq i_l} W_{i_l j_l}$. Let ${\bf i}=(i_1, \cdots i_k)$, and ${\bf j}=(j_1, \cdots j_k)$; put $({\bf i}, {\bf j})=((i_1, j_1), \cdots, (i_k, j_k))$. We call $({\bf i}, {\bf j})$ a {\em length-$k$ string of pairs} or a {\em paired length-$k$ string}.
With this substitution, we conclude that
\begin{align}
   \mathbb{E}\left[\frac{1}{n} \tr\left(\left(\frac{W_n}{\sqrt{n}}\right)^k\right)\right]
&=\frac{1}{n^{k/2}+1} \sum_{({\bf i}, {\bf j}) \in \mathcal{S}} \mathbb{E}\left[W_{i_1j_1} W_{i_2 j_2} \cdots W_{i_kj_k}\right] \label{eq:dediagonalized_product}
\end{align}
where $\mathcal{S}$ is the set consisting of all possible length-$k$ strings of pairs $({\bf i}, {\bf j})=((i_1, j_1), \cdots, (i_k, j_k))$ satisfying the following requirements:
\begin{enumerate}
\item For every $l \in \{1, \cdots, k\}$, $i_l \in \{1, \cdots,  n\}$ and $ j_l \in \{1, \cdots, n\}$;
\item For every $l \in \{1, \cdots, k\}$, $i_l \neq j_l$; 
\item For every $1 \leq l \leq k-1$, either $i_{l+1}=j_l$ (that is, the $l$th factor in the product in Eq. \ref{eq:product_of_matrix_entries_expected_k_moment} was not a diagonal entry) or $i_{l+1}=i_l$ (that is, the $l$th factor in the product in Eq. \ref{eq:product_of_matrix_entries_expected_k_moment} was a diagonal entry), and either $i_1=j_k$ or $i_1=i_k$.
\end{enumerate}
We call the sum
$$\sum_{({\bf i}, {\bf j}) \in \mathcal{S}} \left[W_{i_1j_1} W_{i_2 j_2} \cdots W_{i_kj_k}\right] $$
the {\em de-diagonalized} representation of the trace of the $k$th power of $W_n$.
To compute the expectation
$$\mathbb{E}\left[W_{i_1j_1} W_{i_2 j_2} \cdots W_{i_kj_k}\right]$$
in the de-diagonalized representation, observe that if there is any pair $(i_l, j_l)$ that is not repeated among the $k$ pairs (when taking symmetry of $W$ into account), then there is a mean-zero term $W_{i_l j_l}$ that is independent of all remaining factors, rendering the expectation of the product zero. Hence it suffices only to consider sets of length-$k$ strings of pairs belonging to $\mathcal{S}$ in which every pair $(i_l, j_l)$ is repeated. For a given pair $(i_l, j_l)$, call its {\em flip} $(j_l, i_l)$. We call two pairs {\em distinct} if they differ in at least one entry and are not flips of one another. Define $\mathcal{S}'$ via
$$\mathcal{S'}=\{\{({\bf i}, {\bf j}) \in \mathcal{S}: \forall l, \, \exists l' \textrm{ with } (i_l, j_l)=(i_{l'}, j_{l'}) \textrm{ or }(i_l, j_l)=(j_{l'}, i_{l'})\}$$
Define $\mathcal{S}'_{u}$ by
\begin{equation}\label{eq:sprime_u}
\mathcal{S}'_u=\{({\bf i}, {\bf j}) \in \mathcal{S}': \textrm{ there are } u \textrm{ distinct pairs } (i_l, j_l)\}
\end{equation}
Note that $u$ is some integer satisfying $1 \leq  u \leq \lfloor k/2 \rfloor$; for any such $u$, the size of $\mathcal{S}'_u$ is at most 
\begin{equation}\label{eq:S'u_cardinality}
|\mathcal{S}'_u| \leq n^{u+1} (u+1)^{u-1} u^k
\end{equation}
For any paired length-$k$ string in $\mathcal{S}'_u$ with distinct pairs denoted by $(i_1', j_1'), \cdots (i_u', j_u')$ and the $t$h pair $(i'_t, j_t')$ repeated with multiplicity $r_t$, we have
$$\mathbb{E}\left[W_{i_1 j_1} W_{i_2 j_2} \cdots W_{i_k j_k}\right] =\mathbb{E}[W^{r_1}_{i'_1 j'_1}\cdots W^{r_u}_{i'_u j'_u}]=\Pi_{t=1}^u\mathbb{E}[W^{r_t}_{i'_t j'_t}]\leq \prod_{t=1}^u \Omega_{r_t}\leq \Omega_k^u$$
where the penultimate bound follows from uniform boundedness of all $r_t$th moments of the entries of $W_n$, and there is no loss of generality in supposing that the $\Omega_k$ are increasing. As a consequence, for any paired length-$k$ string in $S'_u$, we have
\begin{align}\label{eq:bound_on_exp_Su}
\frac{1}{n^{k/2+1}}\left(\sum_{({\bf i}, {\bf j}) \in \mathcal{S}'_u}\mathbb{E}\left[W_{i_1 j_1} W_{i_2 j_2} \cdots W_{i_k j_k}\right] \right)&\leq\frac{|\mathcal{S}'_u|\Omega_k^u}{n^{k/2 +1}} \leq \frac{n^{u+1} \Omega_k^u (u+1)^{u-1}u^k}{n^{k/2+1}}
\end{align}
Thus, if $u < k/2$, 
\begin{align}\label{eq:conv_to_zero_sum_terms_u_small}
\lim\limits_{n \rightarrow \infty}\left(\frac{1}{n^{k/2+1}}\sum_{({\bf i}, {\bf j})\in S'_u}\mathbb{E}\left[W_{i_1 j_1} W_{i_2 j_2} \cdots W_{i_k j_k}\right]\right)=0
\end{align}
 When $k$ is odd, $u$ automatically satisfies $u<k/2$, so we conclude that for odd $k$, $$\lim\limits_{n \rightarrow \infty} \mathbb{E}[\hat{L}_n^k]=0$$
For even integers $k$, Equation~(\ref{eq:conv_to_zero_sum_terms_u_small}) implies that to compute the limit of $\mathbb{E}[\hat{L}_n^k]$ as $n \rightarrow \infty$, we need only consider the sum over paired length-$k$ strings in $\mathcal{S}'_u$ for $u=k/2$. 
For notational convenience and to avoid factors of the form $k/2$, we consider an even power represented by $2k$ and put $u=k$. Because we have exactly $k$ distinct pairs for any $({\bf i}, {\bf j})$ in $\mathcal{S}'_k$, each length-$2k$ string of pairs of the form $(i_1, j_2), \cdots, (i_{2k}, j_{2k})$ in $\mathcal{S}'_k$ is a string of pairs belonging to $\mathcal{S}$ for which each pair has a replicate (either the pair itself or its flip), so for every $l$, we have a unique $l'$ with $(i_l, j_l)=(i_{l'}, j_{l'})$ or $(i_l, j_l)=(j_{l'}, i_{l'})$. 

We define two length-$2k$ strings of pairs  $({\bf i}, {\bf j})$ and $({\bf i'}, {\bf j'})$ in $\mathcal{S}'_k$ to be {\em equivalent} if there exists a map $q:\{1, \cdots, n\} \rightarrow \{1, \cdots, n\}$ with $i_l'=q(i_l)$ and $j_l'=q(j_l)$ for every $l \in \{1, \cdots, 2k\}$. That is, two length-$2k$ strings of pairs are equivalent if they are the same up to a relabeling. Under this equivalence, we can restrict our attention to strings of pairs with $i_1=1$. 

We can upper bound the number of distinct equivalence classes $C_k$ as follows: consider the representative from each equivalence class where the indices are labelled by the order in which they appear, e.g., $(1,2), (2,3),$ etc. Since there are only $k$ distinct pairs, and each has $i_l=i_{l-1}$ or $j_{l-1}$, we see that all $i_l, j_l$ belong to $\{1,\ldots,k+1\}$. Then a loose upper bound on the number of distinct equivalence classes is given by the number of ways to choose $k$ distinct pairs from the $(k+1)k$ possible pairs, times $2^k$ options for each distinct pair having its second appearance in the same order as its first appearance or its flip. We can arrange the $2k$ pairs in $(2k)!$ ways, but this over-counts the orderings of the indistinct pairs by a factor of $2^k$, and the equivalence relation allows us to divide by another factor of $k!$. In other words, we have  the upper bound $C_k\leq \binom{(k+1)k}{k} 2^k\frac{(2k)!}{2^k k!}\leq ((k+1)k)^k \binom{2k}{k} \leq (2k^2)^k (2e)^k$, meaning the $C_k$ satisfy Carleman's condition.

We call $(i_l, j_l)$ an {\em anchored pair} from 1 if $i_l=1, j_l \neq 1$, and it appears twice in $(\bf{i},\bf{j})$ in this same order. We call $(i_l, j_l)$ a {\em returning pair} to 1 if $i_l=1, j_l\neq 1$, and its flip appears in $(\bf{i},\bf{j})$. 
Let $a$ denote the number of distinct anchored pairs and $r$ the number of distinct returning pairs. For any returning pair $(i_l,j_l), (i_{l'},j_{l'})=(j_l,i_l)$, we define its {\em internal substring} to be the length-$(l'-l)$ string of pairs $(i_{l+1}, j_{l+1}), \cdots (i_{l'-1}, j_{l'-1})$. If all pairs are anchored pairs, there are no internal substrings. Let $m_t$ be the number of internal substrings of length $2(t-1)$, $1\leq t\leq k$ (noting that an internal substring must have an even length since it begins and ends at $j_l$).

We next count the number of distinct equivalence classes $C_k$ in $\mathcal{S}'_k$ under this equivalence relation. We assert that $C_k$ satisfies the recurrence relation in which $C_0=1$, and for $k\geq 1$,
\begin{equation}\label{eq:C_k_recursion}
C_k=\frac{(2k)!}{2^k k!} + \sum_{a=0}^{k-1} \sum_{r=1}^{k-a} \sum_{\substack{m_1, \cdots, m_k\geq0,\\ \sum m_i =r,\\ \sum im_i=k-a}} \frac{(2a+r)!}{2^a (a! r!)} \frac{r!}{m_1! \cdots m_k!}C_0^{m_1} \cdots C_{k-1}^{m_k}
\end{equation}

To justify this, we consider the possibilities for the number of anchored pairs $a$. If $a=k$, we have $k$ distinct anchored pairs in our $2k$ total pairs. This leads to $(2k)!/2^k k!$ different equivalence classes for such strings: there are $(2k)!$ ways to arrange the $2k$ pairs; each pair can be interchanged with its replicate, prompting the division by $2^k$, and finally there are $k!$ ways to rearrange the $k$ distinct pairs, which accounts for the equivalence relation.

When $a<k$, there must be $1\leq r\leq k-a$ returning pairs, and the total length of their internal substrings must be $2(k-a-r)$. Then $\sum_{i=1}^k m_i = r$, since each returning pair has an internal substring with nonnegative length, and $\sum_{i=1}^k 2(i-1)m_i = 2(k-a-r)$, or equivalently $\sum_{i=1}^k im_i =k-a$. Considering each returning pair with its internal substring as a single object, there are $(2a+r)!$ arrangements of the $2a$ anchored pairs and $r$ returning pairs. Since we can interchange the two copies of any anchored pair without changing the string, we must divide by $2^a$, and the equivalence relation prompts the division by $(a! r!)$. Given any arrangement of these anchored and returning pairs, we may choose the locations of the internal substrings of various lengths from the $r$ possible positions in $r!/m_1!\cdots m_k!$ ways. We then have $C_0^{m_1}=1$ ways to fill the length-0 internal substring, $C_1^{m_2}$ ways to fill the length-2 internal substrings, and so on up to $C_{k-1}^{m_k}$ ways to fill the (at most one) length-$2(k-1)$ internal substring. Here we make use of the equivalence relation to recognize that the number of internal substrings of length $2t$ inside some particular returning pair $(1,j),\ldots,(j,1)$ is just given by $C_t$. The interested reader can verify that the moments $L_{2k}$ give the unique solution to the recurrence relation in Equation~(\ref{eq:C_k_recursion}).

It remains to prove Equation~(\ref{eq:random_moment_close_to_mean}), which we will establish by proving that for each $k\geq 1$,
\begin{equation}
\label{eq:kappaLbound}
\kappa(\hat{L}_n^k)\leq \frac{B_k}{n^2},
\end{equation} where $\kappa$ is the fourth central moment. Applying Markov's inequality and the Borel-Cantelli Lemma then yields Equation~(\ref{eq:random_moment_close_to_mean}).

Let $\{X_i\}$ be a collection of centered random variables. The fourth moment of their sum has the following representation:
\begin{align}
\kappa\left(\sum X_i\right) &= \EE\left[\left(\sum X_i\right)^4\right]\notag\\
&= \EE[ \sum_{i_1,i_2,i_3,i_4}X_{i_1}X_{i_2}X_{i_3}X_{i_4}]\notag\\
&=\sum_{i_1} \kappa(X_{i_1})+4\sum_{i_1\neq i_2} \EE[X_{i_1}X_{i_2}^3]+6\sum_{i_1<i_2}\EE[X_{i_1}^2X_{i_2}^2]\notag\\
&+12\sum_{i_1<i_2, i_3\neq i_1,i_2}\EE[X_{i_1}X_{i_2}X_{i_3}^2]+24\sum_{i_1<i_2<i_3<i_4}\EE[X_{i_1}X_{i_2}X_{i_3}X_{i_4}]\notag\\
&=A_1+A_2+A_3+A_4+A_5.\label{eq:kappabound}
\end{align}

For any of these sums, using H\"{o}lder's inequality and/or Cauchy-Schwarz (potentially multiple times) allows us to say that if all $X_i$ satisfy $\kappa(X_i)\leq \kappa$, then $|A_j|\leq \#\{\text{nonzero terms in }A_j\}\kappa.$ 

We now write the fourth central moment of $\hat{L}_n^k$ using the de-diagonalized representation as before. Let the set of length-$k$ strings of pairs, $\mathcal{S}$, be written as $\{s_1,\ldots,s_{m_k}\}$, noting that $m_k \sim n^{k+1}$. For $1\leq l\leq m_k$, if $s_l=((i_1,j_1),\ldots,(i_k,j_k)),$ then let the corresponding product be written as $\pi_l'=\prod_{t=1}^k W_{i_t,j_t}.$ Let $\pi_l=\pi_l'-\EE[\pi_l']$. Then we have
$$\kappa(\hat{L}_n^k)=\kappa\left(\frac{1}{n^{k/2+1}}\sum_{({\bf i},{\bf j})\in\mathcal{S}} W_{i_1 j_1}\cdots W_{i_k j_k}\right)=\frac{1}{n^{2k+4}}\,\kappa\left(\sum_{l=1}^{m_k} \pi_l\right).$$ It is easy to show that because of the uniform bound on the $4k$th moments of the $W_{ij}$, we have $\kappa(\pi_l)\leq \kappa$ for some constant $\kappa\leq \Omega_{4k}^k$, for all $1\leq l\leq m_k$. For the first three terms in the representation (\ref{eq:kappabound}), we have
$$A_1\leq m_k \kappa,\quad A_2\leq 4m_k^2\kappa,\quad A_3\leq 6m_k^2\kappa,$$ so these terms are all of the correct order with respect to $n$. 

There are two key requirements for a term in $A_4$ to be nonzero: If either sequence $s_{l_1}$ or $s_{l_2}$ has no pairs in common with any other sequence, it is independent and factors out. Since $\EE[\pi_{l}]=0$ for all $l$, these terms must vanish. Secondly, if some pair $(i,j)$ appears in $s_{l_1}$ and is distinct from all other pairs in that sequence, $\EE[\pi_{l_1}']=0$, so $\EE[\pi_{l_1}\pi_{l_2}\pi_{l_3}^2]=\EE[\pi_{l_1}'\pi_{l_2}\pi_{l_3}^2]$, and the mean of the independent variable $W_{ij}$ factors out of this expectation if $(i,j)$ does not appear in one of $s_{l_2}, s_{l_3}$. Since $\EE[W_{ij}]=0$ for all $(i,j)$, these terms must also vanish. Moreover, this argument is identical for any pairs $(i,j)$ in $s_{l_2}$ that are distinct from all other pairs in that sequence.

There are three cases to consider: (i) Either $s_{l_1}\cap s_{l_3}\neq\varnothing$, and $s_{l_2}\cap (s_{l_1}\cup s_{l_3})\neq \varnothing$; or (ii) $s_{l_2}\cap s_{l_3}\neq \varnothing$, and $s_{l_1}\cap(s_{l_2}\cup s_{l_3})\neq\varnothing$; or (iii) $s_{l_1}\cap s_{l_2}\neq\varnothing$ and $(s_{l_1}\cup s_{l_2})\cap s_{l_3}=\varnothing$. Suppose we have
\begin{itemize}
\item $a_{i}$ distinct pairs in $s_{l_1}\cap s_{l_2}\cap s_{l_3}$ that are repeated in $s_{l_i}$, $i=1,2$;
\item $b_i$ distinct pairs in $s_{l_i}\cap s_{l_3}\setminus s_{l_{3-i}}$ that are repeated in $s_{l_i}$, $i=1,2$;
\item $c_i$ distinct pairs in $s_{l_1}\cap s_{l_2}\setminus s_{l_3}$ that are repeated in $s_{l_i}$, $i=1,2$;
\item $d_i$ distinct pairs in $s_{l_i}\setminus(s_{l_3}\cup s_{l_{3-i}})$ that are repeated in $s_{l_i}$, $i=1,2$;
\item $e_i$ distinct pairs in $s_{l_1}\cap s_{l_2}\cap s_{l_3}$ that are not repeated in $s_{l_i}$, $i=1,2$;
\item $f_i$ distinct pairs in $(s_{l_i}\cap s_{l_3})\setminus s_{l_{3-i}}$ that are not repeated in $s_{l_i}$, $i=1,2$;
\item $g_i$ distinct pairs in $(s_{l_1}\cap s_{l_2})\setminus s_{l_3}$ that are not repeated in $s_{l_i}$, $i=1,2$;
\item There cannot be any pairs that appear only once, in only one of either $s_{l_1}$ or $s_{l_2}$.
\end{itemize}
We have the following bounds on these variables:
\begin{align*}
2 a_1+2b_1+2c_1+2d_1+e_1+f_1+g_1&\leq k\\
2 a_2+2b_2+2c_2+2d_2+e_2+f_2+g_2&\leq k\\
a_1+c_1+e_1+g_1 -a_2-c_2-e_2-g_2 &=0
\end{align*}
In any case, there are $m_k\sim n^{k+1}$ ways to choose $s_{l_3}$. Once we choose the $k$ pairs for each of $s_{l_1}$ and $s_{l_2}$, there are $\leq k^{k}\times k^k$ ways to arrange these two strings of pairs into some orderings, allowing repeats. We will now bound the number of ways to choose the pairs for $s_{l_1}$ and $s_{l_2}$ with these values for the variables above, letting $p_k$ denote potentially different constant factors that depend on $k$ but not $n$. For case (i), first enumerating the number of ways to choose $s_{l_1}$ given the choice of $s_{l_3}$, and then enumerating the choice of $s_{l_2}$ given these choices:
$$(p_k)(p_k)(p_k n^{c_1})(p_k n^{d_1})(p_k)(p_k)(p_k n^{g_1})(p_k)(p_k)(p_k)(p_k n^{d_2})(p_k)(p_k)(p_k)\sim n^{c_1+d_1+d_2+g_1}. $$
Adding the equation to the second inequality,
$$a_1+c_1+e_1+g_1+a_2+2b_2+c_2+2d_2+f_2\leq k,$$
so
$$ c_1+d_1+g_1/2 \leq k/2,\quad g_1/2+d_2  \leq k/2,$$
and thus we get an upper bound of $n^{k+1}\times n^{k}=n^{2k+1}$ nonzero terms in this case. Case (ii) is symmetric to case (i), so we only need to consider case (iii). Here $a_1=a_2=0, b_1=b_2=0, e_1=e_2=0,$ and $f_1=f_2=0$, so we get
$$n(1)(1)(p_k n^{c_1})(p_k n^{d_1})(1)(1)(p_k n^{g_1})(1)(1)(p_k)(p_k n^{d_2})(1)(1)(p_k)\sim n^{c_1+d_1+d_2+g_1+1}. $$
The additional leading factor of $n$ comes from the fact that the first pair in $s_{l_1}$ has $n^2$ options when $s_{l_1}\cap s_{l_3}=\varnothing$, whereas when $s_{l_1}\cap s_{l_3}\neq \varnothing$, there is at least one pair $(i_l,j_l)$ that has already been chosen, so the next pair $(i_{l+1},j_{l+1})$ has $i_{l+1}\in\{i_l,j_l\}$, and similarly for all subsequent pairs in the string (and taking into account the fact that $i_1\in\{i_k,j_k\}$). Arguing as in cases (i) and (ii), this is bounded above by $n^{k+1}$, so the total number of nonzero terms in this case is bounded above by $n^{2k+2}$. Finally, we note that these three cases and all possible values of the variables $a_1,a_2,\cdots$ amount to $\leq 3k^{14}$ total possibilities, which is constant with respect to $n$.

The two key requirements for a term in $A_5$ to be nonzero are equivalent to those for a term in $A_4$ to be nonzero, and the analysis of this term is quite similar, though notationally more cumbersome. As expected, when $s_{l_1}\cap s_{l_2}\neq\varnothing$, $s_{l_3}\cap(s_{l_1}\cup s_{l_2})\neq\varnothing,$ and $s_{l_4}\cap(s_{l_1}\cup s_{l_2}\cup s_{l_3})\neq\varnothing$, or by permuting the order of $l_1,l_2,l_3,l_4$, a similar chain can be constructed, we find that there are only $n^{2k+1}$ nonzero terms. When $s_{l_1}\cap s_{l_2}\neq \varnothing, s_{l_3}\cap s_{l_4}\neq\varnothing,$ but $(s_{l_1}\cup s_{l_2})\cap(s_{l_3}\cup s_{l_4})=\varnothing,$ then we recover the additional factor of $n$ as in the analysis of $A_4$'s case (iii), and find a total of $n^{2k+2}$ nonzero terms. This completes the proof of Claim~(\ref{eq:kappaLbound}).

So for any $k\geq 1$, we have 
$$ \PP\left[|\hat{L}_n^k-\EE[\hat{L}_n^k]|>\frac{1}{\log^{1/4}(n)}\right]\leq \log(n)\kappa(\hat{L}_n^k)\leq \frac{B_k\log(n)}{n^2}.$$
This sequence is summable, so by the Borel-Cantelli Lemma, $|\hat{L}_n^k-\EE[\hat{L}_n^k]|\rightarrow0$ as $n\rightarrow\infty$ with probability 1, which is the content of Equation~(\ref{eq:random_moment_close_to_mean}). This completes the proof.
\end{proof}

\begin{proof}[Proof of Corollary~\ref{cor:esd}]
As in the proof of Theorem~\ref{theorem:topeig}, we may write $A(L(G))+2I= B^TB,$ while $A(G)+D=BB^T$, and thus these two matrices have the same nonzero eigenvalues. By the rank inequality, the maximum difference between the ESD of $A(G)+D$ and the ESD of $A(G)+D-pJ$ is at most $1/n$, so if either of these matrices has a limiting ESD, they both must have the same one. Subtracting by $(n-2)pI$ merely shifts the distribution by $(n-2)p$, so $W_n(x)$ must have the same limit as the ESD of $A(G)+D-pJ-(n-2)pI$. But this last matrix is a WignerDD matrix, so Theorem~\ref{theorem:esd} applies.
\end{proof}

\begin{proof}[Proof of Proposition~\ref{prop:qproj}]
Each column $(r,s)$ of $Q$ has nonzeroes only in the block of the induced partition $C_{(r,s)},$ where the value equals $1/\sqrt{m_{r,s}}.$ Then $$(Q^{T}Q)_{(r,s),(t,u)}=\begin{cases}m_{r,s}/\sqrt{m_{r,s}}\sqrt{m_{r,s}}&\text{ if }(r,s)=(t,u),\\ 0&\text{ otherwise}\end{cases}=I,$$ so $(QQ^{T})(QQ^{T})=QQ^{T},$ and by symmetry, this is an orthogonal projection.

We first observe that while the columns of $Q$ are not eigenvectors of $A(L(K_{n})),$ they span an invariant subspace of $A(L(K_{n}))$ of dimension $\binom{k+1}{2}$, meaning that $\sigma(M)\subseteq\sigma(A(L(K_{n}))).$ \\

\emph{Claim 1:} $\mathrm{range}(A(L(K_{n}))Q)\subseteq \mathrm{range}(Q).$\\

Indeed, let's compute $A(L(K_{n}))$ on one of the columns of $Q$ (after removing the normalization). Suppose $1\leq s<t\leq k$.
\begin{align*}
(A(L(K_{n}))(\sqrt{m_{s,t}}Q_{(s,t)}))_{\{i,j\}}&=\sum_{\{\ell,p\}\in C_{(s,t)}} A(L(K_{n}))_{\{i,j\},\{\ell,p\}}\\
&=\begin{cases}
n_s+n_t-2&\text{ if }\{i,j\}\in C_{(s,t)},\\
2n_t&\text{ if }\{i,j\}\in C_{(s,s)},\\
2n_s&\text{ if }\{i,j\}\in C_{(t,t)},\\
n_t&\text{ if }\{i,j\}\in C_{(s,r)}\text{ or }C_{(r,s)}, r\not\in\{s,t\},\\
n_s&\text{ if }\{i,j\}\in C_{(t,r)}\text{ or }C_{(r,t)}, r\not\in \{s,t\},\\
0&\text{ otherwise}.\end{cases}
\end{align*}
These equalities are proved by considering the conditions imposed and summing over the corresponding entries of $A(L(K_{n}))$. For instance, if $i\in C_{s}$ and $j\in C_{t},$ then if $\ell=i\in C_{s},$ there are $|C_{t}\setminus\{j\}|=n_{t}-1$ choices of $p$ for which $|\{i,j\}\cap \{\ell,p\}|=1,$ and if $p=j\in C_{t},$ there are $|C_{s}\setminus\{i\}|=n_{s}-1$ choices of $\ell$ for which this condition holds, giving $n_{s}+n_{t}-2$ in total. Since the value depends only on the block of the induced partition that $\{i,j\}$ belongs to, this has remained in $\mathrm{span}(Q)$.

Similarly, for $1\leq s=t\leq k$, we get
\begin{align*}
(A(L(K_{n}))(\sqrt{m_{s,s}}Q_{(s,s)}))_{\{i,j\}}&=\sum_{\{\ell,p\}\in C_{(s,s)}} A(L(K_{n}))_{\{i,j\},\{\ell,p\}}\\
&=\begin{cases}
2(n_{s}-2)&\text{ if }\{i,j\}\in C_{(s,s)},\\
n_{s}-1&\text{ if }\{i,j\}\in C_{(s,u)}\text{ or }C_{(u,s)}, u\neq s\\
0&\text{ otherwise}.\end{cases}
\end{align*}

This proves Claim 1.\\

Using these calculations, we can compute the entries of $M$, for $1\leq i\leq j\leq k, 1\leq \ell\leq p\leq k$: $$M_{(i,j),(\ell,p)}=\begin{cases} 2(n_{i}-2)&\text{ if }i=j=\ell=p\\ \sqrt{2n_{s}(n_{r}-1)}&\text{ if }(r:=i=j=\ell\neq p=:s)\text{ or }(r:=i=j=p\neq \ell=:s)\\&\text{ or }(r:=i=\ell=p\neq j=:s)\text{ or }(r:=j=\ell=p\neq i=:s)\\ n_{i}+n_{j}-2&\text{ if }i=\ell\neq j=p\\ \sqrt{n_{s}n_{t}}&\text{ if }i\neq j, \ell\neq p, |\{i,j\}\cap\{\ell,p\}|=1,\\&\{s,t\}:=(\{i,j\}\cup\{\ell,p\})\setminus(\{i,j\}\cap\{\ell,p\})\\ 0&\text{ otherwise.}\end{cases}$$ 

For an example, from the proof of the claim,
$$(A(L(K_{n}))Q_{(1,2)})_{\{i,j\}}=\begin{cases}
(n_1+n_2-2)/\sqrt{n_1 n_2}&\text{ if }\{i,j\}\in C_{(1,2)},\\
2\sqrt{n_2/n_1}&\text{ if }\{i,j\}\in C_{(1,1)},\\
2\sqrt{n_1/n_2}&\text{ if }\{i,j\}\in C_{(2,2)},\\
\sqrt{n_2/n_1}&\text{ if }\{i,j\}\in C_{(1,r)}, r\not\in\{1,2\},\\
\sqrt{n_1/n_2}&\text{ if }\{i,j\}\in C_{(2,r)}, r\not\in \{1,2\},\\
0&\text{ otherwise},\end{cases}$$
so $$M_{(1,1),(1,2)}=\binom{n_1}{2}^{-1/2} \left[\binom{n_1}{2} 2\sqrt{n_2/n_1}\right]=\sqrt{2n_2(n_1-1)},$$ matching the second case above with $r=1=i=j=\ell$, $s=2=p$. 

Now that we have $M$, we just need to compute the multiplicities of the eigenvalues $2n-4, n-4,$ and $-2$, since we know these are the only options (see Corollary~\ref{c:lgcompleteeigs}). To see that $2n-4\in \sigma(M),$ we just find the eigenvector, which is $x=[\sqrt{m_{i,j}}]_{(i,j)}.$ Indeed, for $1\leq i\leq k$,
\begin{align*}
(Mx)_{(i,i)}&=\sum_{(\ell,p)}M_{(i,i),(\ell,p)}\sqrt{m_{\ell,p}}\\
&=2(n_i-2)\sqrt{\binom{n_i}{2}}+\sum_{p>i}\sqrt{2n_p(n_i-1)}\sqrt{n_i n_p}+\sum_{\ell<i}\sqrt{2n_\ell(n_i-1)}\sqrt{n_i n_\ell}\\
&=\sqrt{2 n_i(n_i-1)}\left(n_i-2+\sum_{\ell\neq i}n_\ell\right)\\&=2(n-2)\sqrt{\binom{n_i}{2}}=2(n-2) x_{(i,i)}
\end{align*}
and for $i<j$,
\begin{align*}
(Mx)_{(i,j)}&= \sqrt{2n_j(n_i-1)}\sqrt{\binom{n_i}{2}}+\sqrt{2n_i(n_j-1)}\sqrt{\binom{n_j}{2}}+(n_i+n_j-2)\sqrt{n_i n_j}\\&+\sum_{\substack{p>i,\\ p\neq j}}\sqrt{n_p n_j}\sqrt{ n_i n_p}+\sum_{\substack{\ell<j,\\\ell\neq i}}\sqrt{n_i n_\ell}\sqrt{n_j n_\ell}\\
&+\sum_{p>j}\sqrt{n_p n_i}\sqrt{n_j n_p}+\sum_{\ell<i}\sqrt{n_\ell n_j}\sqrt{n_i n_\ell}\\
&=\sqrt{n_i n_j}\left(n_i-1+n_j-1+n_i+n_j-2+2\sum_{\ell\neq i,j}n_\ell\right)\\
&=(2n-4)\sqrt{n_i n_j}=(2n-4) x_{(i,j)} 
\end{align*}

So the only other eigenvalues of $M$ must be either $n-4$ or $-2$. We will compute the nullity of $M+2I$, which will give us the multiplicity of the eigenvalue $-2$, which will complete the proof.\\

\emph{Claim 2:} $\mathrm{nullity}(M+2I)=\binom{k+1}{2}-k=\binom{k}{2}.$\\

We choose an ordering for the indices of $M$ which puts $(1,1),(2,2),\ldots,(k,k)$ first, followed by $(1,2), \ldots, (1,k),$ $(2,3), \ldots, (k-1,k)$ in lexicographical order. Then we may write $$M+2I=\begin{bmatrix}D&X\\ X^{T}&Z\end{bmatrix},$$ where $D\in M_{k}(\RR)$ is diagonal with entries $\{2 n_i-2\}_{i=1}^{k}$ on the diagonal, $X\in M_{k,\binom{k}{2}}(\RR)$ has entries $$X_{(i,i),(\ell,p)}=\begin{cases}\sqrt{2n_s(n_r-1)}& \text{ if }(r:=i=\ell\neq p=:s)\text{ or }(r:=i=p\neq \ell=:s)\\0&\text{otherwise,}\end{cases}$$ and $Z\in M_{\binom{k}{2}}(\RR)$ has entries $$Z_{(i,j),(\ell,p)}=\begin{cases}n_i+n_j&\text{ if }i=\ell,j=p\\ \sqrt{n_s n_t}& \text{ if }|\{i,j\}\cap\{\ell,p\}|=1, \{s,t\}:=(\{i,j\}\cup\{\ell,p\})\setminus(\{i,j\}\cap\{\ell,p\})\\0&\text{ otherwise.}\end{cases}$$

Consider a vector partitioned as $[u^{T},v^{T}]^{T}$, and belonging to the nullspace of $M+2I$. Then $$0=\begin{bmatrix} D&X\\ X^{T}&Z\end{bmatrix}\begin{bmatrix}u\\ v\end{bmatrix}=\begin{bmatrix}Du+Xv\\ X^{T}u+Zv\end{bmatrix}.$$ From the first set of equations, and since $2 n_i-2\geq 2$ for all $1\leq i\leq k$, we see that $u=-D^{-1}Xv.$ Plugging this into the second set of equations, we get $(Z-X^{T}D^{-1}X)v=0,$ so $\mathrm{nullity}(M+2I)\geq \mathrm{nullity}(Z-X^{T}D^{-1}X),$ which we now just have to show is $\binom{k}{2},$ which will follow since $Z=X^{T}D^{-1}X$. This will complete the proof, since $D$ is an order-$k$ submatrix which is clearly invertible, which along with the rank-nullity theorem gives 
\begin{align*}
\mathrm{nullity}(M+2I)&\geq \binom{k}{2},\\
\mathrm{rank}(M+2I)&\geq k,\\
\mathrm{rank}(M+2I)+\mathrm{nullity}(M+2I)&=\binom{k+1}{2},
\end{align*}
meaning the first two inequalities must be equalities.

Now we prove that $X^{T}D^{-1}X=Z$ to complete the proof of the claim (and thus the proof). From the definitions, $$(X^{T}D^{-1}X)_{(i,j),(\ell,p)}=\sum_{r=1}^{k}\frac{X_{(r,r),(i,j)} X_{(r,r),(\ell,p)}}{2n_r-2}.$$ Since $X_{(r,r),(i,j)}$ is only nonzero when $r\in\{i,j\},$ this sum can only be nonzero if $\{i,j\}\cap\{\ell,p\}\neq\varnothing,$ which matches the third case for the entries of $Z$. When $i=\ell$ and $j=p$, this sum equals $$\frac{2 n_j(n_i-1)}{2(n_i-1)}+\frac{2n_i(n_j-1)}{2(n_j-1)}=n_i+n_j=Z_{(i,j),(i,j)},$$ and considering the case $i=\ell,$ $j\neq p$ as prototypical of the case where $|\{i,j\}\cap\{\ell,p\}|=1$ (the other subcases are very similar), this sum equals $$\frac{\sqrt{2 n_j(n_i-1)}\sqrt{2 n_\ell(n_i-1)}}{2(n_i-1)}=\sqrt{n_j n_\ell}=Z_{(i,j),(i,\ell)}.$$ This proves the claim, so completes the proof.

\end{proof}

\begin{proof}[Proof of Proposition~\ref{prop:hmat}]
Using the formula for $\overline{\hat{Q}}$ in Equation~(\ref{eq:qmat}), and the definition of $Q$, we see that
\begin{align*}
H_{\{i,j\},(r,s)}&=[A(L(K_n))(\mathrm{diag}(Y_e)-\mathrm{diag}(\EE[Y_e]))Q]_{\{i,j\},(r,s)}\\
&=\frac{1}{\sqrt{m_{r,s}}}\sum_{\{\ell,p\}\in C_{(r,s)}}A(L(K_n))_{\{i,j\},\{\ell,p\}}(Y_{\{\ell,p\}}-\mu_{r,s}),
\end{align*}
which gives the formula above when we use the definition of $A(L(K_n))_{\{i,j\},\{\ell,p\}}$. This is a sum of dependent, mean-0 random variables, where the exact number of terms $\eta(\{i,j\},(r,s))$ can be determined by a counting argument. This quantity is (letting $t\in[k]\setminus\{r,s\}$):
\begin{center}
\begin{tabular}{c|c|c}
\multirow{5}{*}{$r<s$} & $\{i,j\}\in C_{(r,r)}$ & $2n_s$\\
\cline{2-3}
& $\{i,j\}\in C_{(s,s)}$ & $2n_r$\\
\cline{2-3}
& $\{i,j\}\in C_{(r,s)}$ & $n_r+n_s-2$\\
\cline{2-3}
& $\{i,j\}\in C_{(r,t)}$ or $C_{(t,r)}$ & $n_s$\\
\cline{2-3}
& $\{i,j\}\in C_{(s,t)}$ or $C_{(t,s)}$ & $n_r$\\
\hline
\multirow{2}{*}{$r=s$}& $\{i,j\}\in C_{(r,r)}$ & $2(n_r-2)$\\
\cline{2-3}
& $\{i,j\}\in C_{(r,t)}$ or $C_{(t,r)}$ & $n_r-1$\\
\hline
&All other cases&0
\end{tabular}
\end{center}

For example, when $\{i,j\}\in C_{(r,t)}$ for $r<s<t$, if $i,\ell\in C_r$, $p\in C_s$, $j\in C_t$, then $|\{i,j\}\cap \{\ell,p\}|=1$ if and only if $i=\ell$, so we get $n_s$ terms as $p$ varies in $C_s$. Since this sum only includes terms $Y_e$ where $e\in C_{(r,s)},$ we have independence between the entries in a given row of $H$, but we may have dependence between rows of $H$, since continuing the example above, if $i\in C_r$, $j_1,j_2\in C_t$, then for all $p\in C_s$, $Y_{\{i,p\}}$ appears in the formulas for $H_{\{i,j_1\},(r,s)}=H_{\{i,j_2\},(r,s)}$. In general, the rows $\{i_1,j_1\}$ and $\{i_2,j_2\}$ will be dependent whenever there is some $r\leq s$ and $\{\ell,p\}\in C_{(r,s)}$ such that $|\{i_1,j_1\}\cap\{\ell,p\}|=1$ and $|\{i_2,j_2\}\cap\{\ell,p\}|=1$: but so long as $\{i_1,j_1\}$ and $\{i_2,j_2\}$ are not identical, any $\{\ell,p\}\in\{\{i_1,i_2\},\{i_1,j_2\},\{j_1,i_2\},\{j_1,j_2\}\}$ will fill this role (and at least one of the proposed edges in this set will be well-defined, since $|\{i_1,j_1,i_2,j_2\}|\geq 3$). Thus between rows, there will always be at least weak dependence, and in some cases, certain entries will be exactly equal.

Now we want to prove the result about the size of the entries of $H$. Approaching this naively will not succeed: the entries $Y_e$ in the sum are dependent, since they have a common factor of $1/\sqrt{\widehat{m}_{r,s}}$. However, we may view this sum as $$\frac{\sum_{\{\ell,p\}}\delta_{\{\ell,p\}}}{\sqrt{\widehat{m}_{r,s}}}-\frac{\eta(\{i,j\},(r,s))\mu_{r,s}}{\sqrt{m_{r,s}}}=:\frac{X}{\sqrt{X+Y}}-\frac{\eta(\{i,j\},(r,s))\mu_{r,s}}{\sqrt{m_{r,s}}}.$$ On this view, $|H_{\{i,j\},(r,s)}|$ may be viewed as the absolute deviation of $W=\frac{X}{\sqrt{X+Y}}$ from its mean, where $X$ and $Y$ are independent Binomial($\eta$,$B_{r,s}$) and Binomial($m_{r,s}-\eta$,$B_{r,s}$) random variables, respectively, and letting $\eta:=\eta(\{i,j\},(r,s))$, which is fixed for the time being. 
We say that $W\sim\mathrm{DampedBinomial}(\eta,m_{r,s}-\eta,B_{r,s})$, and prove several properties of $W$ in the supplementary material (\cite{lubberts_et_al_supplement}). In particular, $\mathrm{Var}(W)=O(1/n)$ and $\mathrm{Var}((W-\EE[W])^2)=O(1/n^2)$ when the cluster sizes are all proportional to $n$. 

Indeed, when all $n_i\propto n$, we have $\eta\approx cn$ for each $\{i,j\}$ and $(r,s)$, so 
$$\PP[|H_{\{i,j\},(r,s)}|>t]\leq \frac{1}{t^2}\left(\frac{\eta(m_{r,s}-\eta)(1-B_{r,s})}{m_{r,s}(m_{r,s}-1)}+\frac{\eta^2(1-B_{r,s})}{m_{r,s}^2}\right).$$
In particular, each entry of $H$ converges to $0$ in probability as $n$ grows.

This upper bound for $\PP[|H_{\{i,j\},(r,s)}|>t]$ follows immediately from applying Chebyshev's inequality and using the upper bound on $\mathrm{Var}(W)$ given in the proof of Theorem~\ref{theorem:dampedbinomial}. Now we find an upper bound on $\mathrm{Var}(\|H\|_F^2)$ to apply Chebyshev's inequality to $\|H\|_F$. By the independence between columns of $H$, and applying multiple Cauchy-Schwarz inequalities:
\begin{align*}
\mathrm{Var}(\|H\|_F^2)&=\sum_{\{i,j\},\{\ell,p\},(r,s),(t,u)}\mathrm{Cov}(H_{\{i,j\},(r,s)}^2,H_{\{\ell,p\},(t,u)}^2)\\
&=\sum_{(r,s)}\sum_{\{i,j\},\{\ell,p\}}\mathrm{Cov}(H_{\{i,j\},(r,s)}^2,H_{\{\ell,p\},(r,s)}^2)\\
&\leq \sum_{(r,s)}\sum_{\{i,j\},\{\ell,p\}}\sqrt{\mathrm{Var}(H_{\{i,j\},(r,s)}^2)\mathrm{Var}(H_{\{\ell,p\},(r,s)}^2)}\\
&= \sum_{(r,s)}\left(\sum_{\{i,j\}}\sqrt{\mathrm{Var}(H_{\{i,j\},(r,s)}^2)}\right)^2\\
&\leq \sum_{(r,s)}\binom{n}{2}\sum_{\{i,j\}}\mathrm{Var}(H_{\{i,j\},(r,s)}^2)\\
&\leq \binom{n}{2}^2\binom{k+1}{2}\frac{C}{n^2}\leq C n^2.
\end{align*}
For some constants $C>0$. Now since \begin{align*}\EE[\|H\|_F^2]&=\sum_{\{i,j\},(r,s)}\EE[H_{\{i,j\},(r,s)}^2]\\&=\sum_{\{i,j\},(r,s)}\mathrm{Var}(H_{\{i,j\},(r,s)})\leq C\cdot n\leq n^{3/2},\end{align*} we see that
\begin{align*}
\PP[\|H\|_F>\sqrt{2}n^{3/4}]&\leq\PP[\|H\|_F^2>\EE[\|H\|_F^2]+n^{3/2}]\\
&\leq \frac{\mathrm{Var}(\|H\|_F^2)}{n^3}\\
&\leq \frac{C n^2}{n^3}=\frac{C}{n}.
\end{align*}
\end{proof}

\begin{proof}[Proof of Corollary~\ref{c:mhat}]
We begin by computing
\begin{align*}
\hat{Q}^T A(L(G)) \hat{Q}&=\bar{\hat{Q}}^TA(L(K_n))\bar{\hat{Q}}\\
&=\bar{\hat{Q}}^TA(L(K_n))(\bar{\hat{Q}}-Q\mathrm{diag}(\mu_{i,j}))+\bar{\hat{Q}}^TA(L(K_n))Q\mathrm{diag}(\mu_{i,j})\\
&=\bar{\hat{Q}}^T H+Q^T\mathrm{diag}(Y_e)QM\mathrm{diag}(\mu_{i,j}).
\end{align*}
The first term has mean $0$ and satisfies $\|\bar{\hat{Q}}^TH\|_F\leq \|H\|_F.$ We note that $Q^T\mathrm{diag}(Y_e)Q$ is a diagonal matrix with entry $(i,j)$ being $$\frac{1}{m_{i,j}}\sum_{e\in C_{(i,j)}} Y_e=\sqrt{\frac{\widehat{m}_{i,j}}{m_{i,j}}}=Z_{i,j}.$$ Since $\EE[Z_{i,j}]=\mu_{i,j}$, the second term's mean is $\mathrm{diag}(\mu_{i,j})M\mathrm{diag}(\mu_{i,j})$, and $$\|\hat{Q}^TA(L(G))\hat{Q}-\mathrm{diag}(\mu_{i,j})M\mathrm{diag}(\mu_{i,j})\|_F\leq \|H\|_F+(2n-4)\|\mathrm{diag}(Z_{i,j}-\mu_{i,j})\|_F.$$ The last Frobenius norm may be bounded above by $\sqrt{\binom{k+1}{2}}\max_{i,j}|Z_{i,j}-\mu_{i,j}|,$ so we will apply a union bound to complete the proof after showing that for each $(i,j)$, this last quantity is $\leq n^{-1/2}$ with probability $1-O(1/n)$. But since $\PP[|Z_{i,j}-\mu_{i,j}|>n^{-1/2}]\leq n\mathrm{Var}(Z_{i,j}),$ we can apply Equation~(\ref{eq:mubounds}) to get $$\mathrm{Var}(Z_{i,j})=\EE[Z_{i,j}^2]-\mu_{i,j}^2=B_{i,j}-\mu_{i,j}^2=2\sqrt{B_{i,j}}(\sqrt{B_{i,j}}-\mu_{i,j})\leq \frac{1-B_{i,j}}{m_{i,j}}\leq C n^{-2},$$ which gives $\PP[|Z_{i,j}-\mu_{i,j}|>n^{-1/2}]\leq C/n.$
\end{proof}

\begin{proof}[Proof of Theorem~\ref{theorem:singvecs}]
We begin by computing $A(L(G))\hat{Q}-\hat{Q}\mathrm{diag}(\mu_{i,j})M\mathrm{diag}(\mu_{i,j})$, which equals 
\begin{align*}
\hat{Q}&\hat{Q}^TA(L(G))\hat{Q}+(I-\hat{Q}\hat{Q}^T)A(L(G))\hat{Q}-\hat{Q}\mathrm{diag}(\mu_{i,j})M\mathrm{diag}(\mu_{i,j})\\
&=\hat{Q}(\hat{Q}^TA(L(G))\hat{Q}-\mathrm{diag}(\mu_{i,j})M\mathrm{diag}(\mu_{i,j}))+(I-\hat{Q}\hat{Q}^T)A(L(G))\hat{Q}.
\end{align*}
The first term may be bounded by Corollary~\ref{c:mhat}, since $\hat{Q}$ is an isometry. We observe that the latter term has the same nonzero rows as $(P-\bar{\hat{Q}}\bar{\hat{Q}}^T)A(L(K_n))\bar{\hat{Q}}.$ Writing $A(L(K_n))\bar{\hat{Q}}$ as $QM\mathrm{diag}(\mu_{i,j})+H,$ the contractivity of $P-\bar{\hat{Q}}\bar{\hat{Q}}^T$ and bounds on $H$ from Proposition~\ref{prop:hmat} allow us to restrict our attention to $(P-\bar{\hat{Q}}\bar{\hat{Q}}^T)QM\mathrm{diag}(\mu_{i,j}).$ Now we see that 
\begin{align*}
PQ&=\bar{\hat{Q}}\mathrm{diag}\left(\sqrt{\frac{\widehat{m}_{i,j}}{m_{i,j}}}\right),\\
\bar{\hat{Q}}\bar{\hat{Q}}^TQ&=\bar{\hat{Q}}Q^T\mathrm{diag}(Y_e)Q,
\end{align*}
and the proof of Corollary~\ref{c:mhat} tells us that these are the same, so $(P-\bar{\hat{Q}}\bar{\hat{Q}}^T)Q=0.$ This gives $$\|A(L(G))\hat{Q}-\hat{Q}\mathrm{diag}(\mu_{i,j})M\mathrm{diag}(\mu_{i,j})\|_F\leq Cn^{3/4}\;\text{with probability} \;1-O(1/n).$$

Let $H'=A(L(G))\hat{Q}-\hat{Q}\mathrm{diag}(\mu_{i,j})M\mathrm{diag}(\mu_{i,j}).$ By Theorem 3 of \cite{DK_usefulvariant}, since $\|H'\|_2\leq \|H'\|_F$, on the set where $\|H'\|_F\leq C n^{3/4}$ (which has probability $\geq 1-O(1/n)$), there is an orthogonal matrix $\mathcal{O}$ such that
\begin{align*}\|\hat{U}-U\mathcal{O}\|_F &\leq \frac{2^{3/2}(\sigma_1(\hat{Q}\mathrm{diag}(\mu_{i,j})M\mathrm{diag}(\mu_{i,j}))+\|H'\|_2)\|H'\|_F}{\sigma_{k}^2(\hat{Q}\mathrm{diag}(\mu_{i,j})M\mathrm{diag}(\mu_{i,j}))-\sigma_{k+1}^2(\hat{Q}\mathrm{diag}(\mu_{i,j})M\mathrm{diag}(\mu_{i,j}))}\\&\leq\frac{C(n+n^{3/4})n^{3/4}}{n^2}.\end{align*}
Here we bound $\sigma_1(\hat{Q}\mathrm{diag}(\mu_{i,j})M\mathrm{diag}(\mu_{i,j}))$ by Result~\ref{res:ostrowski}, using $\sigma_1(M)=2n-4$ and the bounds in Equation~(\ref{eq:mubounds}) to obtain $$\sigma_1(\hat{Q}\mathrm{diag}(\mu_{i,j})M\mathrm{diag}(\mu_{i,j})=\sigma_1(\mathrm{diag}(\mu_{i,j})M\mathrm{diag}(\mu_{i,j}))\in \sigma_1(M)[\min_{i,j}\mu_{i,j}^2,\max_{i,j}\mu_{i,j}^2]\approx C n.$$ Proceeding similarly using $\sigma_k(M)=n-4$ and $\sigma_{k+1}(M)=2$, the denominator satisfies $$\sigma_k^2(\hat{Q}\mathrm{diag}(\mu_{i,j})M\mathrm{diag}(\mu_{i,j}))-\sigma_{k+1}^2(\hat{Q}\mathrm{diag}(\mu_{i,j})M\mathrm{diag}(\mu_{i,j}))\geq C n^2.$$
\end{proof}

We close this section by proving Lemma~\ref{lem:lemma3analogue}, which we used previously in the proof of Theorem~\ref{theorem:topeigclt}.\\

\noindent \textbf{Proof of Lemma~\ref{lem:lemma3analogue}.} We prove Equation~(\ref{eq:clt1}) using Chebyshev's inequality. $S_i - L =2\sum_{j\neq i} (A_{ij}-p)$, so
$$ \EE[\|S-Lj\|^2] = 4\sum_i \sum_{j,k\neq i} \EE[(A_{ij}-p)(A_{ik}-p)]=4n(n-1)p(1-p),$$ since these expectations are only nonzero when $k=j$.

$$
\EE[\|S-Lj\|^4]=
\EE\left[ 16 \sum_{i,i'} \sum_{j,k\neq i}\sum_{j',k'\neq i'} (A_{ij}-p)(A_{ik}-p)(A_{i'j'}-p)(A_{i'k'}-p) \right].
$$
Applying linearity, we must consider several cases for the expectation of this product with different values of $i,i',j,k,j',k'$. When $i=i'$, the only nonzero terms appear when $j,k,j',k'$ arrange themselves into two pairs: if these pairs have the same value, we get $\mu_4:= p(1-p)^4+(1-p)p^4$; otherwise we get $p^2(1-p)^2$. When $i\neq i'$, the only nonzero terms appear when $j=k$ and $j'=k'$: if $j=k=i'$ and $j'=k'=i$, then we get $\mu_4$, and otherwise we get $p^2(1-p)^2$. Counting the respective terms yields 
$$ \EE[\|S-Lj\|^4] = 16n(n-1) [2\mu_4 + 5(n-2)p^2(1-p)^2+(n-2)^2p^2(1-p)^2].$$ When we compute the variance, the $n^4$ terms cancel, which means that 
$$\mathrm{Var}(\|S-Lj\|^2)= \EE[\|S-Lj\|^4]- \EE[\|S-Lj\|^2]^2 = 32 n^3 p^2(1-p)^2+ O(n^2).$$

We prove Equation~(\ref{eq:clt2}) using Chebyshev's inequality. Define $B_{ij} = \tilde{A}_{ij}(S_i-L)(S_j-L),$ so that the quantity of interest in Equation~(\ref{eq:clt2}), $X= \sum_{i,j} B_{ij}$. Note the symmetry $B_{ij}=B_{ji}$. By linearity,
$$\EE[X] = \sum_i \EE[B_{ii}] + \sum_{i\neq j} \EE[B_{ij}].$$ Let us take these terms individually. For the diagonal terms,
\begin{align*}
\EE[B_{ii}]&= 4\sum_{q\neq i}\sum_{r\neq i}\sum_{s\neq i}\EE[(A_{iq}-p)(A_{ir}-p)(A_{is}-p)]\\
&=4(n-1)p(1-p)(1-2p),
\end{align*}
since the expectation is only nonzero when $q=r=s$. For the off-diagonal terms,
$$
\EE[B_{ij}]= 4\sum_{q\neq i}\sum_{r\neq j}\EE[A_{ij}(A_{ir}-p)(A_{js}-p)]=4p(1-p)^2,
$$
since the expectation is only nonzero when $r=j, s=i$.
This gives
$$\EE[X] = 4n(n-1)[p(1-p)(1-2p)+p(1-p)^2] = 4n(n-1)p(1-p)(2-3p).$$

Now we compute the variance of $X$.
\begin{align*}
\mathrm{Var}(X)&= \sum_{i,j}\sum_{i',j'} \mathrm{Cov}(B_{ij},B_{i'j'})\\
&= \sum_i \mathrm{Var}(B_{ii})+ 2\sum_{i\neq j} \mathrm{Var}(B_{ij})\\
&+\sum_{i\neq j} \mathrm{Cov}(B_{ii},B_{jj})+4\sum_i\sum_{j\neq i} \mathrm{Cov}(B_{ii},B_{ij})\\
&+2\sum_i \sum_{\substack{j\neq j'\\ j,j'\neq i}}\mathrm{Cov}(B_{ij},B_{ij'})+2\sum_i\sum_{\substack{i'\neq j'\\ i',j'\neq i}} \mathrm{Cov}(B_{ii},B_{i'j'})\\
&+\sum_{|\{i,j,i',j'\}|=4} \mathrm{Cov}(B_{ij},B_{i'j'}).
\end{align*}

We have already computed $\EE[B_{ii}]=4(n-1)p(1-p)(1-2p)$ in the proof of the expectation, so we just need 
\begin{align*}
&\EE[B_{ii}^2] = 16\sum_{q\neq i} \sum_{r\neq i} \sum_{s\neq i} \sum_{t\neq i}\sum_{u\neq i}\sum_{v\neq i} \EE[(A_{iq}-p)(A_{ir}-p)(A_{is}-p)(A_{it}-p)(A_{iu}-p)(A_{iv}-p)]\\
&= 16(n-1)
 [ 15(n-2)(n-3) p^3(1-p)^3]\\
 &+ 16(n-1)[10(n-2)p^2(1-p)^2(1-2p)^2 +15(n-2)p(1-p)\mu_4+\mu_6],
\end{align*}
since the only nonzero terms arise when we have 3 distinct pairs (yielding $p^3(1-p)^3$), 2 distinct triples (yielding $p^2(1-p)^2(1-2p)^2$), a pair and a quartet with different values (yielding $p(1-p)\mu_4$), or a sextet (yielding $\mu_6=p(1-p)^6+(1-p)p^6$). We then just have to count how many sets of indices fall into each of the given patterns. This shows that $\mathrm{Var}(B_{ii})=O(n^3)$.

We already computed $\EE[B_{ij}]=4p(1-p)^2$ in the proof of the expectation, so we just need
\begin{align*}
\EE[B_{ij}^2] &= 16\sum_{\substack{k\neq i\\ \ell\neq j}}\sum_{\substack{k'\neq i\\ \ell'\neq j}} \EE[A_{ij}(A_{ik}-p)(A_{i\ell}-p)(A_{ik'}-p)(A_{i\ell'}-p)]\\
&= 16\sum_{\substack{k\neq i\\ \ell\neq j}} \EE[A_{ij}(A_{ik}-p)^2(A_{i\ell}-p)^2]\\
&= 16p(1-p)^2[(n-2)^2p^2+2(n-2)p(1-p)+(1-p)^2],
\end{align*}
since nonzero terms only appear when $k'=k, \ell'=\ell$. When $k\neq j, \ell \neq i$, we get $p^3(1-p)^2$; when exactly one of $k=j$ or $\ell=i$, we get $p^2(1-p)^3$; and when $k=j, \ell=i$, we get $p(1-p)^4$. Since the mean has constant order, we see that $\mathrm{Var}(B_{ij})=O(n^2)$.

The remaining covariances may be computed similarly by finding $\EE[B_{ij}B_{i'j'}]$ for each of the different sets of conditions on $i,j,i',j'$.
\begin{align*}
\EE[B_{ii}B_{jj}] &= 16[(n-2)^2p^2(1-p)^2[(1-2p)^2+9p(1-p)]+6(n-2)p(1-p)\mu_4+\mu_6]\\
\EE[B_{ii}B_{ij}]&= 16p(1-p)[3(n-2)(n-3)p^2(1-p)^2]\\
&+16p(1-p)[(n-2)[4p(1-p)^2(1-2p)+6p(1-p)^3+\mu_4]
+(1-p)^4]\\
\EE[B_{ij}B_{ij'}]&= 16p^2(1-p)^2[(n-3)p+4(1-p)]\\
\EE[B_{ii}B_{i'j'}]&= 16p(1-p) [(n-1)p(1-p)^2(1-2p)]\\
&+16p(1-p)[6(n-2)p^2(1-p)^2+2\mu_4 +6p^2(1-p)(1-2p)]\\
\EE[B_{ij}B_{i'j'}]&= 16 p^2(1-p)^2 [ (1-p)^2+4p(1-p)+2p^2 ]
\end{align*}
Since $\EE[B_{ii}]=O(n), \EE[B_{ij}]=O(1)$, this shows that each of the sums in the expansion of the variance above has size at most $O(n^4)$, as desired.

We may write $\sum S_i = 4\sum_{i< j}A_{i,j} -n(n-1)p$. The sum is just a Binomial($n(n-1)/2,p$) random variable, so Equation~(\ref{eq:clt3}) is immediate. Equation~(\ref{eq:clt4}) follows as in \cite{furedi1981eigenvalues}. \hfill $\square$\\

\noindent \textbf{Bounds for damped binomial distribution}\\

Next, we prove our results on the damped binomial distribution. Let $n,m$ be positive integers and let $p\in(0,1)$. Then $Z\sim\mathrm{DampedBinomial}(n,m,p)$ when $$Z=\frac{X}{\sqrt{X+Y}},$$ where $X$ and $Y$ are independent binomial random variables with $n$ and $m$ trials respectively and common success probability $p$. $Z$ should be understood to take the value $0$ when $X=Y=0$. We will require the following easily-verified lemma:

\begin{lemma}
\label{lem:rootbounds}
For all $x\geq0,$
\begin{align*}
1+\frac{1}{2}(x-1)-\frac{1}{2}(x-1)^2 \leq x^{1/2}&\leq 1+\frac{1}{2}(x-1), \textrm{ and }\\
1+\frac{3}{2}(x-1)+\frac{3}{8}(x-1)^2-\frac{1}{16}(x-1)^3\leq x^{3/2}&\leq 1+\frac{3}{2}(x-1)+\frac{1}{2}(x-1)^2, \textrm{ and }
\end{align*}
\begin{align*}
    1+\frac{5}{2}(x-1)+\frac{3}{2}(x-1)^2\leq x^{5/2}&\leq 1+ \frac{5}{2}(x-1)+\frac{15}{8}(x-1)^2+\frac{15}{48}(x-1)^3.
\end{align*}
So for a random variable $V$ with mean $\mu$, variance $\sigma^2$, and third central moment $\kappa$, \begin{align*}
\mu^{1/2}-\frac{1}{2\mu^{3/2}}\sigma^2 \leq \EE[V^{1/2}] &\leq \mu^{1/2}, \textrm{ and }\\
\mu^{3/2}+\frac{3}{8\mu^{1/2}}\sigma^2-\frac{1}{16\mu^{3/2}}\kappa\leq \EE[V^{3/2}]&\leq \mu^{3/2}+\frac{1}{2\mu^{1/2}}\sigma^2, \textrm{ and }\\ 
\mu^{5/2}+\frac{3\mu^{1/2}}{2}\sigma^2\leq \EE[V^{5/2}]&\leq \mu^{5/2}+\frac{15\mu^{1/2}}{8}\sigma^2+\frac{15}{48\mu^{1/2}}\kappa.
\end{align*}
For $x\geq1$, 
\begin{align*}
(x+1)^{-1/2}\leq x^{-1/2}&\leq (x+1)^{-1/2}+\frac{2}{(x+1)\sqrt{x+2}}\\
(x+1)^{-1}\leq x^{-1}&\leq (x+1)^{-1}+\frac{3}{(x+1)(x+2)}
\end{align*}

\end{lemma}

\begin{theorem}[Moments of the Damped Binomial]
\label{theorem:dampedbinomial}
Suppose $m\sim\Theta(n^2)$, and let\\ $Z\sim\mathrm{DampedBinomial}(n,m,p)$. Then with $(n)_k=n(n-1)\cdots(n-k+1)$,
\begin{align*}
\EE[Z] & = \frac{n}{m+n}\sqrt{(m+n)p}+O(1/n^2),\\
\EE[Z^2]&=\frac{(n)_2}{(m+n)_2}\left[(m+n)p+\frac{m}{n-1}\right]+O(1/n^2),\\
\EE[Z^3]&=\frac{(n)_3}{(m+n)_3}\sqrt{(m+n)p}\left[(m+n)p+\frac{3m}{n-2}\right]+O(1/n^2),\\
\EE[Z^4]&=\frac{(n)_4}{(m+n)_4}(m+n)p\left[(m+n)p+\frac{6m}{n-3}\right]+O(1/n^2).
\end{align*}
Moreover,
\begin{align*}
\mathrm{Var}(Z)&=\frac{nm(1-p)}{(m+n)_2}+O(1/n^2),\\
\mathrm{Var}((Z-\EE[Z])^2)&=O(1/n^2).
\end{align*}
\end{theorem}

\begin{proof}
\noindent\textbf{Analysis of the first moment $\EE[Z]$:}
\begin{align*}
\EE[Z]&=\sum_{j=0}^{n}\sum_{k=0}^{m}\binom{n}{j}\binom{m}{k}\frac{j}{\sqrt{j+k}}p^{j+k}(1-p)^{m+n-j-k}\\
&=\sum_{t=1}^{m+n}\sum_{j=1}^{\min\{n,t\}}\binom{n}{j}\binom{m}{t-j}\frac{j}{\sqrt{t}}p^{t}(1-p)^{m+n-t}\\
&=\sum_{t=1}^{m+n}n\left[\sum_{j=1}^{\min\{n,t\}}\binom{n-1}{j-1}\binom{m}{t-j}\right]\frac{1}{\sqrt{t}}p^{t}(1-p)^{m+n-t}\\
&=\sum_{t=1}^{m+n}n\binom{m+n-1}{t-1}\frac{1}{\sqrt{t}}p^t(1-p)^{m+n-t}\\
&=\sum_{t=1}^{m+n}\frac{n}{m+n}\binom{m+n}{t}\sqrt{t}p^t(1-p)^{m+n-t}\\
&=\frac{n}{m+n}\sum_{t=0}^{m+n}\binom{m+n}{t}\sqrt{t}p^t(1-p)^{m+n-t}.
\end{align*}
This last expression gives $\EE[Z]=\frac{n}{m+n}\EE[T^{1/2}],$ where $T\sim\mathrm{Binomial}(m+n,p)$. Applying the bounds of Lemma~\ref{lem:rootbounds} yields $$\frac{n}{m+n}\left[\sqrt{(m+n)p}-\frac{(m+n)p(1-p)}{2((m+n)p)^{3/2}}\right]\leq \EE[Z]\leq \frac{n}{m+n}\sqrt{(m+n)p}.$$ The lower bound may be simplified to $$\frac{n}{m+n}\sqrt{(m+n)p}-\frac{n(1-p)}{2(m+n)^{3/2}\sqrt{p}}\leq \EE[Z] \leq \frac{n}{m+n}\sqrt{(m+n)p},$$ so the discrepancy between $\EE[Z]$ and its leading term when $m\sim\Theta(n^2)$ is $O(1/n^2)$.\\

\noindent\textbf{Analysis of the second moment $\EE[Z^2]$:}
We proceed similarly to $\EE[Z]$, using $j^2=(j)_2+j$.
\begingroup
\allowdisplaybreaks
\begin{align*}
\EE&[Z^2]=\sum_{j=0}^{n}\sum_{k=0}^{m}\binom{n}{j}\binom{m}{k}\frac{j^2}{j+k}p^{j+k}(1-p)^{m+n-j-k}\\
&=\sum_{t=1}^{m+n}\sum_{j=1}^{\min\{n,t\}}\binom{n}{j}\binom{m}{t-j}\frac{j^2}{t}p^t(1-p)^{m+n-t}\\
&=\sum_{t=2}^{m+n}\sum_{j=2}^{\min\{n,t\}}(n)_2\binom{n-2}{j-2}\binom{m}{t-j}\frac{1}{t}p^t(1-p)^{m+n-t}\\
&\quad+\sum_{t=1}^{m+n}\sum_{j=1}^{\min\{n,t\}}n\binom{n-1}{j-1}\binom{m}{t-j}\frac{1}{t}p^t(1-p)^{m+n-t}\\
&=\sum_{t=2}^{m+n}(n)_2\binom{m+n-2}{t-2}\frac{1}{t}p^t(1-p)^{m+n-t}+\sum_{t=1}^{m+n}n\binom{m+n-1}{t-1}\frac{1}{t}p^t(1-p)^{m+n-t}\\
&=\frac{(n)_2}{(m+n)_2}\sum_{t=2}^{m+n}\binom{m+n}{t}(t-1)p^t(1-p)^{m+n-t}+\frac{n}{m+n}\sum_{t=1}^{m+n}\binom{m+n}{t}p^t(1-p)^{m+n-t}\\
&=\frac{n}{m+n}\sum_{t=1}^{m+n}\binom{m+n}{t}\left(\frac{n-1}{m+n-1}t+\frac{m}{m+n-1}\right)p^t(1-p)^{m+n-t}\\
&=\frac{(n)_2}{(m+n)_2}(m+n)p+\frac{nm}{(m+n)_2}(1-(1-p)^{m+n}).
\end{align*}
\endgroup
In the penultimate step, we used the fact that when $t=1$, $t-1=0$, allowing us to add the $t=1$ term and bring the two sums together. In the final step, we use properties of the $\mathrm{Binomial}(m+n,p)$ distribution. This gives the desired bounds for $\EE[Z^2]$ immediately.\\

\noindent\textbf{Analysis of $\mathrm{Var}(Z)$:}
Applying the lower bound for $\EE[Z]$ and using $1-(1-p)^{m+n}\leq 1$, we obtain
\begin{align*}
\mathrm{Var}(Z)&\leq \frac{(n)_2}{(m+n)_2}(m+n)p+\frac{nm}{(m+n)_2}-\left(\frac{n}{m+n}\sqrt{(m+n)p}-\frac{n(1-p)}{2(m+n)^{3/2}\sqrt{p}}\right)^2\\
&=\frac{nm(1-p)}{(m+n)_2}+\frac{n^2(1-p)}{(m+n)^2}-\frac{n^2(1-p)^2}{4(m+n)^3p}\\
&=\frac{nm(1-p)}{(m+n)_2}+\frac{n^2(1-p)}{(m+n)^2}\left(1-\frac{1-p}{4(m+n)p}\right).
\end{align*}
The lower bound is given by
\begin{align*}
\mathrm{Var}(Z)&\geq \frac{(n)_2}{(m+n)_2}(m+n)p+\frac{nm}{(m+n)_2}(1-(1-p)^{m+n})-\left(\frac{n}{m+n}\sqrt{(m+n)p}\right)^2\\
&=\frac{nm(1-p)}{(m+n)_2}(1-(1-p)^{m+n-1}),
\end{align*}
whence the stated bounds follow since $(1-p)^{m+n}\sim o(1/n)$.\\

\noindent\textbf{Analysis of $\EE[Z^3]$:}
We use the formula $j^3=(j)_3+3(j)_2+j$ as in the calculation of $\EE[Z^2]$ to obtain:
\begingroup
\allowdisplaybreaks
\begin{align*}
\EE[Z^3]&=\sum_{j=0}^{n}\sum_{k=0}^{m}\binom{n}{j}\binom{m}{k}\frac{j^3}{(j+k)^{3/2}}p^{j+k}(1-p)^{m+n-j-k}\\
&=\sum_{t=1}^{m+n}\sum_{j=1}^{\min\{n,t\}}\binom{n}{j}\binom{m}{t-j}\frac{j^3}{t^{3/2}}p^t(1-p)^{m+n-t}\\
&=\sum_{t=3}^{m+n}(n)_3\binom{m+n-3}{t-3}\frac{1}{t^{3/2}}p^t(1-p)^{m+n-t}\\
&\quad+\sum_{t=2}^{m+n}3(n)_2\binom{m+n-2}{t-2}\frac{1}{t^{3/2}}p^t(1-p)^{m+n-t}\\
&\quad+\sum_{t=1}^{m+n}n\binom{m+n-1}{t-1}\frac{1}{t^{3/2}}p^t(1-p)^{m+n-t}\\
&=\sum_{t=1}^{m+n}\binom{m+n}{t}\left[\frac{(n)_3(t-1)(t-2)}{(m+n)_3}+\frac{3(n)_2(t-1)}{(m+n)_2}+\frac{n}{m+n}\right]\frac{1}{\sqrt{t}}p^t(1-p)^{m+n-t}\\
&=\frac{n}{(m+n)_3}\sum_{t=1}^{m+n}\binom{m+n}{t}[(n-1)_2t^2+3tm(n-1)+m(m-n)]\frac{1}{\sqrt{t}}p^t(1-p)^{m+n-t}\\
&=\frac{n}{(m+n)_3}\sum_{t=0}^{m+n}\binom{m+n}{t}[(n-1)_2 t^{3/2}+ 3m(n-1)t^{1/2}]p^t(1-p)^{m+n-t} \tag{$L_1$}\\
&\quad+\frac{n}{(m+n)_3}\sum_{t=1}^{m+n}\binom{m+n}{t}\frac{m(m-n)}{\sqrt{t}}p^t(1-p)^{m+n-t}. \tag{$L_2$}
\end{align*}
\endgroup
We bound the first line $L_1$ using Lemma~\ref{lem:rootbounds} for $\EE[T^{3/2}]$ and $\EE[T^{1/2}]$ with $T\sim\mathrm{Binomial}(m+n,p)$. After simplifying, this yields
\begin{align*}
\frac{(n)_3}{(m+n)_3}&\left[\sqrt{(m+n)p}\left(\frac{3(1-p)}{8}\right)-\frac{1-p}{\sqrt{(m+n)p}}\left[\frac{3m}{2(n-2)}+\frac{1-2p}{16}\right]\right]\\
&\leq L_1-\frac{(n)_3}{(m+n)_3}\left[((m+n)p)^{3/2}+\sqrt{(m+n)p}\frac{3m}{n-2}\right]\\
&\leq \frac{(n)_3}{(m+n)_3}\sqrt{(m+n)p}\left(\frac{1-p}{2}\right).
\end{align*}
We note that the upper and lower bounds are both $O(1/n^2)$. The line $L_2$ is bounded by applying the inequalities for $x^{-1/2}$ in Lemma~\ref{lem:rootbounds}, giving
\begin{align*}
L_2&\geq \frac{nm(m-n)}{(m+n+1)_4}\sum_{t=1}^{m+n}\binom{m+n+1}{t+1}\sqrt{t+1}p^t(1-p)^{m+n-t}\\
&=\frac{nm(m-n)}{(m+n+1)_4}\left[\sum_{t=0}^{m+n+1}\binom{m+n+1}{t}\sqrt{t}p^{t-1}(1-p)^{m+n+1-t}-(m+n+1)(1-p)^{m+n}\right]\\
&\geq \frac{nm(m-n)}{(m+n+1)_4 p} \left[\sqrt{(m+n+1)p}-\frac{(m+n+1)p(1-p)}{2((m+n+1)p)^{3/2}}\right]-\frac{nm(m-n)}{(m+n)_3}(1-p)^{m+n}
\end{align*}
and
\begin{align*}
L_2&\leq \frac{nm(m-n)}{(m+n+1)_4}\sum_{t=1}^{m+n}\binom{m+n+1}{t+1}\sqrt{t+1}p^t(1-p)^{m+n-t}\\
&\quad+ \frac{2nm(m-n)}{(m+n+2)_5}\sum_{t=1}^{m+n}\binom{m+n+2}{t+2}\sqrt{t+2}p^t(1-p)^{m+n-t}\\
&\leq \frac{nm(m-n)}{(m+n+1)_4}\sum_{t=0}^{m+n+1}\binom{m+n+1}{t}\sqrt{t}p^{t-1}(1-p)^{m+n+1-t}\\
&\quad+\frac{2nm(m-n)}{(m+n+2)_5}\sum_{t=0}^{m+n+2}\binom{m+n+2}{t}\sqrt{t} p^{t-2}(1-p)^{m+n+2-t}\\
&\leq \frac{nm(m-n)}{(m+n+1)_4 p}\sqrt{(m+n+1)p}+\frac{2nm(m-n)}{(m+n+2)_5 p^2}\sqrt{(m+n+2)p}.
\end{align*}
Together, these bounds show that $L_2=\Theta(1/n^2),$ so combining the bounds on $L_1$ and $L_2$, we obtain the desired bound on $\EE[Z^3]$.\\

\noindent\textbf{Analysis of $\EE[Z^4]$:} Here we use the formula $j^4=(j)_4+6(j)_3+7(j)_2+j$ as in the analysis of $\EE[Z^2]$ to obtain the following calculation:
\begin{align*}
\EE[Z^4]&=\sum_{j=0}^{n}\sum_{k=0}^{m}\binom{n}{j}\binom{m}{k}\frac{j^4}{(j+k)^2}p^{j+k}(1-p)^{m+n-j-k}\\
&=\sum_{t=1}^{m+n}\sum_{j=1}^{\min\{n,t\}}\binom{n}{j}\binom{m}{t-j}\frac{j^4}{t^2}p^t(1-p)^{m+n-t}\\
&=\sum_{t=4}^{m+n}(n)_4\binom{m+n-4}{t-4}\frac{p^t(1-p)^{m+n-t}}{t^2}+\sum_{t=3}^{m+n}6(n)_3\binom{m+n-3}{t-3}\frac{p^t(1-p)^{m+n-t}}{t^2}\\
&\quad+\sum_{t=2}^{m+n}7(n)_2\binom{m+n-2}{t-2}\frac{p^t(1-p)^{m+n-t}}{t^2}+\sum_{t=1}^{m+n}n\binom{m+n-1}{t-1}\frac{p^t(1-p)^{m+n-t}}{t^2}
\end{align*}
This gives
\begin{align*}
\EE[Z^4]&=\sum_{t=1}^{m+n}\binom{m+n}{t}\frac{p^t(1-p)^{m+n-t}}{t}\\
&\times\left[\frac{(n)_4}{(m+n)_4}(t-1)_3+\frac{6(n)_3}{(m+n)_3}(t-1)_2+\frac{7(n)_2}{(m+n)_2}(t-1)+\frac{n}{m+n}\right]\\
&=\frac{(n)_4}{(m+n)_4}\sum_{t=1}^{m+n}\binom{m+n}{t}\left[t^3+\frac{6m}{n-3}t^2+\frac{m(7m-4n+1)}{(n-2)_2}t\right]\frac{p^t(1-p)^{m+n-t}}{t}\tag{$L_1$}\\
&\quad+\frac{n}{(m+n)_4}\sum_{t=1}^{m+n}\binom{m+n}{t}m((m+n+1)_2-6mn)\frac{p^t(1-p)^{m+n-t}}{t}\tag{$L_2$}
\end{align*}

We can use the moments of the $\mathrm{Binomial}(m+n,p)$ distribution to compute $L_1$ as
$$\frac{(n)_4}{(m+n)_4}\left[(m+n)p\left((m+n)p+\frac{6m}{n-3}+1-p\right)\right]+\frac{(n)_2 m(7m-4n+1)}{(m+n)_4}(1-(1-p)^{m+n}),$$ the last term of which is $O(1/n^2)$. We apply the bounds on $x^{-1}$ from Lemma~\ref{lem:rootbounds} to bound $L_2$, using a similar approach to remove terms from the denominator as we did in the analysis of the $L_2$ term in $\EE[Z^3]$.
\begin{align*}
L_2&\geq \frac{nm((m+n+1)_2-6mn)}{(m+n+1)_5}\sum_{t=1}^{m+n}\binom{m+n+1}{t+1}p^t(1-p)^{m+n-t}\\
&=\frac{nm((m+n+1)_2-6mn)}{(m+n+1)_5 p}\PP[T_1\geq 2]=:\mathrm{LB},
\end{align*}
where $T_1\sim\mathrm{Binomial}(m+n+1,p)$. Note that this quantity is $O(1/n^3)$.
\begin{align*}
L_2-\mathrm{LB}&\leq \frac{3nm((m+n+1)_2-6mn)}{(m+n+2)_6}\sum_{t=1}^{m+n}\binom{m+n+2}{t+2}p^t(1-p)^{m+n-t}\\
&=\frac{3nm((m+n+1)_2-6mn)}{(m+n+2)_6 p^2}\PP[T_2\geq 3],
\end{align*}
where $T_2\sim\mathrm{Binomial}(m+n+2,p)$. This quantity is $O(1/n^5)$, which shows that $L_2=O(1/n^3)$, and completes the analysis of this term.\\

\noindent\textbf{Analysis of $\mathrm{Var}((Z-\EE[Z])^2)$:} The following formula is valid for any random variable $Z$, as can be verified with simple algebra:
$$\mathrm{Var}((Z-\EE[Z])^2)=\mathrm{Var}(Z^2)+4\EE[Z]^2\mathrm{Var}(Z)-4\EE[Z](\EE[Z^3]-\EE[Z^2]\EE[Z]).$$
We analyze these three terms separately using the results above. Noting that $(C+O(1/n^2))^2=C^2+O(1/n^2)$,
\begingroup
\allowdisplaybreaks
\begin{align*}
\phantom{=}&\mathrm{Var}(Z^2)=\EE[Z^4]-\EE[Z^2]^2\\
=&\frac{(n)_4}{(m+n)_4}\left[((m+n)p)^2+(m+n)p\frac{6m}{n-3}\right]-\left(\frac{(n)_2}{(m+n)_2}\left[(m+n)p+\frac{m}{n-1}\right]\right)^2+O(n^{-2})\\
=&\frac{(n)_4}{(m+n)_4}\left[((m+n)p)^2+(m+n)p\frac{6m}{n-3}-\frac{(n)_2(m+n-2)_2}{(n-2)_2(m+n)_2}\left[(m+n)p+\frac{m}{n-1}\right]^2\right]\\
&\quad+O(n^{-2})\\
=&\frac{(n)_4}{(m+n)_4}\left[((m+n)p)^2\left(1-\frac{(n)_2(m+n-2)_2}{(n-2)_2(m+n)_2}\right)\right.\\
&\qquad\qquad\qquad+\left.2(m+n)mp\left(\frac{3}{n-3}-\frac{n(m+n-2)_2}{(n-2)_2(m+n)_2}\right)\right]+O(1/n^2)\\
=&\frac{(n)_4}{(m+n)_4}\left[((m+n)p)^2\frac{-4m^2n}{(n-2)_2(m+n)_2}+4(m+n)mp\frac{m^2n}{(n-2)_2(m+n)_2}\right]+O(1/n^2)\\
=&\frac{(n)_4}{(m+n)_4}\left[4(m+n)mp\frac{m^2n(1-p)}{(n-2)_2(m+n)_2}\right]+O(1/n^2),
\end{align*}
\endgroup
where we discarded any higher-order terms into the $O(1/n^2)$ in the last two lines. This is $O(1/n)$. Since $\EE[Z]=O(1)$ and $\mathrm{Var}(Z)=O(1/n)$, $4\EE[Z]^2\mathrm{Var}(Z)=O(1/n)$. Finally,
\begingroup
\allowdisplaybreaks
\begin{align*}
\EE&[Z^3]-\EE[Z^2]\EE[Z]\\
&=\frac{(n)_3}{(m+n)_3}\sqrt{(m+n)p}\left[(m+n)p+\frac{3m}{n-2}\right]\\
&\quad-\frac{(n)_2}{(m+n)_2}\left[(m+n)p+\frac{m}{n-1}\right]\frac{n}{m+n}\sqrt{(m+n)p}+O(1/n^2)\\
&=\frac{(n)_3}{(m+n)_3}\left[((m+n)p)^{3/2}\left(1-\frac{n(m+n-2)}{(n-2)(m+n)}\right)\right.\\
&\qquad\qquad\qquad+\left.m\sqrt{(m+n)p}\left(\frac{3}{n-2}-\frac{n(m+n-2)}{(n-1)_2(m+n)}\right)\right]+O(1/n^2)\\
&=\frac{(n)_3}{(m+n)_3}\left[((m+n)p)^{3/2}\frac{-2m}{(n-2)(m+n)}+m\sqrt{(m+n)p}\frac{2mn}{(n-1)_2(m+n)}\right]+O(n^{-2})\\
&=\frac{(n)_3}{(m+n)_3}\left[m\sqrt{(m+n)p}\frac{2mn(1-p)}{(n-1)_2(m+n)}\right]+O(1/n^2),
\end{align*}
\endgroup
where we again discarded higher-order terms in the last two lines. This is $O(1/n)$. Inspecting the leading terms and using $(n^k+O(n^{k-1}))^{-1}=n^{-k}(1+O(1/n))$, we see that
\begin{align*}
\mathrm{Var}((Z-\EE[Z])^2)&=\frac{4m^3n^3p(1-p)}{m^5}+\frac{4mn^3p(1-p)}{m^3}-\frac{8m^2n^3p(1-p)}{m^4}+O(1/n^2)\\
&=O(1/n^2).
\end{align*}
\end{proof}

\begin{theorem}[Limiting distribution of the Damped Binomial]
\label{theorem:limitdist}
Let\\ $Z_n\sim\mathrm{DampedBinomial}(n,m,p),$ and suppose $m/n^2 \rightarrow \alpha>0$. Then $Z_n$ converges in distribution to a Normal random variable as $n\rightarrow\infty$: 
$$\sqrt{n}\;\frac{Z_n-\frac{n}{\sqrt{m}}\sqrt{p}}{\sqrt{(1-p)/\alpha}}\rightarrow \mathcal{N}(0,1).$$
If $(m-\alpha n^2)\sim o(n^{3/2})$, then 
$$\sqrt{n}\;\frac{Z_n-\sqrt{p/\alpha}}{\sqrt{(1-p)/\alpha}}\rightarrow \mathcal{N}(0,1).$$
\end{theorem}

\begin{proof}
From the definition of the DampedBinomial distribution, we may write $$Z_n = \frac{X_n}{\sqrt{X_n+Y_n}},$$ where $X_n\sim\mathrm{Binomial}(n,p)$, $Y_n\sim \mathrm{Binomial}(m,p)$ are independent. Let $\mathcal{E}_n$ be the event that $|Y_n-mp|\leq n\log(n)$. Clearly $\PP[\mathcal{E}_n]\rightarrow 1$ as $n\rightarrow\infty$. Moreover, with probability 1 for each $n$, we have
$$Y_n\leq Y_n+X_n\leq Y_n+n.$$
Then 
$$ \frac{X_n}{\sqrt{mp}\sqrt{1+\frac{Y_n-mp+n}{mp}}} \leq Z_n \leq \frac{X_n}{\sqrt{mp}\sqrt{1+\frac{Y_n-mp}{mp}}},$$ and on the set $\mathcal{E}_n$,
$$ \frac{X_n}{\sqrt{mp}\sqrt{1+\frac{n(\log(n)+1)}{mp}}}\leq Z_n\leq \frac{X_n}{\sqrt{mp}\sqrt{1-\frac{n\log(n)}{mp}}}.$$ Setting $x=\frac{n(\log(n)+1)}{mp}\sim O(\log(n)/n)$, we apply the bounds
$$ 1-x/2 \leq \frac{1}{\sqrt{1+x}} \leq 1-x/2+ 3x^2/4,\quad x\geq 0;$$
$$ 1+x/2 \leq \frac{1}{\sqrt{1-x}} \leq 1+x/2+3x^2/4,\quad x\in[0,1/2];$$
to see that 
$$\frac{1}{\sqrt{1+\frac{n(\log(n)+1)}{mp}}}=1-\frac{n(\log(n)+1)}{2mp}+O((\log(n)/n)^{2}),$$ and
$$\frac{1}{\sqrt{1-\frac{n\log(n)}{mp}}}= 1+\frac{n\log(n)}{2mp}+O((\log(n)/n)^{2}).$$

Since $X_n/ \sqrt{mp}$ is bounded above by a constant, the previous bounds show that $Z_n - \frac{X_n}{\sqrt{mp}} \sim O(log(n)/n)$. So by Slutzky's theorem, if $\sqrt{n}(X_n/\sqrt{mp}-c_n)\rightarrow F$ in distribution as $n\rightarrow\infty$, then $\sqrt{n}(Z_n-c_n)$ also converges to $F$. But since $X\sim\mathrm{Binomial}(n,p)$, we may apply the ordinary central limit theorem to say that 
$$\frac{X_n-np}{\sqrt{np(1-p)}}\rightarrow \mathcal{N}(0,1). $$ Using $m/n^2\rightarrow \alpha$, we get
$$ \sqrt{n}\frac{X_n-np}{\sqrt{mp}}\rightarrow \mathcal{N}\left(0,\frac{1-p}{\alpha}\right).$$ Dividing by $\sqrt{(1-p)/\alpha}$ gives the first conclusion of the theorem.

When $(m-\alpha n^2)\sim o(n^{3/2}),$ then we may write $n/\sqrt{m}= \frac{n}{\sqrt{n^2\alpha}\sqrt{1+ \frac{m-n^2\alpha}{n^2\alpha}}}.$ By virtue of the bounds on $(1+x)^{-1/2}$ and $(1-x)^{-1/2}$ above, 
$\frac{n}{\sqrt{m}}-\frac{1}{\sqrt{\alpha}}\sim o(n^{-1/2}),$ so we may apply Slutzky's theorem a second time to prove the latter bound.
\end{proof}
\end{document}